\documentclass[a4paper,12pt,twoside]{report}

\usepackage[ngerman,english]{babel}
\usepackage[T1]{fontenc} 
\usepackage[latin1]{inputenc}
\usepackage{ae,aecompl}

\usepackage[intlimits]{amsmath}
\usepackage{amssymb}

\usepackage{fancyhdr}
\usepackage{booktabs}
\usepackage[bf,small]{caption}
\usepackage[dvips]{graphicx}
\usepackage[dvips]{color}

\usepackage{axodraw}
\usepackage{pstricks}

\usepackage{mathrsfs}
\usepackage{slashed}

\usepackage[dvips]{hyperref}

\title{
\vspace*{-3.95cm}
{\normalsize\normalfont
DESY-THESIS-2008-043\hfill\mbox{}\\
December 2008\hfill\mbox{}\\}
\vspace{2.95cm}
\textbf{Neutrino Signals from Gravitino Dark Matter with Broken \textit{R}-Parity} \vspace{10mm}}

\author{\textsc{Diplomarbeit}\\[1.5mm] zur Erlangung des akademischen Grades\\[1.5mm] \textsc{Diplom-Physiker}\\[1.5mm] des Departments Physik der Universität Hamburg\\[13mm] durchgeführt unter der Betreuung von\\[1.5mm] Dr. Laura Covi\\[1.5mm] in der Theory Group am DESY, Hamburg\\[13mm] vorgelegt von\\[1.5mm] \textsc{Michael Grefe}\\[1.5mm] aus Lüneburg\\[20mm] Hamburg\\[1.5mm] September 2008}
 
\date{}

\definecolor{bordeauxred}{rgb}{0.4,0,0}

\hypersetup{
 linktocpage=true,
 bookmarksnumbered=true,
 pdftitle = {Neutrino Signals from Gravitino Dark Matter with Broken R-Parity},
 pdfauthor = {Michael Grefe},
 pdfsubject = {Gravitino Dark Matter},
 pdfkeywords = {Gravitino, Dark Matter, Neutrino, R-Parity Violation},
 colorlinks=true,
 linkcolor=bordeauxred,
 citecolor=bordeauxred,
 urlcolor=bordeauxred
}

\numberwithin{equation}{chapter}
\numberwithin{figure}{chapter}
\numberwithin{table}{chapter}

\providecommand{\unit}[1]{\ensuremath{\mathrm{#1}}}
\providecommand{\usk}{\ensuremath{\,}}
\providecommand{\abs}[1]{\ensuremath{\left\lvert #1\right\rvert }}
\providecommand{\MP}{\ensuremath{M_{\text{Pl}}}}
\DeclareMathOperator{\BR}{BR}
\DeclareMathOperator{\Tr}{Tr}

\begin{document}

\maketitle

\newpage
\thispagestyle{empty}
\section*{}
\vfill
\subsubsection*{Gutachter der Diplomarbeit} 
Erstgutachter: Dr. Laura Covi \\
Zweitgutachter: Prof. Dr. Jan Louis 
\newpage

\begin{minipage}[t]{\linewidth}
 \begin{abstract}
 The gravitino is a promising supersymmetric dark matter candidate, even without strict $ R $-parity conservation. In fact, with some small $ R $-parity violation, gravitinos are sufficiently long-lived to constitute the dark matter of the universe, while the resulting cosmological scenario is consistent with primordial nucleosynthesis and the high reheating temperature needed for thermal leptogenesis. Furthermore, in this scenario the gravitino is unstable and might thus be accessible by indirect detection via its decay products. \bigskip

 We compute in this thesis the partial decay widths for the gravitino in models with bilinear $ R $-parity breaking. In addition, we determine the neutrino signal from astrophysical gravitino dark matter decays. Finally, we discuss the feasibility of detecting these neutrino signals in present and future neutrino experiments, and conclude that it will be a challenging task. Albeit, if detected, this distinctive signal might bring considerable support to the scenario of decaying gravitino dark matter. 
\end{abstract}
 \vspace{-12mm}
 {\selectlanguage{ngerman}
 \begin{abstract}
 Das Gravitino ist auch ohne strikte $ R $-Parität-Erhaltung ein vielversprechender Kandidat für supersymmetrische dunkle Materie. Tatsächlich sind Gravitinos bei leichter $ R $-Parität-Verletzung genügend langlebig um die dunkle Materie im Universum darzustellen, während das daraus resultierende kosmologische Szenario mit primordialer Nukleosynthese und der hohen Reheating-Temperatur, die für thermische Leptogenese benötigt wird, vereinbar ist. Darüber hinaus ist das Gravitino in diesem Szenario instabil und könnte daher durch seine Zerfallsprodukte für eine indirekte Entdeckung zugänglich sein. \bigskip

 In dieser Diplomarbeit berechnen wir die partiellen Zerfallsbreiten des Gravitinos in Modellen mit bilinearer $ R $-Parität-Brechung. Zudem bestimmen wir das Neutrinosignal astrophysikalischer Zerfälle von dunkler Materie, die aus Gravitinos besteht. Schließlich diskutieren wir die Möglichkeit diese Neutrinosignale in heutigen und zukünftigen Neutrinoexperimenten zu beobachten und kommen zu dem Schluss, dass es eine sehr herausfordernde Aufgabe ist. Dessen ungeachtet könnte dieses charakteristische Signal eine deutliche Unterstützung für das Szenario zerfallender Gravitinos als dunkler Materie darstellen. 
\end{abstract}
 }
\end{minipage}

\newpage
\thispagestyle{empty}
\section*{}
\newpage

\pagenumbering{roman} 

\phantomsection 
\addcontentsline{toc}{chapter}{Contents}
\tableofcontents

\newpage

\phantomsection 
\addcontentsline{toc}{chapter}{List of Figures}
\listoffigures

\newpage

\phantomsection 
\addcontentsline{toc}{chapter}{List of Tables}
\listoftables

\newpage 

{\thispagestyle{empty}
\cleardoublepage}

\pagenumbering{arabic} 

\phantomsection
\addcontentsline{toc}{chapter}{Preface}
\chapter*{Preface}
\markboth{\textsc{Preface}}{\textsc{Preface}}

The nature of the dark matter (DM) in the universe is still one of the unsolved mysteries in modern cosmology. Its existence is strong evidence for physics beyond the Standard Model, but until now the only indication for dark matter is based on its gravitational interaction. 

One of the most favored solutions to the dark matter problem is the introduction of supersymmetry with conserved $ R $-parity. In this framework, the lightest supersymmetric particle (LSP) is stable and therefore a natural DM candidate, if it is also neutral and only weakly interacting. The most extensively studied case is where the LSPs are Weakly Interacting Massive Particles (WIMPs) and the dominant contribution to the dark matter in the universe. One of the most interesting candidates is, in fact, the lightest neutralino, since it can be tested in accelerator, direct detection and indirect detection experiments in the near future~\cite{Bertone:2004pz}. 

On the other hand, it is also possible that dark matter only interacts gravitationally, and supersymmetry offers also candidates of this type. In local supersymmetry (i.e. supergravity) the gravitino arises as a natural candidate for dark matter. It is the superpartner of the graviton and has been the first supersymmetric dark matter candidate proposed~\cite{Pagels:1981ke}. The gravitino is also one of the most elusive dark matter candidates, since, as part of the gravity multiplet, all its interactions are suppressed either by the Planck scale (for the spin-3/2 component) or by the supersymmetry breaking scale (for the goldstino component). Nonetheless, a high reheating temperature after the inflationary phase in the early universe leads via thermal gravitino production to a relic abundance that can account for the observed dark matter density in the universe and is compatible with baryogenesis via thermal leptogenesis. 

Typically, the extremely weak interactions of the gravitino cause severe trouble for the standard cosmological scenario, since they make either the gravitino---if it is not the LSP---or the next-to-lightest supersymmetric particle (NLSP) so long-lived that it decays during or after Big Bang Nucleosynthesis (BBN). Therefore, the successful prediction of the light element abundances can be spoiled (see for instance~\cite{Sarkar:1995dd}). 

However, it was recently proposed that these problems are easily circumvented if a small breaking of $ R $-parity is introduced~\cite{Buchmuller:2007ui}. In that case, the neutralino is too short-lived to play the role of dark matter, whereas the gravitino---if it is the LSP---can still have a sufficiently long lifetime, even exceeding the age of the universe by many orders of magnitude. Since the NLSP decays into Standard Model particles via $ R $-parity violating couplings, the cosmology reduces to the standard, non-supersymmetric case well before the onset of BBN. 

In collider experiments, late decays of the NLSP can be a signature of this gravitino LSP scenario. Another advantage of models with $ R $-parity violation is the fact that gravitino dark matter is no more invisible in astrophysical experiments: Even if the strongly suppressed couplings rule out an observation in direct dark matter detection experiments, the tremendous number of gravitinos in the Milky Way and nearby galaxies compensates for the suppressed decay rate, possibly making gravitino decay products visible in indirect detection experiments. 

In fact, it has been found in~\cite{Ibarra:2007wg, Ibarra:2008qg, Ishiwata:2008cu} that gravitino dark matter with a lifetime of $ \sim 10^{26} $\usk s and a mass of $ \sim 150 $\usk GeV can account simultaneously for the anomalies in the extragalactic gamma-ray flux, suggested by the Energetic Gamma Ray Experiment Telescope (EGRET)~\cite{Strong:2004ry}, and in the positron fraction, suggested by the High Energy Antimatter Telescope (HEAT) balloon experiment~\cite{Barwick:1997ig}. \smallskip

Motivated by this coincidence, we compute in this thesis the neutrino signal in the same scenario, both as a consistency check and to find out whether in this scenario an anomalous contribution to the neutrino flux may be expected in present and future neutrino experiments. While such signal is certainly much more difficult to disentangle from the background, it would provide significant support to the scenario of decaying dark matter,
possibly consisting of gravitinos that are unstable due to bilinear $ R $-parity violation. \smallskip

The thesis is organized as follows: In the first chapter we will briefly review Big Bang cosmology and the astrophysical evidence for dark matter in the universe. In Chapter~\ref{susygravitino} we will give a short introduction to supersymmetry and supergravity, and discuss more thoroughly the gravitino field, its interactions and its consequences  for cosmology. In Chapter~\ref{model} we will introduce models with bilinear $ R $-parity breaking, discuss in detail the decay channels of the LSP gravitino in this framework and present the resulting neutrino spectrum from gravitino decay. 

In Chapter~\ref{neutrino} we will start with the computation of the neutrino flux from extragalactic sources and from our own galaxy, considering also the effect of neutrino oscillations on the signal expected at the Earth. Then we will discuss the different neutrino backgrounds in the relevant energy range, compare them to our signal and propose strategies to disentangle the signal from the background. Subsequently, we will examine the prospects for a detection of the signal from gravitino dark matter decay in Super-Kamiokande and future neutrino detectors. Finally, we will present constraints on the gravitino lifetime from the non-observation of a neutrino signal, and then conclude in the last chapter. 

In Appendix~\ref{constants} we introduce units with $ \hbar=c=k=1 $ and present a list of the physical and astrophysical constants used in the course of the thesis. In Appendix~\ref{notation} we introduce the used notation and conventions. There, we also list numerous formulae used in the course of the thesis and derive some relevant identities. In Appendix~\ref{decaywidths} we present the Feynman rules for gravitino interactions and the detailed computations of the gravitino decay widths. In Appendix~\ref{pythia} we briefly present an example of the PYTHIA program used to simulate the fragmentation of the gravitino decay products. \bigskip

\noindent
The results from this work have been published in~\cite{Covi:2008jy}.  

\chapter{Dark Matter and the Early Universe}
In this chapter we want to review shortly the basic cosmological concepts that are relevant for the problem of dark matter. More comprehensive reviews on this topic can be found for instance in~\cite{Bertone:2004pz, Trodden:2004st}. 

\section{Big Bang Cosmology}
\label{cosmology}
In this section we will first introduce the dynamics of our universe and then highlight several important stages of its thermal history. 

\subsubsection*{Dynamics of the Universe}
The geometry of space-time is determined by the energy content of the universe. This is expressed in Einstein's equations from general relativity 
\begin{equation}
 R_{\mu\nu}-\frac{1}{2}\,g_{\mu\nu}R=8\pi\,G_NT_{\mu\nu}\,. 
\end{equation}
Here, $ R_{\mu\nu} $ and $ R $ are the Ricci tensor and Ricci scalar, respectively, while $ g_{\mu\nu} $ is the space-time metric and $ T_{\mu\nu} $ the energy-momentum tensor. To solve this set of coupled equations we have to assume symmetries of the universe. Measurements of the Cosmic Microwave Background (CMB) show that the universe is highly \textit{isotropic}. In addition, galaxy surveys indicate that the universe is also \textit{homogeneous} on large scales ($ \mathcal{O}(100)\usk\unit{Mpc} $). 

The most general space-time metric compatible with isotropy and homogeneity is the Friedmann--Robertson--Walker metric. The line element reads 
\begin{equation}
 g_{\mu\nu}dx^{\mu}dx^{\nu}=ds^2=dt^2-a^2(t)\left[ \frac{dr^2}{1-kr^2}+r^2\left( d\theta^2+\sin^2\theta\, d\phi^2\right) \right] , 
\end{equation}
where $ a(t) $ is the scale factor, $ r $, $ \theta $ and $ \phi $ are the comoving spatial coordinates and the constant $ k $ characterizes the spatial curvature of the universe: $ k=-1 $ corresponds to an open, $ k=0 $ to a flat and $ k=+1 $ to a closed universe. Substituting $ d\chi\equiv dr/\sqrt{1-kr^2} $, the line element can be written as 
\begin{equation}
 ds^2=dt^2-a^2(t)\left[ d\chi^2+S^2(\chi)\left( d\theta^2+\sin^2\theta\, d\phi^2\right) \right] , 
 \label{FRW}
\end{equation}
where $ S(\chi)=\sinh\chi $ for $ k=-1 $, $ S(\chi)=\chi $ for $ k=0 $ and $ S(\chi)=\sin\chi $ for $ k=+1 $. 

A simplifying assumption---compatible with the above discussed symmetries of the universe---is that the matter and energy content of the universe can be described by a perfect fluid. The energy-momentum tensor for a perfect fluid in its rest frame reads 
\begin{equation}
 T^{\mu}_{\;\;\nu}=
 \begin{pmatrix}
  +\varrho & 0 & 0 & 0 \\
   0 & -p & 0 & 0 \\
   0 & 0 & -p & 0 \\
   0 & 0 & 0 & -p
 \end{pmatrix}. 
\end{equation}

Solving Einstein's equations with the above assumptions results in the Friedmann equation 
\begin{equation}
 \left( \frac{\dot{a}}{a}\right) ^2\equiv H^2=\frac{8\pi\,G_N}{3}\sum_i\varrho_i-\frac{k}{a^2} 
\end{equation}
and the acceleration equation 
\begin{equation}
 \frac{\ddot{a}}{a}=\dot{H}+H^2=-\frac{4\pi\,G_N}{3}\sum_i\left( \varrho_i+3\,p_i\right) . 
\end{equation}
These equations determine the dynamics of the universe. Here we introduced the Hubble parameter $ H=\dot{a}/a $ that characterizes the expansion rate of the universe. From the Friedmann equation and the acceleration equation we can derive the continuity equation 
\begin{equation}
 \dot{\rho}_i+3\,H\left( \varrho_i+p_i\right) =0\,, 
\end{equation}
which is equivalent to the covariant conservation of the energy-momentum tensor. 

There are several kinds of energy content in the universe, characterized by the equation of state 
\begin{equation}
 p_i=w_i\varrho_i\,. 
\end{equation}
Radiation and relativistic particles have $ w_r=1/3 $ whereas pressureless matter or dust (i.e. non-relativistic particles) has $ w_m=0 $. A cosmological constant $ \Lambda $ can be described by an energy component with negative pressure $ w_{\Lambda}=-1 $. 

The energy of photons and other relativistic particles decreases during their propagation through an expanding universe. This is expressed in terms of the redshift parameter $ z $ that is defined to be 
\begin{equation}
 \frac{\lambda_o}{\lambda_e}=\frac{a_0}{a_e}\equiv 1+z\,, 
 \label{redshift}
\end{equation}
where $ \lambda_e $ and $ a_e $ are the wavelength and scale factor at emission, and $ \lambda_o $ and $ a_0 $ are the observed wavelength and the present-day scale factor, respectively. 

From the continuity equation we can derive the dependence of the energy density on the redshift parameter 
\begin{equation}
 \varrho_i(z)=\varrho_i\left( 1+z\right) ^{3\left( 1+w_i\right) }. 
\end{equation}
The energy density of non-relativistic matter decreases with $ \left( 1+z\right) ^3 $ due to the dilution of the number density with the expansion of the universe. By contrast, the energy density of relativistic matter decreases with an additional factor of $ \left( 1+z\right) $ because of the energy redshift in an expanding universe. The cosmological constant is equivalent to an intrinsic energy of the vacuum and is independent of the dynamics of the universe. 

We can rewrite the Friedmann and acceleration equations using the density parameter 
\begin{equation}
 \Omega_i(z)=\frac{\varrho_i(z)}{\varrho_c(z)} 
\end{equation}
that gives the energy density with respect to the critical density $ \varrho_c(z)=3H^2/(8\pi\,G_N) $, that corresponds to a spatially flat universe. The present-day critical density is given by 
\begin{equation}
 \varrho_c=\frac{3H_0^2}{8\pi\,G_N}\simeq 1.05\times 10^{-5}\usk h^2\usk\unit{GeV}\usk\unit{cm}^{-3}, 
\end{equation}
 where the Hubble constant $ H_0 $ is the present-day Hubble parameter and is usually parameterized as 
\begin{equation}
 H_0=100\usk h\usk\unit{km}\usk\unit{s}^{-1}\usk\unit{Mpc}^{-1}, 
\end{equation}
with $ h\simeq 0.7 $. With the above definitions the Friedmann equation and the acceleration equation can be rewritten as 
\begin{equation}
 \begin{split}
  1 &=\Omega_{\text{tot}}(z)-\frac{k}{a^2H^2}\equiv\Omega_{r}(z)+\Omega_m(z)+\Omega_{\Lambda}(z)-\frac{k}{a^2H^2} \\
  &=\frac{H_0^2}{H^2}\left[ \Omega_{r}\left( 1+z\right) ^4+\Omega_m\left( 1+z\right) ^3+\Omega_{\Lambda}-\frac{k}{a_0^2H_0^2}\left( 1+z\right) ^2\right] ,
 \end{split}
 \label{friedmann}
\end{equation}
\begin{equation}
 \frac{\ddot{a}}{aH^2}=-\frac{1}{2}\,\Omega_{\text{\text{tot}}}(z)\left( 1+3\,w_{\text{\text{eff}}}(z)\right) \equiv-\frac{1}{2}\sum_i\Omega_i(z)\left( 1+3\,w_i\right) .
\end{equation}
Thus, the present-day Friedmann equation gives a cosmic sum rule 
\begin{equation}
 1=\Omega_{\text{tot}}-\frac{k}{a_0^2H_0^2}\equiv\Omega_{r}+\Omega_m+\Omega_{\Lambda}-\frac{k}{a_0^2H_0^2}\,. 
 \label{cosmicsum}
\end{equation}
Cosmological observations suggest that the universe is spatially flat. Additionally, the present-day radiation density is negligible, giving the relation 
\begin{equation}
 1\simeq\Omega_{\text{tot}}\simeq\Omega_m+\Omega_{\Lambda}\,. 
\end{equation}

\subsubsection*{Inflation}
Standard cosmology requires very specific initial conditions in order to explain the observed cosmological parameters. The main two problems are the \textit{flatness problem} and the \textit{homogeneity problem}. 

Differentiating equation (\ref{friedmann}) with respect to the time, we obtain 
\begin{equation}
 \frac{d\left( \Omega_{\text{tot}}-1\right) }{dt}=-2\,\frac{\ddot{a}}{aH}\left( \Omega_{\text{tot}}-1\right) =\left( 1+3\,w_{\text{eff}}\right) H\Omega_{\text{tot}}\left( \Omega_{\text{tot}}-1\right) . 
 \label{flatness}
\end{equation}
Since the Hubble parameter is always positive in an expanding universe, we see that $ \Omega_{\text{tot}} $ departs from 1 in a matter or radiation dominated universe, if it is not exactly 1 in the beginning. Thus, in order to obtain the observed value $ \Omega_{\text{tot}}\simeq 1 $ today, the initial value must be extremely fine-tuned. This dilemma is known as the flatness problem. 

Cosmic microwave background observations indicate that the universe was highly isotropic before structure formation. However, the observed CMB sky is many orders of magnitude larger than the causal horizon at the time of photon decoupling. Thus, the homogeneity of the temperature could not be achieved by physical interactions. Instead, it could only be achieved by extremely fine-tuned initial conditions. This dilemma is known as the homogeneity problem. 

Both problems can be solved by the introduction of an \textit{inflationary phase} (where $ w_{\text{eff}}\simeq-1 $) that lasts for about 60 $ e $-folds. This means that the scale factor grows by a factor of $ e^{60} $ in this phase. During inflation, $ \Omega_{\text{tot}}=1 $ becomes an attractor as we can see from equation (\ref{flatness}). Therefore, the universe can arrive at $ \Omega_{\text{tot}}\simeq 1 $ regardless of the initial conditions and stay close to that value until today. The isotropy of the CMB sky can also be explained: The entire observed universe had initially been a small causally connected region and had expanded tremendously during inflation. 

Such an inflationary phase can be realized by a scalar inflaton field that enters a so-called slow-roll phase. Apart from solving the above issues, inflation theories predict large-scale density perturbations that arise from quantum fluctuations of the inflaton field. These are observed in the form of temperature anisotropies in the CMB and finally lead to the formation of structures like galaxies and stars in the universe. 

After the inflationary phase, the density of all particles that initially were in the universe is diluted. However, the decay of the inflaton field at the end of the inflationary phase transfers its energy density into a hot thermal plasma of elementary particles. This process is known as the reheating of the universe, and the equilibrium temperature of the thermal plasma is therefore called the \textit{reheating temperature}. After this phase the universe is described by standard thermal cosmology. Energies in the early universe are usually given according to the characteristic temperature of the thermal plasma. Due to the adiabatic expansion of the universe the plasma temperature decreases as 
\begin{equation}
 T=T_0\left( 1+z\right) , 
\end{equation}
according to the redshift of relativistic particles. Here, $ T_0 $ is the present-day CMB radiation temperature.

\subsubsection*{Baryogenesis via Thermal Leptogenesis}
A crucial question in cosmology and particle physics is why there is more matter than antimatter in the universe. This problem shows up in the baryon-to-photon ratio $ \eta\equiv (n_b-n_{\bar{b}})/n_{\gamma}=n_b/n_{\gamma} $ that is different from zero. In order to generate a baryon asymmetry it is necessary to satisfy the Sakharov conditions~\cite{Sakharov:1967dj}: 
\begin{itemize}
 \item Baryon number ($ B $) violation, 
 \item $ C $-symmetry and $ CP $-symmetry violation, 
 \item Departure from thermal equilibrium. 
\end{itemize}
One of the models proposed to solve the problem of baryon asymmetry is \textit{baryogenesis via thermal leptogenesis}~\cite{Fukugita:1986hr}. In this model baryon asymmetry is generated from a lepton asymmetry. A nonvanishing lepton number $ L $ can be converted into a nonvanishing baryon number through non-perturbative sphaleron processes. These processes violate the linear combination $ B+L $ but conserve $ B-L $. If the sphaleron processes are in thermal equilibrium in the reheating phase of the universe, they lead to the relation~\cite{Barbier:2004ez} 
\begin{equation}
 B=\frac{24+4\,N_H}{66+13\,N_H}\left( B-L\right) ,  
\end{equation}
where $ N_H $ is the number of Higgs doublets and a theory with three fermion generations is assumed. In the minimal supersymmetric Standard Model with two Higgs doublets (see Section~\ref{susy}) one has $ B=8/23\left( B-L\right) =-8/15\,L $. 

In thermal leptogenesis the needed lepton number $ L $ can be created in $ CP $ violating out-of-equilibrium decays of heavy right-handed Majorana neutrinos. This mechanism is closely related to the problem of neutrino masses, since heavy right-handed Majorana neutrinos can also explain small nonvanishing masses for the light neutrinos via the seesaw mechanism. The actual observation of nonvanishing neutrino masses in the last years strongly supports the existence of heavy right-handed neutrinos and therefore also the mechanism of thermal leptogenesis.
 
To achieve the observed baryon asymmetry, the model of baryogenesis via thermal leptogenesis needs a high reheating temperature in the early universe of $ T_R\gtrsim 10^9\usk\unit{GeV} $~\cite{Davidson:2002qv, Buchmuller:2004nz}.

\subsubsection*{Primordial Nucleosynthesis}
Big Bang nucleosynthesis takes place at $ T\simeq 1 $--0.1\usk MeV and is thus based on well understood Standard Model physics. Moreover, BBN offers the deepest reliable probe of the early universe. The predictions of the abundances of the light elements D, $ ^3 $He, $ ^4 $He and $ ^7 $Li are very sensitive to the physical conditions at that temperature. Thermal equilibrium of neutrons and protons is achieved through weak interactions like 
\begin{equation}
 n\,\nu_e\leftrightarrow p\,e^-,\quad n\,e^+\leftrightarrow p\,\bar{\nu}_e\quad\text{and}\quad n\leftrightarrow p\,e^-\,\bar{\nu}_e\,, 
\end{equation}
and leads to a neutron-to-proton ratio of 
\begin{equation}
 \frac{n_n}{n_p}=e^{-\frac{m_n-m_p}{T}}, 
\end{equation}
where we neglected the chemical potential. Especially the number of relativistic particle species (e.g. the number of light neutrino species) and the baryon-to-photon ratio $ \eta\equiv n_b/n_{\gamma} $ determine the freeze-out time of weak interactions and thereby fix the initial neutron-to-proton ratio to be $ n_n/n_p\approx 1/6 $. This ratio slightly decreases to about $ 1/7 $ due to neutron decay until the neutrons are stabilized in bound states. Regardless of the detailed interaction processes, virtually all neutrons combine with protons to form $ ^4\text{He} $. The relative abundance by weight of $ ^4\text{He} $ can then easily be estimated: 
\begin{equation}
 Y_p\equiv\frac{\varrho_{^4\!\text{He}}}{\varrho_{p}+\varrho_{^4\!\text{He}}}\approx\frac{2\,n_n}{n_p+n_n}=\frac{2\,n_n/n_p}{1+n_n/n_p}\approx 25\usk\%. 
\end{equation}
The complete calculation of the light element abundances involves all the details of nuclear interactions and is able to predict the abundances at the $ 10^{-10} $ level. 

\begin{figure}
 \centering
 \includegraphics[scale=0.787]{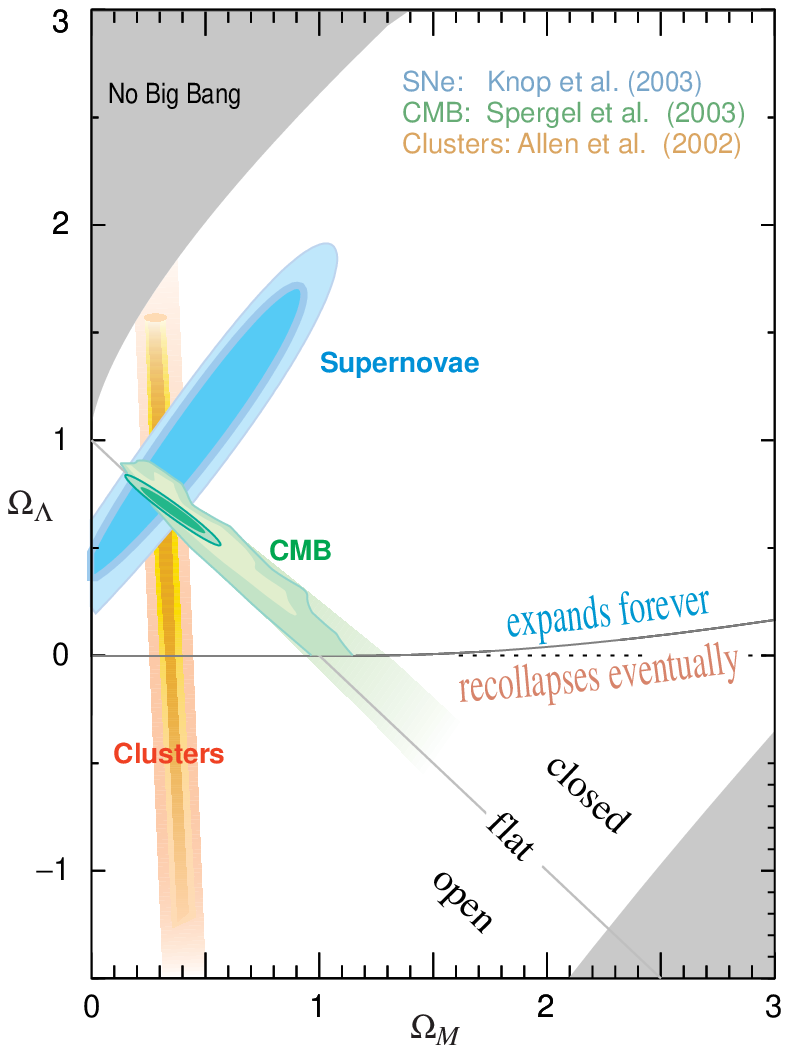} 
 \includegraphics[scale=0.67]{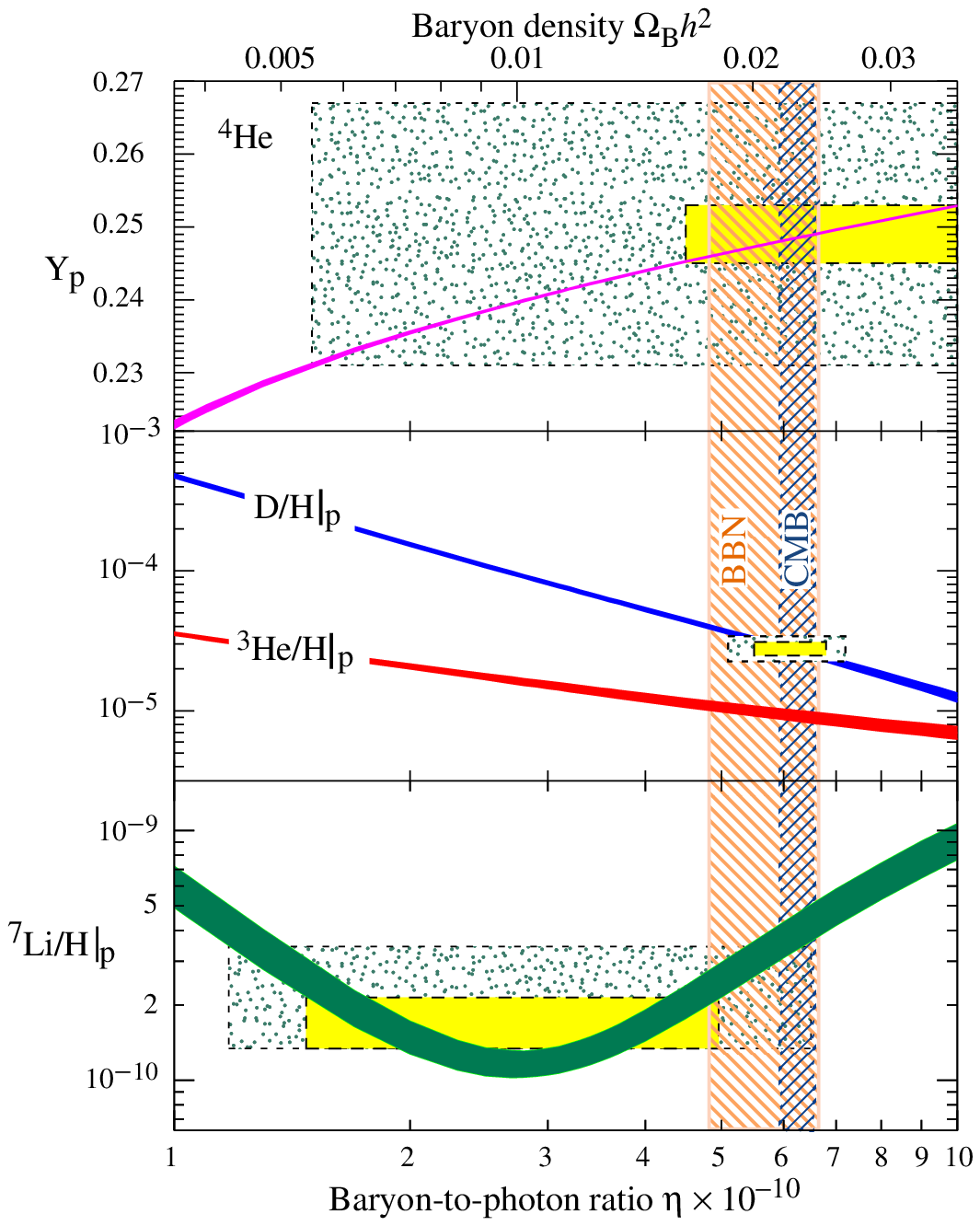} 
 \caption[Concordance model of cosmology and BBN predictions of the light element abundances.]{$ \Omega_{\Lambda}-\Omega_m $ plane of the concordance model of cosmology (left). The combination of several astrophysical observations suggests $ \Omega_{\Lambda}\simeq 0.75 $ and $ \Omega_m\simeq 0.25 $. BBN predictions of the light element abundances (right). All bands show the $ 2\,\sigma $ uncertainties. The boxes show the primordial abundances derived from astrophysical observations ($ 2\,\sigma $ statistical and $ 2\,\sigma $ statistical + systematic errors, respectively). Both figures are taken from~\cite{Amsler:2008zz}.}
 \label{ConcordanceBBN}
\end{figure}

The Standard Model particle content and a baryon-to-photon ratio $ \eta\simeq 6\times 10^{-10} $, as determined from the CMB measurements~\cite{Amsler:2008zz}, yield abundances of the light elements that are in good agreement with data from astrophysical observations (see Figure~\ref{ConcordanceBBN}). 

The agreement of the predictions and measurements of the light element abundances from BBN constrains deviations imposed by physics beyond the Standard Model. 

\subsubsection*{Cosmic Microwave Background}
The cosmic microwave background is a relic from the time when the photons decoupled from the thermal plasma of electrons and light elements at $ T\simeq 0.25\usk\unit{eV} $ (i.e. $ z\simeq 1100 $). After the temperature of the photons dropped below the energy to ionize hydrogen, the universe became transparent for them. The Cosmic Background Explorer (COBE) satellite mission found that the CMB is highly isotropic and corresponds to an almost perfect black body radiation spectrum with a temperature of $ T_0\simeq 2.7\usk\unit{K} $ (i.e. $ T_0\simeq 2.3\times 10^{-4}\usk\unit{eV} $). Furthermore, COBE observed temperature anisotropies of the CMB at the $ 10^{-5} $ level. 

The Wilkinson Microwave Anisotropy Probe (WMAP) satellite mission investigated the CMB temperature anisotropies in detail. Expansion of the temperature anisotropy map into spherical harmonics 
\begin{equation}
 \frac{\delta T}{T}(\theta,\,\phi)=\sum_{l=2}^{\infty}\sum_{m=-l}^la_{lm}Y_{lm}(\theta,\,\phi)
\end{equation}
gives the CMB power spectrum $ l(l+1)C_l/(2\pi) $ in terms of the multipole moment $ l $ with 
\begin{equation}
 C_l\equiv\left\langle \abs{a_{lm}}^2\right\rangle =\frac{1}{2l+1}\sum_{m=-l}^l\abs{a_{lm}}^2 . 
\end{equation}
Using a cosmological model with a limited number of parameters, it is possible to obtain best-fit values for the cosmological parameters from the observed CMB power spectrum. Generally, one uses a $ \Lambda $CDM model that assumes a cosmological constant $ \Lambda $ and Cold Dark Matter (CDM, i.e. particles that were already non-relativistic before structure formation) as the dominant dark matter component. Some of these parameters are given in Appendix~\ref{constants} (see also Figure~\ref{ConcordanceBBN}). In particular, from the position of the first acoustic peak one finds that the universe is approximately flat ($ \Omega_{\text{tot}}\simeq 1 $).

\section{Dark Matter}
\label{darkmatter}
In the following we will discuss the astrophysical evidence for dark matter in the universe, briefly touch upon the proposed particle physics candidates for dark matter and introduce two astrophysical particle dark matter detection techniques: Direct and indirect detection of dark matter. 

\subsection{Evidence for Dark Matter}
Astrophysical observations suggest the existence of dark matter: There is evidence from observations on galactic scales up to cosmological scales through the gravitational force. On the other hand, there is no experimental proof of DM at microscopic scales yet. In this section we summarize the astrophysical DM evidence following the review in~\cite{Bertone:2004pz}. 

\paragraph{Galactic Scale}
A very convicing evidence for dark matter on galactic scales comes from the observation of \textit{rotation curves of galaxies}, i.e. the circular velocity distribution of stars and gas as a function of the distance to the galactic center. 

\begin{figure}
 \centering
 \includegraphics[scale=0.329]{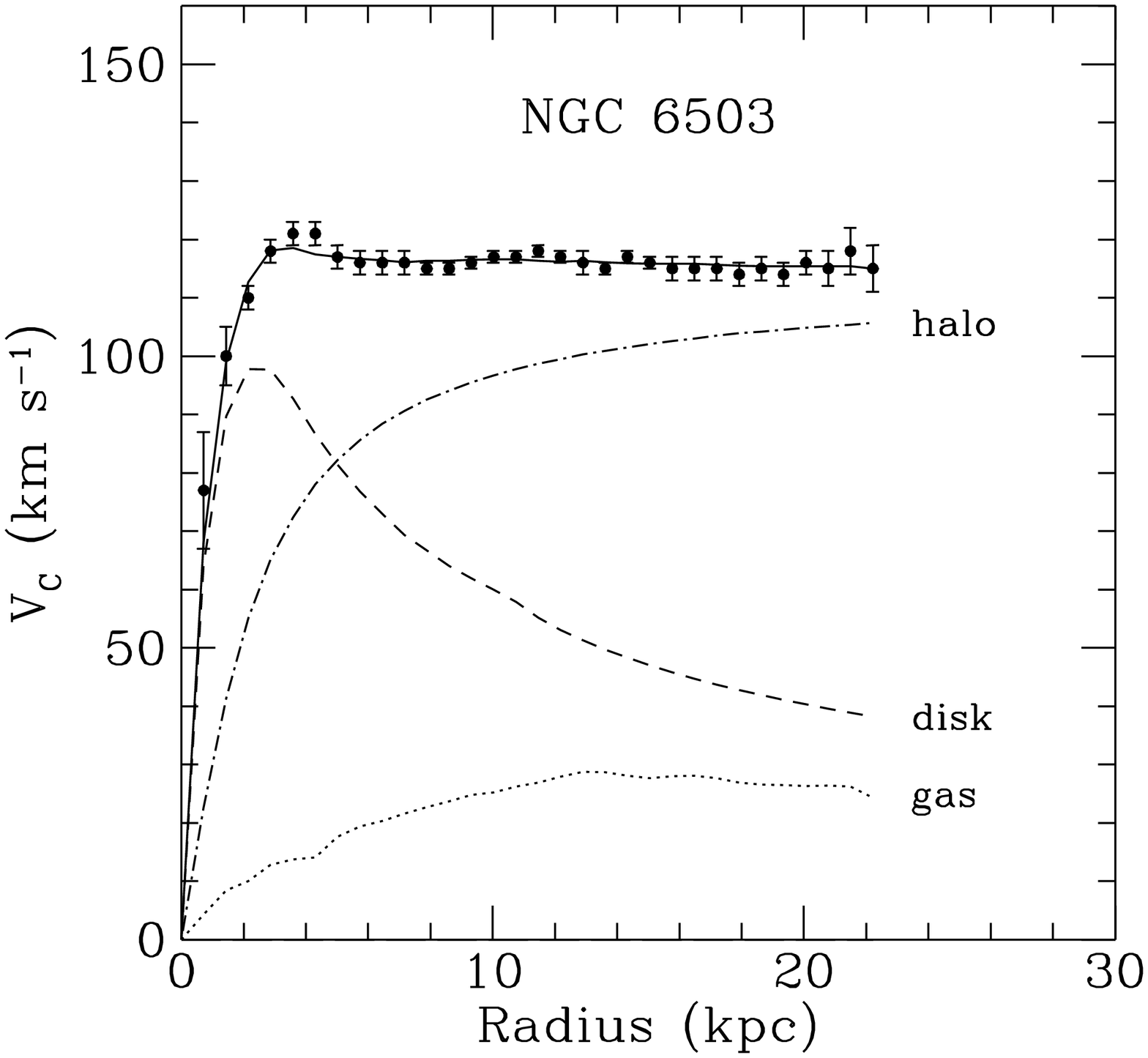} 
 \includegraphics[scale=1.08]{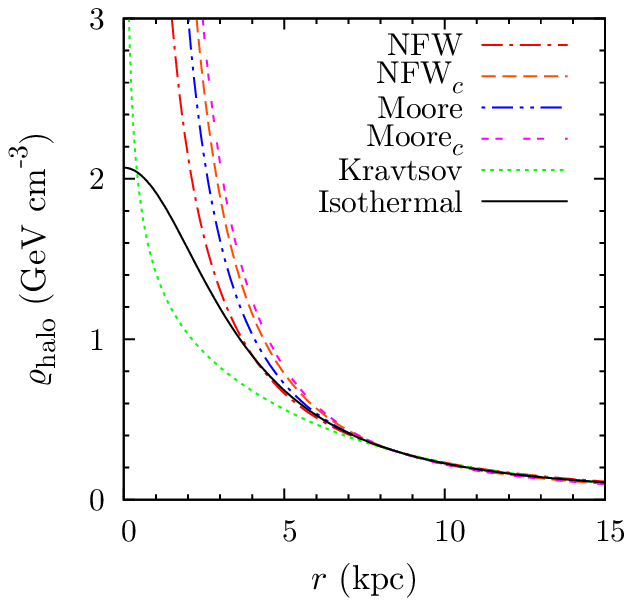} 
 \caption[Typical rotation curve of a galaxy and miscellaneous halo density profiles.]{Typical rotation curve of a galaxy (left). The solid line shows the fit to the data (dots with error bars) and the dashed, dotted and dash-dotted lines show the decomposition into the contributions from the luminous disk, gas and the dark matter halo, respectively. Figure taken from~\cite{Begeman:1991iy}. Shapes of the halo density profiles listed in the text (right). All profiles except for the isothermal are singular at the galactic center.}
 \label{rotationcurve}
\end{figure}

From Newtonian dynamics we expect the circular velocity to be given by 
\begin{equation}
 v(r)=\sqrt{\frac{G_NM(r)}{r}}\,, 
\end{equation}
with $ M(r)=4\pi\int_0^r\varrho(r)r^2dr $. In the outer regions of a galaxy (where there is no luminous matter anymore) one would expect the velocity to fall off as $ v\propto r^{-1/2} $. However, observations of galactic rotation curves show that the velocity remains constant even far beyond the luminous disk (see Figure \ref{rotationcurve}). 
This can be explained by the existence of a spherical halo of dark matter with a density profile $ \varrho_{halo}\propto 1/r^2 $ in the outer regions. Different halo profiles can be parameterized in the following way: 
\begin{equation}
 \varrho_{halo}(r)=\frac{\varrho_0}{\left( r/r_c\right) ^{\gamma}\left[ 1+\left( r/r_c\right) ^{\alpha}\right] ^{(\beta-\gamma)/\alpha}}\,. 
\end{equation}
Profiles obtained from N-body simulations include the profiles of Navarro, Frenk and White (NFW)~\cite{Navarro:1995iw}, Moore \textit{et al.}~\cite{Moore:1999gc} and Kravtsov \textit{et al.}~\cite{Kravtsov:1997dp}. Additionally, we list two profiles including adiabatic compression (i.e. including the effect of the baryonic gas in the halo): The NFW compressed and the Moore compressed profile. For comparison we also list the simple isothermal profile. The parameters of these profiles are presented in Table~\ref{profiles} and taken from~\cite{Bertone:2004pz, Mambrini:2005vk}. The density has been normalized using a local dark matter density of $ \varrho_{loc}=0.3\usk\unit{GeV\usk cm^{-3}} $ for the Milky Way~\cite{Bertone:2004pz}. While the slope in the outer regions is strongly constrained by observations, the slope of the halo density profile in the inner parts of galaxies is not very well known: All halo profiles from N-body simulations have a singular behavior at the galactic center, whereas the isothermal profile is finite (cf. Figure~\ref{rotationcurve}). 
\begin{table}
 \centering
 \begin{tabular}{lccccc}
  \toprule 
  Profile & $ \alpha $ & $ \beta $ & $ \gamma $ & $ r_c\usk (\unit{kpc}) $ & $ \varrho_0\usk (\unit{GeV\usk cm^{-3}}) $ \\
  \midrule 
  NFW & $ 1.0 $ & $ 3.0 $ & $ 1.0 $ & $ 20.0 $ & $ 0.26 $ \\
  NFW$ _c $ & $ 0.8 $ & $ 2.7 $ & $ 1.45 $ & $ 20.0 $ & $ 0.16 $ \\
  Moore & $ 1.5 $ & $ 3.0 $ & $ 1.5 $ & $ 28.0 $ & $ 0.059 $ \\
  Moore$ _c $ & $ 0.8 $ & $ 2.7 $ & $ 1.65 $ & $ 28.0 $ & $ 0.064 $ \\
  Kra & $ 2.0 $ & $ 3.0 $ & $ 0.4 $ & $ 10.0 $ & $ 0.57 $ \\
  iso & $ 2.0 $ & $ 2.0 $ & $ 0 $ & $ 3.5 $ & $ 2.1 $ \\
  \bottomrule 
 \end{tabular}
 \caption[Parameters for miscellaneous halo models.]{Parameters for the halo models listed in the text. The normalization is given according to the local halo density in the Milky Way.}
 \label{profiles}
\end{table}\smallskip

It has been proposed that part of the galactic DM halos could be composed of non-luminous Massive Compact Halo Objects (MACHOs). These could be for instance stellar-mass black holes, feint neutron stars, white dwarfs or stars, or Jupiter-sized planets. There have been searches for these objects through the microlensing effect, where several collaborations monitored the luminosity of millions of stars in the Large and Small Magellanic Clouds for several years. However, these experiments found that MACHOs can only contribute a subdominant part of the galactic DM. Thus, we need non-baryonic dark matter to explain the galactic dynamics.

\paragraph{Scale of Galaxy Clusters}
Evidence for non-baryonic dark matter comes also from the scale of galaxy clusters. In order to indicate DM in clusters, one has to show that the cluster mass inferred from the luminous components is smaller than the mass obtained independently using the gravitational force. Several methods provide this possibility. 

Application of the virial theorem to the observed \textit{velocity dispersion of galaxies in clusters} allows an estimation of the cluster mass. Observation of the \textit{X-ray spectrum of the hot gas inside clusters} gives the temperature of the gas, which is correlated to the cluster mass under the assumption of the gas being in hydrostatic equilibrium. Another method is \textit{gravitational lensing}: Light from distant sources is distorted due to the mass of galaxy clusters in the line of sight. Analysis of a combination of many background sources allows to determine the shape of the potential well of the cluster. 

Comparison of the cluster mass estimated from luminous matter and gas to the mass determined using the above listed methods gives a clear evidence for DM in galaxy clusters.

\paragraph{Cosmological Scale}
In contrast to the previously discussed observations, the measurement of the cosmic microwave background can be used to determine the total amount of dark matter in the universe. The values found for the matter density and the baryon density of the universe in analyses of the CMB power spectrum strongly hint on the existence of a non-baryonic dark matter. \smallskip

The combination of all astrophysical observations gives striking evidence for a cold non-baryonic dark matter component of the universe. The value of $ \Omega_m $ from cluster observations, CMB measurements and supernova data (concordance model, see Figure~\ref{ConcordanceBBN}) together with the baryon density $ \Omega_b $ from BBN predictions and CMB measurements leads to an energy density of non-baryonic dark matter of $ \Omega_{DM}h^2=\Omega_mh^2-\Omega_bh^2\simeq 0.1 $ (cf. Appendix~\ref{constants}).

\subsection{Particle Dark Matter Candidates}
We have seen that the dark matter in the universe must be non-baryonic. The observations also imply that the dark matter is electromagnetically neutral and has interactions comparable to the strength of weak interactions or even smaller. 

The Standard Model (SM) of particle physics offers only one candidate for particle dark matter: \textit{neutrinos}. They are neutral and weakly interacting, and due to their small mass they would contribute to hot dark matter (i.e. they are still relativistic at the time of structure formation). Experimentally, however, simulations of structure formation favor cold dark matter in order to be in agreement with the observed large scale structure. In addition, it was found that the three light neutrinos cannot amount to a significant energy density: $ 0.0005<\Omega_{\nu}h^2<0.023 $~\cite{Amsler:2008zz}. Thus, SM neutrinos do not play a dominant role for the dark matter problem. However, there is still the possibility that \textit{sterile neutrinos}, that are neutral with respect to the SM gauge groups, account for the dark matter of the universe. 

Apart from that, several theories beyond the Standard Model propose natural dark matter candidates (see for instance the reviews in~\cite{Bertone:2004pz, Jungman:1995df}). The most prominent and most extensively studied is the lightest \textit{neutralino} of supersymmetry. With conserved $ R $-parity---if the neutralino is the lightest supersymmetric particle---it would be absolutely stable. Additionally, the neutralino can have naturally a relic density on the correct order of magnitude. Other supersymmetric candidates are the superpartner of the neutrino, the \textit{sneutrino}, and the superpartner of the axion, the \textit{axino}. Sneutrino dark matter has already been ruled out by direct dark matter detection experiments, while the axino is still a viable dark matter candidate. 

In supergravity the \textit{gravitino} arises as a new dark matter candidate with extremely weak interactions. Thus, if it is not the LSP, it is very long-lived and thereby the origin of many problems in standard cosmology. On the other hand, if the gravitino is the LSP, it is stable due to $ R $-parity conservation and the NLSP becomes long-lived, causing similar cosmological problems. However, this work will deal with the consequences for indirect detection of gravitino dark matter in models, where $ R $-parity is slightly violated. In these models, all the cosmological obstacles can be evaded (see Section~\ref{gravitino}). 

Even though there are a lot more particle dark matter candidates apart from supersymmetry, we will restrict to this list since we are primarily interested in the gravitino in this work.

\subsection{Direct and Indirect Detection of Dark Matter}
Besides collider experiments that can determine the mass spectrum of particles beyond the Standard Model, there are two astrophysical techniques to detect dark matter particles: \textit{Direct} and \textit{indirect detection} of dark matter (see for instance~\cite{Bertone:2004pz}). 

\subsubsection*{Direct Detection}
Direct detection experiments try to detect particles from the dark matter halo that cross the Earth. They observe the recoil of target nuclei induced by elastic scatterings of weakly interacting massive particles off those nuclei. The expected signal is determined assuming that the dark matter particles are distributed in the halo according to an isothermal profile and have a Maxwell--Boltzmann velocity distribution with a characteristic velocity of $ v_0\sim 270\usk\unit{km\usk s^{-1}} $~\cite{Bertone:2004pz}. 

Gravitinos have much weaker interactions than WIMPs and thus no signal from gravitino dark matter can be observed in direct detection experiments. 

\subsubsection*{Indirect Detection}
The method of indirect dark matter detection uses the observation of cosmic rays produced in dark matter annihilations or decays. The cosmic ray flux originating from dark matter is proportional to the annihilation rate of the dark matter particle or---if it is unstable---to its decay rate. Annihilations require the collision of two dark matter particles, so the resulting flux depends on the square of the dark matter density. The flux from decays, on the other hand, is proportional to the density. Fluxes from annihilations therefore depend strongly on the dark matter distribution. 

\begin{figure}
 \centering
 \includegraphics[scale=0.8]{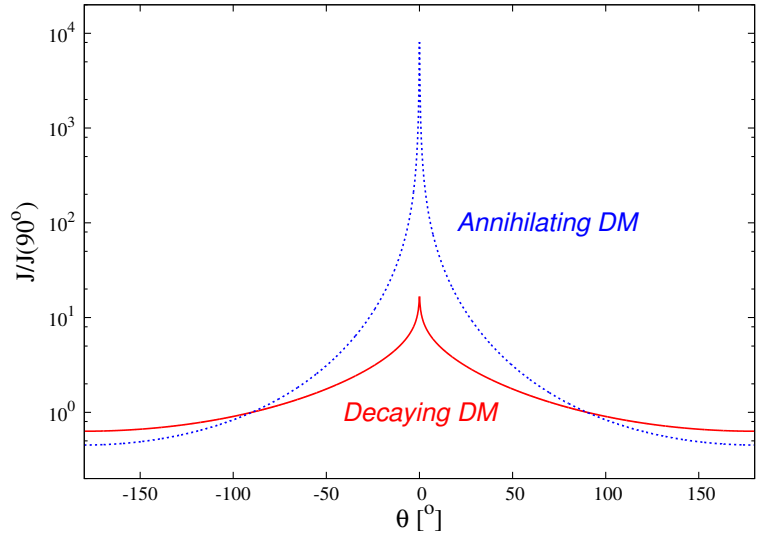} 
 \caption[Comparison of the angular profile of the flux from decaying and annihilating DM.]{Comparison of the angular profile of the flux from decaying DM (solid line) and annihilating DM (dashed line) for a NFW halo profile. Here, $ \theta $ is the angle to the galactic center. The fluxes are normalized to their values at the galactic poles ($ \theta=\pm\, 90^{\circ} $). Figure taken from~\cite{Bertone:2007aw}. }
 \label{angular}
\end{figure}
A comparison of the fluxes $ J $ from annihilating and decaying dark matter in a galactic halo with NFW density distribution is shown in Figure~\ref{angular}. While the signals from the galactic center and anticenter differ by less than two orders of magnitude for the case of decaying dark matter, they differ by more than four orders of magnitude for annihilating dark matter. The exact numbers depend on the angular resolution used to regularize the flux from the galactic center, since the NFW profile is singular at that position. 

Cosmic rays from annihilating dark matter are therefore most likely observed from very dense regions like the galactic center. Other astrophysical objects like the Sun or the Earth could also amplify the dark matter density, since weakly interacting dark matter particles are captured in the center of these objects via energy loss through scattering off nucleons. However, only neutrinos can escape these dense objects and could be observed. 

In the case of gravitino dark matter there is no accumulation inside astrophysical objects since the gravitational interaction is too weak. Thus, we have a gravitino distribution according to the galaxy density profile. There is no chance to observe gravitino annihilations, since these are suppressed by higher orders of the Planck scale. So, if the gravitino is stable, there is no observable signal in indirect detection experiments. On the other hand, if the gravitino is unstable, it could be possible to detect signals from its decay. In this thesis we want to discuss the prospects for the indirect detection of neutrino signals from the decays of an unstable gravitino.  

\chapter{Supersymmetry, Supergravity and the Gravitino}
\label{susygravitino}
In this chapter we want to give a brief introduction to the relevant topics of supersymmetry and supergravity and then discuss in more detail the gravitino. We will mainly follow the reviews on supersymmetry and the introductions to supergravity given in~\cite{Amsler:2008zz, Martin:1997ns, Chung:2003fi, Collins:1989kn, Nilles:1983ge}. For the discussion of the gravitino field we refer to~\cite{Pradler:2007ne, Bolz:2000xi, Moroi:1995fs}. 

\section{Supersymmetry}
\label{susy}
Supersymmetry (SUSY) is a generalization of the space-time symmetries of quantum field theory (QFT) that relates bosons and fermions. It introduces new fermionic generators $ Q $ that transform fermions into bosons and vice versa: 
\begin{equation*}
 Q\left| \text{boson}\right\rangle \simeq\left| \text{fermion}\right\rangle ,\qquad Q\left| \text{fermion}\right\rangle \simeq\left| \text{boson}\right\rangle . 
\end{equation*}
This is a nontrivial extension to the Poincar\'e symmetry of ordinary QFT and its structure is highly constrained by the theorem of Haag, Lopuszanski and Sohnius~\cite{Haag:1974qh}. 

If SUSY were an exact symmetry of nature, particles and their superpartners would be degenerate in mass. However, since no superpartners have been observed yet, SUSY must be a broken symmetry. 

Although not yet confirmed experimentally, there are several theoretical motivations for interest in this additional symmetry. The first is the \textit{hierarchy problem} of the Standard Model. This problem stems from the huge difference of the electroweak scale ($ \mathcal{O}(100)\usk\unit{GeV} $) and the (reduced) Planck scale 
\begin{equation}
 \MP=\frac{1}{\sqrt{8\pi\, G_N}}\simeq2.4\times 10^{18}\usk\unit{GeV}\,, 
\end{equation}
where gravitational interactions become comparable in magnitude to gauge interactions. The only scalar particle in the Standard Model, the Higgs boson, receives quadratic radiative corrections to its mass due to fermion loops. These quadratic divergences can be cancelled by the contributions of the bosonic superpartners to the radiative corrections. However, this cancellation works only for softly broken supersymmetry. If the Higgs mass is to be at the TeV scale, the soft SUSY breaking parameters can be no larger than a few TeV. 

\begin{figure}
 \centering
 \includegraphics[scale=0.52]{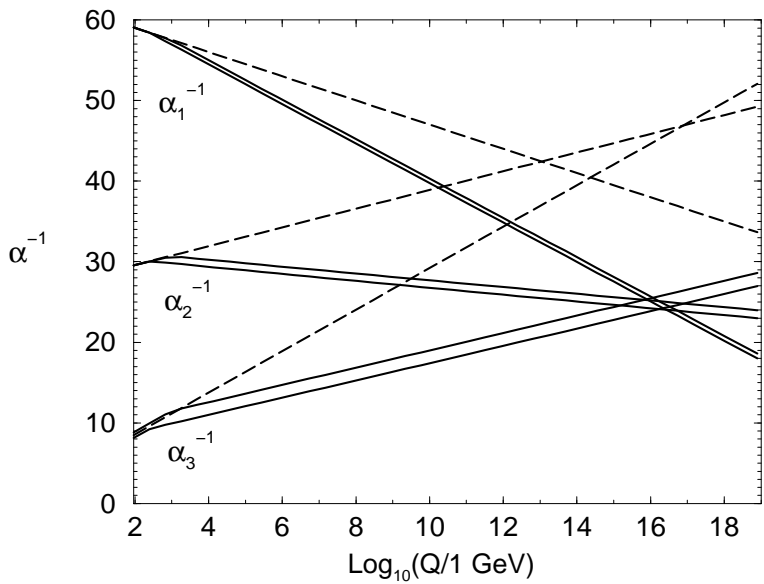} 
 \caption[Renormalization group evolution of the inverse gauge couplings in the SM and the MSSM.]{Renormalization group evolution of the inverse gauge couplings in the Standard Model (dashed) and the MSSM (solid). The sparticle mass thresholds are varied between 250\usk GeV and 1\usk TeV. Figure taken from~\cite{Martin:1997ns}.}
 \label{Unification}
\end{figure}

The second motivation is the \textit{unification of gauge couplings} $ \alpha_{\alpha}=g_{\alpha}^2/4\pi $, $ \alpha=1,\,2,\,3 $, where $ g_1=\sqrt{5/3}\,g' $ and $ g_2=g $ are the electroweak coupling constants and $ g_3=g_s $ is the strong coupling constant of the unbroken Standard Model gauge group 
\begin{equation*}
 SU(3)_c\times SU(2)_L\times U(1)_Y\,. 
\end{equation*}
The renormalization group equations, that determine the evolution of the gauge couplings, depend on the particle content of the theory. In the Standard Model the gauge couplings do not unify. By contrast, with the altered particle content of a supersymmetric theory at the TeV scale, the gauge couplings unify at the unification scale $ M_U\simeq2\times10^{16}\usk\unit{GeV} $ (see Figure~\ref{Unification}). 

The third motivation is that SUSY provides a promising candidate for particle dark matter: The lightest supersymmetric particle. With conserved \textit{R}-parity the LSP is absolutely stable. In order to constitute the dark matter of the universe, the LSP must be colorless and electromagnetically neutral. In that case, it only interacts gravitationally and---if it is not neutral with respect to weak interactions---through weak elastic scatterings with Standard Model particles.

\subsection*{The Minimal Supersymmetric Standard Model}
The Minimal Supersymmetric extension of the Standard Model (MSSM) introduces superpartners of the Standard Model particles as well as two Higgs-doublets and their superpartners. Standard Model particles and their superpartners are combined into supermultiplets. The \textit{gauge supermultiplets} (or vector supermultiplets) consist of a spin-1 vector boson $ A_{\mu}^a $, a spin-1/2 Majorana fermion $ \lambda^a $ and a scalar auxiliary field $ D^a $, where $ a $ labels the gauge group generators. In particular, we have the electroweak gauge bosons and the corresponding fermionic gauginos as well as the gluons and the fermionic gluinos (see Table~\ref{GaugeMSSM}). The table gives the notation of the particles in Lagrangians and Feynman rules, and their transformation properties under the Standard Model gauge groups. 
\begin{table}
 \centering
 \begin{tabular}{lccc}
  \toprule 
  Name & Gauge bosons & Gauginos & $ \left( SU(3)_c\,,\,SU(2)_L\right) _Y $ \\
  \midrule 
  $ B $ boson, bino & $ A_{\mu}^{(1)}=B_{\mu} $ & $ \lambda^{(1)}=\tilde{B} $ & $ \left( \mathbf{1},\,\mathbf{1}\right) _0 $ \\
  $ W $ bosons, winos & $ A_{\mu}^{(2)\, a}=W_{\mu}^a $ & $ \lambda^{(2)\, a}=\tilde{W}^a $ & $ \left( \mathbf{1},\,\mathbf{3}\right) _0 $ \\
  gluons, gluinos & $ A_{\mu}^{(3)\, a}=G_{\mu}^a $ & $ \lambda^{(3)\, a}=\tilde{g}^a $ & $ \left( \mathbf{8},\,\mathbf{1}\right) _0 $ \\
  \bottomrule 
 \end{tabular}
 \caption{Gauge supermultiplets of the MSSM.}
 \label{GaugeMSSM}
\end{table}
  
\textit{Chiral supermultiplets} consist of one complex scalar $ \phi $, a two-component chiral fermion $ \chi $ and an auxiliary scalar field $ F $. Namely, there are three generations of left-handed and right-handed leptons and quarks, the scalar sleptons and squarks, and the corresponding antiparticles. Additionally, there are two Higgs doublets, the fermionic higgsinos and their antiparticles (see Table~\ref{ChiralMSSM}). Here, $ i=1,\,2,\,3 $ is the generation index of (s)leptons and (s)quarks and $ h=r,\,g,\,b $ is the color index of (s)quarks. The table gives the notation of the particles in Lagrangians and Feynman rules, and their transformation properties under the Standard Model gauge groups. 

The chiral matter fermions carry spin 1/2 and their bosonic partners are scalars, while the Higgs bosons are complex scalars and the higgsinos are spin-1/2 fermions. The enlarged Higgs sector is needed to guarantee the cancellation of anomalies from the introduction of the higgsino superpartners. In addition, the two Higgs doublets are needed to generate masses for 'up'- and 'down'-type quarks as well as charged leptons after electroweak symmetry breaking. 
\begin{table}
 \centering
 \begin{tabular}{lccc}
  \toprule 
  Name & Scalars $ \phi^i $ & Fermions $ \chi_L^i $ & $ \left( SU(3)_c\,,\,SU(2)_L\right) _Y $ \\
  \midrule 
  Sleptons, leptons & $ \tilde{L}^i=
  \begin{pmatrix}
  \tilde{\nu}_L^i \\
  \tilde{e}_L^{-\, i}
  \end{pmatrix} $ & $ L^i=
  \begin{pmatrix}
  \nu_L^i \\
  e_L^{-\, i}
  \end{pmatrix} $ & $ \left( \mathbf{1},\,\mathbf{2}\right) _{-\frac{1}{2}} $ \\
  & $ \tilde{E}^{*\, i}=\tilde{e}_R^{-\, *\, i} $ & $ E^{c\, i}=e_R^{-\, c\, i} $ & $ \left( \mathbf{1},\,\mathbf{1}\right) _{+1} $ \\[2pt]
  Squarks, quarks & $ \tilde{Q}_h^i=
  \begin{pmatrix}
  \tilde{u}_{L,\,h}^i \\
  \tilde{d}_{L,\,h}^i
  \end{pmatrix} $ & $ Q_h^i=
  \begin{pmatrix}
  u_{L,\,h}^i \\
  d_{L,\,h}^i
  \end{pmatrix} $ & $ \left( \mathbf{3},\,\mathbf{2}\right) _{+\frac{1}{6}} $ \\
  & $ \tilde{U}_h^{*\, i}=\tilde{u}_{R,\,h}^{*\, i} $ & $ U_h^{c\, i}=u_{R,\,h}^{c\, i} $ & $ \left( \bar{\mathbf{3}},\,\mathbf{1}\right) _{-\frac{2}{3}} $ \\
  & $ \tilde{D}_h^{*\, i}=\tilde{d}_{R,\,h}^{*\, i} $ & $ D_h^{c\, i}=d_{R,\,h}^{c\, i} $ & $ \left( \bar{\mathbf{3}},\,\mathbf{1}\right) _{+\frac{1}{3}} $ \\[2pt]
  Higgs, higgsinos & $ H_d=
  \begin{pmatrix}
  H_d^0 \\
  H_d^-
  \end{pmatrix} $ & $ \tilde{H}_d=
  \begin{pmatrix}
  \tilde{H}_d^0 \\
  \tilde{H}_d^- 
  \end{pmatrix} $ & $ \left( \mathbf{1},\,\mathbf{2}\right) _{-\frac{1}{2}} $ \\
  & $ H_u=
  \begin{pmatrix}
  H_u^+ \\
  H_u^0
  \end{pmatrix} $ & $ \tilde{H}_u=
  \begin{pmatrix}
  \tilde{H}_u^+ \\
  \tilde{H}_u^0 
  \end{pmatrix} $ & $ \left( \mathbf{1},\,\mathbf{2}\right) _{+\frac{1}{2}} $ \\
  \bottomrule 
 \end{tabular}
 \caption{Chiral supermultiplets of the MSSM.}
 \label{ChiralMSSM}
\end{table}

Supersymmetric Lagrangians are determined by the superpotential $ W $, which is a function of the supermultiplets.\footnote{We do not derive the MSSM Lagrangian here. Thus, we do not introduce the superfield formalism and two component notation even if we write some expressions in this form.} In the MSSM with conserved $ R $-parity the superpotential is given by~\cite{Barbier:2004ez} 
\begin{equation}
 W_{\text{MSSM}}=\mu\hat{H}_u\hat{H}_d+\lambda_{ij}^e\hat{H}_d\hat{L}_i\hat{E}_j^c+\lambda_{ij}^d\hat{H}_d\hat{Q}_i\hat{D}_j^c-\lambda_{ij}^u\hat{H}_u\hat{Q}_i\hat{U}_j^c\,. 
\end{equation}
The hats denote superfields and we sum over generation indices $ i,\,j=1,\,2,\,3 $ and the suppressed gauge indices. $ \mu $ is the supersymmetric Higgs mass parameter and $ \lambda_{ij}^{u,\,d,\,e} $ denote the quark and charged lepton Yukawa coupling matrices.

\subsubsection*{\textit{R}-Parity}
As a consequence of observed $ B-L $ conservation one usually introduces an additional parity called $ R $-parity in the MSSM. This parity is added by hand in order to forbid processes that lead to proton decay. Therefore, a new multiplicative quantum number 
\begin{equation}
 R_p=(-1)^{3(B-L)+2S} 
 \label{Rparity}
\end{equation}
is assigned to all the MSSM particles with spin $ S $. Standard Model particles have even ($ R_p=+1 $) and their supersymmetric partners odd $ R $-parity ($ R_p=-1 $). This imposes that supersymmetric particles can only be produced pairwise at colliders and that they cannot decay into Standard Model particles only. 

Thus, the lightest supersymmetric particle would be absolutely stable. Moreover, to be consistent with cosmology, the LSP must be electromagnetically and color neutral and therefore provides a natural candidate for particle dark matter. \smallskip

On the other hand, the most general MSSM superpotential contains also $ R $-parity violating terms~\cite{Barbier:2004ez}: 
\begin{equation}
 W_{\slashed{R}_p}=\mu_i\hat{H}_u\hat{L}_i+\frac{1}{2}\lambda_{ijk}\hat{L}_i\hat{L}_j\hat{E}_k^c+\lambda'_{ijk}\hat{L}_i\hat{Q}_j\hat{D}_k^c+\frac{1}{2}\lambda_{ijk}^{\prime\prime}\hat{U}_i^c\hat{D}_j^c\hat{D}_k^c. 
\end{equation}
As in the case of conserved $ R $-parity, summation over the generation indices $ i,\,j,\,k=1,\,2,\,3 $ and the suppressed gauge indices is assumed. The $ R $-parity breaking mass parameters $ \mu_i $, and the trilinear Yukawa couplings $ \lambda_{ijk} $ and $ \lambda'_{ijk} $ violate lepton number, while the couplings $ \lambda_{ijk}^{\prime\prime} $ violate baryon number. There is no symmetry of the theory that forbids these terms. Experimentally, however, the couplings of these interactions are very restricted: The proton lifetime of $ \tau_{p}>2.1\times 10^{29}\usk\unit{years} $~\cite{Amsler:2008zz} demands, for instance, that either the $ L $ violating or the $ B $ violating couplings vanish, or that all the couplings are extremely suppressed. In addition, the requirement that an existing baryon asymmetry in the early universe is not erased before the electroweak phase transition implies that $ \lambda_{ijk},\,\lambda_{ijk}'<10^{-7} $~\cite{Campbell:1990fa, Fischler:1990gn, Dreiner:1992vm}. 

The distinction between Higgs and matter supermultiplets is lost in $ R $-parity violating models. Therefore, $ R $-parity violation permits the mixing of sleptons and Higgs bosons, the mixing of neutrinos and neutralinos, and the mixing of charged leptons and charginos. In this case, the LSP is not stable anymore. This is a crucial fact for the decays analyzed in this work.

\subsubsection*{Supersymmetry Breaking}
As mentioned before, supersymmetry must be spontaneously broken to explain the mass differences of Standard Model particles and their superpartners. In order to solve the hierarchy problem, the SUSY breaking must be soft, i.e. the SUSY breaking parameters must not reintroduce quadratic divergences. Moreover, these parameters have to be at the TeV scale. 

There are no realistic models of spontaneously broken supersymmetry where the SUSY breaking arises from the particle interactions of the MSSM. Usually one assumes a \textit{hidden sector}, consisting of particles that are neutral with respect to the Standard Model gauge groups, and a \textit{visible sector} that contains the MSSM particles. SUSY breaking is assumed to occur in the hidden sector and to be mediated to the MSSM by some mechanism. Supersymmetry is broken when one of the hidden sector fields obtains a non-vanishing $ F $-term vacuum expectation value (VEV), labeled as $ \left\langle F\right\rangle  $. 

Since the SUSY generators are fermionic, the breaking of supersymmetry generates a massless goldstone fermion, the \textit{goldstino}, analogous to the massless goldstone boson of electroweak symmetry breaking. This is a problem of global supersymmetry since no massless fermion has been observed. 

The arising soft SUSY breaking parameters are, for instance, the mass parameters $ M_1 $, $ M_2 $ and $ M_3 $ of the electroweak and strong gauginos, respectively. Further parameters are, among others, the soft masses of the squarks and sleptons, and the soft masses $ m_{H_u}^2 $ and $ m_{H_d}^2 $ of the Higgs doublets. With broken $ R $-parity numerous additional soft terms arise, for instance the Higgs--slepton mixing parameters $ B_i $ and $ m_{L_iH_d}^2 $.

\subsubsection*{Electroweak Symmetry Breaking}
As we have stated before, in the MSSM there are two complex Higgs doublets. Electroweak symmetry is broken down to electromagnetism, 
\begin{equation*}
 SU(2)_L\times U(1)_Y\rightarrow U(1)_{em}\,, 
\end{equation*}
dynamically through radiative corrections to the soft Higgs masses $ m_{H_u} $ and $ m_{H_d} $. The neutral Higgs fields then acquire vacuum expectation values $ \left\langle H_u^0\right\rangle =v_u $ and $ \left\langle H_d^0\right\rangle =v_d $. The ratio of the Higgs VEVs is usually denoted as 
\begin{equation}
 \tan{\beta}\equiv\frac{v_u}{v_d}\,. 
\end{equation}
The VEVs of the Higgs doublets are related to the Standard Model Higgs VEV $ v\simeq 174\usk\unit{GeV} $ in the following way: 
\begin{equation}
 \begin{split}
  v^2 &=v_u^2+v_d^2\,, \\
  v_u &=v\sin{\beta}\,, \\
  v_d &=v\cos{\beta}\,. 
 \end{split}
 \label{higgsVEV}
\end{equation}
According to electroweak symmetry breaking through the Higgs mechanism, the electroweak gauge bosons absorb the three massless degrees of freedom of the two complex Higgs doublets: The Goldstone bosons $ G^0 $ and $ G^{\pm} $. These become the longitudinal modes of the massive $ Z^0 $ and $ W^{\pm} $ vector bosons. The new mass eigenstates are given by
\begin{align}
 \begin{pmatrix}
  A_{\mu} \\
  Z^0_{\mu}
 \end{pmatrix}
 &=\begin{pmatrix}
  \cos{\theta_W} & \sin{\theta_W} \\
  -\sin{\theta_W} & \cos{\theta_W}
 \end{pmatrix}
 \begin{pmatrix}
  B_{\mu} \\
  W_{\mu}^0
 \end{pmatrix}, \label{neutraleigenstates} \\
 W_{\mu}^{\pm} &=\frac{1}{\sqrt{2}}\left( W_{\mu}^1\mp iW_{\mu}^2\right), \label{chargedeigenstates}
\end{align}
where the weak mixing angle $ \theta_W $ is defined as 
\begin{equation}
 \begin{split}
  \sin{\theta_W} &=\frac{g'}{\sqrt{g^2+g'^2}}\,, \\
  \cos{\theta_W} &=\frac{g}{\sqrt{g^2+g'^2}}\,, 
 \end{split}
 \label{weakangle}
\end{equation}
with the $ SU(2)_L $ gauge coupling $ g $ and the $ U(1)_Y $ gauge coupling $ g' $. The photon is massless, while the other electroweak gauge bosons acquire masses 
\begin{equation}
 \begin{split}
  m_W &=\frac{g\,v}{\sqrt{2}}\,, \\
  m_Z &=\frac{g\,v}{\sqrt{2}\cos{\theta_W}}\,. 
 \end{split}
 \label{WZmass}
\end{equation}
The Higgs gauge eigenstates can be expressed in terms of the Higgs mass eigenstates: The neutral gauge eigenstates are decomposed as 
\begin{equation}
 \begin{pmatrix}
  H_u^0 \\
  H_d^0
 \end{pmatrix}=
 \begin{pmatrix}
  v_u \\
  v_d
 \end{pmatrix}+\frac{1}{\sqrt{2}}
 \begin{pmatrix}
  c_{\alpha} & s_{\alpha} \\
  -s_{\alpha} & c_{\alpha} 
 \end{pmatrix}
\begin{pmatrix}
  h \\
  H
 \end{pmatrix}+\frac{i}{\sqrt{2}}
 \begin{pmatrix}
  s_{\beta_0} & c_{\beta_0} \\
  -c_{\beta_0} & s_{\beta_0} 
 \end{pmatrix}
 \begin{pmatrix}
  G^0 \\
  A^0
 \end{pmatrix}
 \label{HiggsStates}
\end{equation}
and the charged gauge eigenstates read 
\begin{equation}
 \begin{pmatrix}
  H_u^+ \\
  H_d^{-\,*}
 \end{pmatrix}=
 \begin{pmatrix}
  s_{\beta_{\pm}} & c_{\beta_{\pm}} \\
  -c_{\beta_{\pm}} & s_{\beta_{\pm}} 
 \end{pmatrix}
 \begin{pmatrix}
  G^+ \\
  H^+
 \end{pmatrix}. 
\end{equation}
Here and in the following parts we use the abbreviations $ s_{\alpha}\equiv\sin{\alpha} $ and $ c_{\alpha}\equiv\cos{\alpha} $ for the mixing angles. In the tree-level approximation one has $ \beta_0=\beta_{\pm}=\beta $ and the masses of the Higgs mass eigenstates are given by 
\begin{equation}
 \begin{split}
  m_{h,H}^2 &=\frac{1}{2}\left( m_{A^0}^2+m_Z^2\mp\sqrt{\left( m_{A^0}^2-m_Z^2\right) ^2+4\,m_Z^2m_{A^0}^2\sin^2{2\beta}}\right), \\
  m_{A^0}^2 &=2\abs{\mu}^2+m_{H_u}^2+m_{H_d}^2\,, \\
  m_{H^{\pm}}^2 &=m_{A^0}^2+m_W^2\,. 
 \end{split}
\end{equation}
In this case, the mixing angle $ \alpha $ is determined by the conditions 
\begin{equation}
 \frac{\sin{2\alpha}}{\sin{2\beta}}=-\left( \frac{m_H^2+m_h^2}{m_H^2-m_h^2}\right) \qquad\!\!\text{and}\qquad\frac{\tan{2\alpha}}{\tan{2\beta}}=\left( \frac{m_{A^0}^2+m_Z^2}{m_{A^0}^2-m_Z^2}\right) , 
 \label{alphaangle}
\end{equation}
and is usually chosen to be negative. \smallskip

In the MSSM decoupling limit, i.e. for a large $ \mu $ parameter, the particles $ H $, $ A^0 $ and $ H^{\pm} $ are very heavy and decouple from low-energy experiments. In this case, using the relations (\ref{alphaangle}), the mixing angle becomes $ \alpha\simeq\beta -\pi/2 $ and the lightest Higgs boson $ h $ obtains the couplings of the ordinary Standard Model Higgs boson. We will use this framework later in this analysis.

\subsubsection*{Neutralinos and Charginos}
Gauginos and higgsinos mix with each other due to electroweak symmetry breaking. The neutral gauginos and the neutral higgsinos combine to form the four mass eigenstates called neutralinos $ \tilde{\chi}_{\alpha}^0\, $, while the charged gauginos and the charged higgsinos mix to form the two mass eigenstates called charginos $ \tilde{\chi}_{\alpha}^{\pm} $. As mentioned before, the lightest neutralino $ \tilde{\chi}_1^0 $ is a thoroughly studied candidate for the cold dark matter in MSSM models. 

In the gauge eigenstate basis $ \psi^0=(\tilde{B},\,\tilde{W}^0,\,\tilde{H}_d^0,\,\tilde{H}_u^0)^T $ the neutralino mass part of the Lagrangian is given by 
\begin{equation}
 \mathscr{L}_{\text{neutralino mass}}=-\frac{1}{2}\,\psi^{0\,T}M_N\psi^0+h.c.\,, 
\end{equation}
where the neutralino mass matrix is symmetric and reads
\begin{equation}
 M_N=
 \begin{pmatrix}
  M_1 & 0 & -\frac{g'v_d}{\sqrt{2}} & \frac{g'v_u}{\sqrt{2}} \\
  0 & M_2 & \frac{g\,v_d}{\sqrt{2}} & -\frac{g\,v_u}{\sqrt{2}} \\
  -\frac{g'v_d}{\sqrt{2}} & \frac{g\,v_d}{\sqrt{2}} & 0 & -\mu \\
  \frac{g'v_u}{\sqrt{2}} & -\frac{g\,v_u}{\sqrt{2}} & -\mu & 0
 \end{pmatrix}. 
\end{equation}
Using the relations (\ref{higgsVEV}), (\ref{weakangle}) and (\ref{WZmass}), the mass matrix can be rewritten in the following way: 
\begin{equation}
 M_N=
 \begin{pmatrix}
  M_1 & 0 & -m_Zs_Wc_{\beta} & m_Zs_Ws_{\beta} \\
  0 & M_2 & m_Zc_Wc_{\beta} & -m_Zc_Ws_{\beta} \\
  -m_Zs_Wc_{\beta} & m_Zc_Wc_{\beta} & 0 & -\mu \\
  m_Zs_Ws_{\beta} & -m_Zc_Ws_{\beta} & -\mu & 0
 \end{pmatrix}, 
\end{equation}
where we used the abbreviations $ s_W\equiv\sin{\theta_W} $ and $ c_W\equiv\cos{\theta_W} $ for the weak mixing angle. Using an orthogonal matrix that includes equation (\ref{neutraleigenstates}), we can change the basis from the gauge eigenstates to the supersymmetric partners of the massive gauge bosons $ \psi^{\prime0}=(\tilde{\gamma},\,\tilde{Z}^0,\,\tilde{H}_d^0,\,\tilde{H}_u^0)^T $: 
\begin{equation}
 \begin{pmatrix}
  \tilde{\gamma} \\
  \tilde{Z}^0 \\
  \tilde{H}_d^0 \\
  \tilde{H}_u^0
 \end{pmatrix}=
 \begin{pmatrix}
  c_W & s_W & 0 & 0 \\
  -s_W & c_W & 0 & 0 \\
  0 & 0 & 1 & 0 \\
  0 & 0 & 0 & 1 
 \end{pmatrix}
 \begin{pmatrix}
  \tilde{B} \\
  \tilde{W}^0 \\
  \tilde{H}_d^0 \\
  \tilde{H}_u^0
 \end{pmatrix}=R
 \begin{pmatrix}
  \tilde{B} \\
  \tilde{W}^0 \\
  \tilde{H}_d^0 \\
  \tilde{H}_u^0
 \end{pmatrix}. 
\end{equation}
The neutralino mass term can then be rewritten in the form 
\begin{equation}
 \mathscr{L}_{\text{neutralino mass}}=-\frac{1}{2}\,\psi^{\prime0\,T}M_N'\psi^{\prime0}+h.c.\,, 
\end{equation}
with 
\begin{equation}
 \begin{split}
  M_N' &=RM_NR^T \\
  &=
  \begin{pmatrix}
   M_1c_W^2+M_2s_W^2 & \left( M_2-M_1\right) s_Wc_W & 0 & 0 \\
   \left( M_2-M_1\right) s_Wc_W & M_1s_W^2+M_2c_W^2 & m_Zc_{\beta} & -m_Zs_{\beta} \\
   0 & m_Zc_{\beta} & 0 & -\mu \\
   0 & -m_Zs_{\beta} & -\mu & 0
  \end{pmatrix}. 
 \end{split}
\end{equation}
Similarly, the basis can be changed to the mass eigenstates, which are the neutralinos. Symmetric matrices can always be diagonalized by a unitary matrix $ S $ and its transposed matrix. Since we are interested in positive eigenvalues, we have to add an additional diagonal unitary phase matrix $ P $ (for the problem of negative eigenvalues see for instance~\cite{Gunion:1984yn}). Using these matrices, the transformation into the neutralino basis reads 
\begin{equation}
 \begin{pmatrix}
  \tilde{\chi}_1^0 \\
  \tilde{\chi}_2^0 \\
  \tilde{\chi}_3^0 \\
  \tilde{\chi}_4^0 
 \end{pmatrix}=PS
 \begin{pmatrix}
  \tilde{\gamma} \\
  \tilde{Z}^0 \\
  \tilde{H}_d^0 \\
  \tilde{H}_u^0
 \end{pmatrix}, 
\end{equation}
so that the neutralino mass matrix is diagonalized in the following way: 
\begin{equation}
 P^*S^*M_N'S^{\dagger}P^{\dagger}=
 \begin{pmatrix}
  m_{\tilde{\chi}_1^0} & 0 & 0 & 0 \\
  0 & m_{\tilde{\chi}_2^0} & 0 & 0 \\
  0 & 0 & m_{\tilde{\chi}_3^0} & 0 \\
  0 & 0 & 0 & m_{\tilde{\chi}_4^0}
 \end{pmatrix}. 
\end{equation}

Similarly, for the chargino mass term in the Lagrangian we have in the gauge eigenstate basis $ \psi^{\pm}=(\tilde{W}^+,\,\tilde{H}_u^+,\,\tilde{W}^-,\,\tilde{H}_d^-)^T $ 
\begin{equation}
 \mathscr{L}_{\text{chargino mass}}=-\frac{1}{2}\,\psi^{\pm\,T}M_C\psi^{\pm}+h.c.\,, 
\end{equation}
where the chargino mass matrix $ M_C $ can be written in $ 2\times 2 $ block form
\begin{equation}
 M_C=
 \begin{pmatrix}
  0 & X^T \\
  X & 0
 \end{pmatrix}, 
\end{equation}
with
\begin{equation}
 X=
 \begin{pmatrix}
  M_2 & gv_u \\
  gv_d & \mu
 \end{pmatrix}=
 \begin{pmatrix}
  M_2 & \sqrt{2}\,m_Ws_{\beta} \\
  \sqrt{2}\,m_Wc_{\beta} & \mu
 \end{pmatrix}. 
\end{equation}
For the last expression we used the relations (\ref{higgsVEV}) and (\ref{WZmass}). We can change the basis to the mass eigenstates using two unitary matrices $ U $ and $ V $: 
\begin{equation}
 \begin{pmatrix}
  \tilde{\chi}_1^+ \\
  \tilde{\chi}_2^+ 
 \end{pmatrix}=V
 \begin{pmatrix}
  \tilde{W}^+ \\
  \tilde{H}_u^+ 
 \end{pmatrix},\qquad
 \begin{pmatrix}
  \tilde{\chi}_1^- \\
  \tilde{\chi}_2^- 
 \end{pmatrix}=U
 \begin{pmatrix}
  \tilde{W}^- \\
  \tilde{H}_d^- 
 \end{pmatrix}. 
\end{equation}
The matrix $ X $ can be diagonalized by $ U $ and $ V $ in the following way: 
\begin{equation}
 U^*XV^{\dagger}=VX^{\dagger}U^T=
 \begin{pmatrix}
  m_{\tilde{\chi}_1^{\pm}} & 0 \\
  0 & m_{\tilde{\chi}_2^{\pm}}
 \end{pmatrix}, 
\end{equation}
where the masses are the positive roots of the eigenvalues of $ X^{\dagger}X $, since 
\begin{equation}
 VX^{\dagger}XV^{\dagger}=U^*XX^{\dagger}U^T=
 \begin{pmatrix}
  m_{\tilde{\chi}_1^{\pm}}^2 & 0 \\
  0 & m_{\tilde{\chi}_2^{\pm}}^2
 \end{pmatrix}, 
\end{equation}
and are given by 
\begin{equation}
 \begin{split}
  m_{\tilde{\chi}_{1,2}^{\pm}}^2=&\:\frac{1}{2}\bigg( \abs{M_2}^2+\abs{\mu}^2+2\,m_W^2 \\
  &\mp\left.\sqrt{\left( \abs{M_2}^2+\abs{\mu}^2+2\,m_W^2\right) ^2-4\abs{\mu M_2-m_W^2\sin{2\beta}}^2}\right) . 
 \end{split}
\end{equation}

In models where the gaugino masses unify at the gauge coupling unification scale $ M_U $, one has the tree-level relation~\cite{Martin:1997ns} 
\begin{equation}
 M_1\approx\frac{5}{3}\tan^2{\theta_W}M_2\approx0.5\,M_2
\end{equation}
between the bino and the wino mass parameter at the electroweak scale. In this work we will use the GUT relation 
\begin{equation}
 M_2\simeq 1.9\,M_1\,,
 \label{GUTrelation}
\end{equation}
obtained from the numerically computed renormalization group evolution of the gaugino masses down to the electroweak scale~\cite{Ibarra:2007wg}.

\section{Supergravity}
If supersymmetry is promoted to a local symmetry, i.e. the parameter in SUSY transformations becomes coordinate-dependent, the theory must incorporate gravity. This is because in order to achieve invariance under local SUSY transformations, one has to add a new supermultiplet to the theory: The gravity supermultiplet, which consists of the spin-2 graviton and the spin-3/2 gravitino (see Table~\ref{Gravitymultiplet}). The resulting locally supersymmetric theory is called \textit{supergravity} (SUGRA). The gravitino, as well as the graviton, is neutral with respect to the SM gauge groups and has vanishing mass (and therefore only two transverse helicity states) in the case where supergravity is unbroken. 
\begin{table}
 \centering
 \begin{tabular}{lccc}
  \toprule 
  Name & Bosons & Fermions & $ \left( SU(3)_c\,,\,SU(2)_L\right) _Y $ \\
  \midrule 
  Graviton, gravitino & $ g_{\mu\nu} $ & $ \psi_{\mu} $ & $ \left( \mathbf{1},\,\mathbf{1}\right) _0 $ \\
  \bottomrule 
 \end{tabular}
 \caption{Gravity supermultiplet.}
 \label{Gravitymultiplet}
\end{table}

Since in supergravity there appear couplings with negative mass dimension, the theory is nonrenormalizable. On the other hand, we assume that SUGRA is an appropriate low-energy approximation of a more general theory. In the flat limit, i.e. $ \MP\rightarrow\infty $, the renormalizability is restored. In this work, however, we are interested in couplings of the gravitino to MSSM particles that are suppressed by the Planck mass. Therefore, we do not work in the flat limit, but take into account only tree-level interactions.

\subsubsection{Supergravity Breaking and the Super-Higgs Mechanism}
Analogous to the Higgs mechanism of electroweak symmetry breaking, in supergravity there exists a \textit{super-Higgs mechanism} of supergravity breaking. The massless Goldstone fermion of supersymmetry breaking, the goldstino, is absorbed by the gravitino, which thereby acquires its longitudinal (helicity $ \pm 1/2 $) components and becomes massive. 

This is because in spontaneously broken supergravity the Lagrangian contains a gravitino--goldstino mixing mass term. Invariance of the gravitino and the goldstino fields under local supersymmetry transformations demands a redefinition of the fields. The redefined gravitino is a linear combination of the gravitino and the goldstino, and therefore gets all four helicity states. In this case the gravitino mass becomes 
\begin{equation}
 m_{3/2}=\frac{\left\langle F\right\rangle }{\sqrt{3}M_P}\,, 
\end{equation}
where $ \left\langle F\right\rangle  $ is the $ F $-term VEV responsible for the spontaneous breaking of supergravity. 

The value of the gravitino mass depends on the particular SUSY breaking scheme and can range from the eV scale to scales beyond TeV. For instance, in gauge-mediated SUSY breaking the gravitino mass is usually much less than 1\usk GeV, while for gravity mediation a mass in the GeV to TeV region is expected~\cite{Steffen:2007sp}. Regardless of the specific SUSY breaking mechanism, we take the gravitino mass as a variable parameter that is expected to be about $ \mathcal{O} $(10--100)\usk\unit{GeV} in order to account for the dark matter density of the universe (cf. Section~\ref{gravitino}).

\section{The Gravitino}
\label{gravitino}
In the following we will discuss the classical gravitino field and its consequences for the standard cosmological scenario. 

\subsection{The Free Massive Gravitino Field}
The massive gravitino is described by the following Lagrangian~\cite{Bolz:2000fu}: 
\begin{equation}
 \mathscr{L}=-\frac{1}{2}\,\varepsilon^{\mu\nu\rho\sigma}\bar{\psi}_{\mu}\gamma^{5}\gamma_{\nu}\partial_{\rho}\psi_{\sigma}-\frac{1}{4}\,m_{3/2}\bar{\psi}_{\mu}\left[ \gamma^{\mu},\,\gamma^{\nu}\right] \psi_{\nu}+\mathscr{L}_{int}\,. 
\end{equation}
For the free gravitino field we discard the interaction part of the Lagrangian. Application of the Euler-Lagrange equations to this Lagrangian results in the equations of motion for the free massive gravitino field: 
\begin{equation}
 \begin{split}
  0 &=\frac{\partial\mathscr{L}}{\partial\bar{\psi}_{\mu}}-\partial_{\nu}\frac{\partial\mathscr{L}}{\partial(\partial_{\nu}\bar{\psi}_{\mu})} \\
  &=-\frac{1}{2}\,\varepsilon^{\mu\nu\rho\sigma}\gamma^{5}\gamma_{\nu}\partial_{\rho}\psi_{\sigma}-\frac{1}{4}\,m_{3/2}\left[ \gamma^{\mu},\,\gamma^{\nu}\right] \psi_{\nu}=0\,. 
  \label{GravitinoEoM}
 \end{split}
\end{equation}
From this expression a simpler set of equations of motion can be derived. To achieve this, we apply either $ \partial_{\mu} $ or the identity 
\begin{equation}
 \gamma_{\mu}\varepsilon^{\mu\nu\rho\sigma}=-i\gamma^5\left( \gamma^{\nu}\gamma^{\rho}\gamma^{\sigma}-g^{\rho\sigma}\gamma^{\nu}+g^{\nu\sigma}\gamma^{\rho}-g^{\nu\rho}\gamma^{\sigma}\right) , 
\end{equation}
that is derived in Appendix~\ref{notation}, to equation (\ref{GravitinoEoM}). We obtain 
\begin{align}
 m_{3/2}\left( \slashed{\partial}\gamma^{\nu}\psi_{\nu}-\gamma^{\nu}\slashed{\partial}\psi_{\nu}\right) &=0\,, \label{EoM1}\\
 i\left( \gamma^{\mu}\slashed{\partial}\gamma^{\nu}\psi_{\nu}-\gamma^{\mu}\partial^{\nu}\psi_{\nu}+\slashed{\partial}\psi^{\mu}-\gamma^{\nu}\partial^{\mu}\psi_{\nu}\right) +m_{3/2}\left( \gamma^{\mu}\gamma^{\nu}\psi_{\nu}-\psi^{\mu}\right)  &=0\,. \label{EoM2}
\end{align}
Application of $ \gamma_{\mu} $ to (\ref{EoM2}) then results in 
\begin{equation}
 i\left( \slashed{\partial}\gamma^{\nu}\psi_{\nu}-\gamma^{\nu}\slashed{\partial}\psi_{\nu}\right) +3\,m_{3/2}\gamma^{\nu}\psi_{\nu}=0\,. \label{EoM3}
\end{equation}
From equations (\ref{EoM1}) -- (\ref{EoM3}) the Rarita--Schwinger equations~\cite{Rarita:1941mf} for the massive gravitino field can be derived: 
\begin{align}
 \gamma^{\mu}\psi_{\mu}(x) &=0\,, \label{Rarita1}\\
 \left( i\slashed{\partial}-m_{3/2}\right) \psi_{\mu}(x) &=0\,. \label{Rarita2}
 \intertext{These equations yield the further constraint}
 \partial^{\mu}\psi_{\mu}(x) &=0\,. \label{Rarita3}
\end{align}
From (\ref{Rarita1}) -- (\ref{Rarita3}) we can also derive the adjoint equations: 
\begin{equation}
 \begin{split}
  \bar{\psi}_{\mu}(x)\,\gamma^{\mu} &=0\,, \\
  i\partial_{\nu}\bar{\psi}_{\mu}(x)\gamma^{\nu}+m_{3/2}\bar{\psi}_{\mu}(x) &=0\,, \\
  \partial^{\mu}\bar{\psi}_{\mu}(x) &=0\,. 
 \end{split}
\end{equation}
In order to find the solutions to the Rarita--Schwinger equations, we go to momentum space and note that similar to the case of Dirac spinors there are positive and negative frequency solutions:\footnote{As stated in~\cite{Bolz:2000xi}, all gravitino processes can be described using only one mode function, since the gravitino is a Majorana particle. However, in order to maintain the freedom to choose the direction of the continuous fermion flow in the Feynman rules (see Appendix~\ref{decaywidths}), we have to discuss both mode functions.} 
\begin{equation}
 \psi_{\mu}(x)=\psi_{\mu}^{+\,s}(p)\,e^{-ip\cdot x}\quad\text{and}\quad\psi_{\mu}(x)=\psi_{\mu}^{-\,s}(p)\,e^{ip\cdot x},\; s=\pm\frac{3}{2}, \pm\frac{1}{2}\,, 
\end{equation}
where the mode functions $ \psi_{\mu}^+ $ and $ \psi_{\mu}^- $ have to obey 
\begin{align}
 \gamma^{\mu}\psi_{\mu}^{+\,s}(p) &=0\,, &\gamma^{\mu}\psi_{\mu}^{-\,s}(p) &=0\,, \label{modeRarita1}\\
 \left( \slashed{p}-m_{3/2}\right) \psi_{\mu}^{+\,s}(p) &=0\,,\qquad\text{and} &\left( \slashed{p}+m_{3/2}\right) \psi_{\mu}^{-\,s}(p) &=0\,, \label{modeRarita2}\\
 p^{\mu}\psi_{\mu}^{+\,s}(p) &=0 &p^{\mu}\psi_{\mu}^{-\,s}(p) &=0\,. \label{modeRarita3}
\end{align}
Following~\cite{Auvil:1966iu} the mode functions can be constructed using the tensor product of the familiar Dirac spinors $ u $ and $ v $ of spin-1/2 particles, and the polarization vector $ \epsilon_{\mu} $ of a massive spin-1 particle. The result is 
\begin{equation}
 \begin{split}
  \psi_{\mu}^{+\,s}(p) &= \sum_{m,\lambda}\left\langle \left( \frac{1}{2},\,m\right) \left( 1,\,\lambda\right) \:\right\vert \left. \left( \frac{3}{2},\,s\right) \right\rangle u^m(p)\,\epsilon_{\mu}^{\lambda}(p)\,, \\
  \psi_{\mu}^{-\,s}(p) &=\sum_{m,\lambda}\left\langle \left( \frac{1}{2},\,m\right) \left( 1,\,\lambda\right) \:\right\vert \left. \left( \frac{3}{2},\,s\right) \right\rangle v^m(p)\,\epsilon_{\mu}^{\lambda}(p)\,, \label{modefunction}
 \end{split}
\end{equation}
where $ \left\langle \left( j_1,\,m_1\right) \left( j_2,\,m_2\right) \right\vert \left. \left( J,\,M\right) \right\rangle $ are Clebsch--Gordan coefficients with $ M=m_1+m_2 $. The values of the coefficients in equation (\ref{modefunction}) are given in Table~\ref{clebschgordan}. 
\begin{table}
 \centering
 \begin{tabular}{cccc}
  \toprule 
  & $ m_2=+1 $ & $ m_2=0 $ & $ m_2=-1 $ \\
  \midrule 
  $ m_1=+\frac{1}{2} $ & $ 1 $ & $ \sqrt{2/3} $ & $ \sqrt{1/3} $ \\
  $ m_1=-\frac{1}{2} $ & $ \sqrt{1/3} $ & $ \sqrt{2/3} $ & $ 1 $ \\
  \bottomrule 
 \end{tabular}
 \caption[Clebsch--Gordan coefficients.]{Clebsch--Gordan coefficients for $ j_1=\frac{1}{2} $, $ j_2=1 $ and $ J=\frac{3}{2} $. Figures taken from~\cite{Amsler:2008zz}. }
 \label{clebschgordan}
\end{table}

Using this form of the mode functions, and the normalization of the Dirac spinors (\ref{spinornorm}) and the polarization vector (\ref{epsilonnorm}), we can derive the normalization of the gravitino mode functions. The result is 
\begin{equation}
 \begin{split}
  \bar{\psi}_{\mu}^{+\,s}(p)\,\psi^{+\,s'\,\mu}(p) &=-2\,m_{3/2}\,\delta^{ss'}, \\
  \bar{\psi}_{\mu}^{-\,s}(p)\,\psi^{-\,s'\,\mu}(p) &=2\,m_{3/2}\,\delta^{ss'}. 
 \end{split}
 \label{gravitinonorm}
\end{equation}
In the calculation of unpolarized decay rates we encounter a summation over the helicity states of the gravitino field. The gravitino polarization tensor is defined as 
\begin{equation}
 P_{\mu\nu}^{\pm}(p)=\sum_s\psi_{\mu}^{\pm\,s}(p)\,\bar{\psi}_{\nu}^{\pm\,s}(p), 
\end{equation}
where the sum is performed over the four gravitino polarizations $ s=\pm\frac{3}{2}, \pm\frac{1}{2} $. Using the normalization of the mode functions (\ref{gravitinonorm}) we obtain 
\begin{equation}
 \begin{split}
  P_{\mu\lambda}^{\pm}(p)\,P_{\quad\:\nu}^{\pm\,\lambda}(p) &=\sum_{s,s'}\psi_{\mu}^{\pm\,s}(p)\,\bar{\psi}_{\lambda}^{\pm\,s}(p)\,\psi^{\pm\,s'\,\lambda}(p)\,\bar{\psi}_{\nu}^{\pm\,s'}(p) \\
  &=\mp\, 2\,m_{3/2}P_{\mu\nu}^{\pm}(p)\,. 
 \end{split}
 \label{projector}
\end{equation}
Thus the polarization tensor is a projector. We find that the polarization tensors for a gravitino with four-momentum $ p $ are given by 
\begin{equation}
 P_{\mu\nu}^+(p)=-\left( \slashed{p}+m_{3/2}\right) \left\lbrace \Pi_{\mu\nu}(p)-\frac{1}{3}\,\Pi_{\mu\sigma}(p)\Pi_{\nu\lambda}(p)\gamma^{\sigma}\gamma^{\lambda}\right\rbrace 
 \label{PolTensor+}
\end{equation}
for the positive frequency mode functions and 
\begin{equation}
 P_{\mu\nu}^-(p)=-\left( \slashed{p}-m_{3/2}\right) \left\lbrace \Pi_{\mu\nu}(p)-\frac{1}{3}\,\Pi_{\mu\sigma}(p)\Pi_{\nu\lambda}(p)\gamma^{\sigma}\gamma^{\lambda}\right\rbrace 
 \label{PolTensor-}
\end{equation}
for the negative frequency mode functions. In the above expressions we use 
\begin{equation}
 \Pi_{\mu\nu}(p) =\left( g_{\mu\nu}-\frac{p_{\mu}p_{\nu}}{m_{3/2}^2}\right) . 
\end{equation}
For a derivation of the polarization tensors see Appendix~\ref{notation}. 

Since the gravitino field is a solution of the Rarita--Schwinger equations of motion, the polarization tensor obeys the following constraints 
\begin{align}
 \gamma^{\mu}P_{\mu\nu}^{\pm}(p) &=0\,, &P_{\mu\nu}^{\pm}(p)\,\gamma^{\nu} &=0\,, \nonumber\\
 p^{\mu}P_{\mu\nu}^{\pm}(p) &=0\,, \qquad\text{and} &P_{\mu\nu}^{\pm}(p)\,p^{\nu} &=0\,, \label{polarizationIDs}\\
 \left( \slashed{p}\mp m_{3/2}\right) P_{\mu\nu}^{\pm}(p) &=0 &P_{\mu\nu}^{\pm}(p)\,\left( \slashed{p}\mp m_{3/2}\right) &=0\,. \nonumber
\end{align}

\subsection{Gravitino Interactions}
In the previous section we discussed the free gravitino field. Now we want to add interactions between the gravitino and the MSSM fields. The gravitino interaction part of the Lagrangian reads~\cite{Pradler:2007ne, Bolz:2000fu}
\begin{equation}
 \begin{split}
  \mathscr{L}_{int}= &-\frac{i}{\sqrt{2}\,\MP}\left[ \left( D_{\mu}^*\phi^{i*}\right) \bar{\psi}_{\nu}\gamma^{\mu}\gamma^{\nu}P_L\chi^i-\left( D_{\mu}\phi^i\right) \bar{\chi}^iP_R\gamma^{\nu}\gamma^{\mu}\psi_{\nu}\right] \\
  &-\frac{i}{8\MP}\,\bar{\psi}_{\mu}\left[ \gamma^{\nu},\,\gamma^{\rho}\right] \gamma^{\mu}\lambda^{(\alpha)\,a}F_{\nu\rho}^{(\alpha)\,a}+\mathcal{O}(\MP^{-2})\,. 
 \end{split}
\end{equation}
We immediately see that all the gravitino interactions are suppressed by the Planck scale. The covariant derivative of scalar fields is given in the MSSM as~\cite{Pradler:2007ne}
\begin{align}
 D_{\mu}\phi_i &=\partial_{\mu}\phi_i+i\sum_{\alpha=1}^3g_{\alpha}A_{\mu}^{(\alpha)\, a}T_{a,\, ij}^{(\alpha)}\phi_j 
 \intertext{and the field strength tensor for the gauge bosons reads}
 F_{\mu\nu}^{(\alpha)\, a} &=\partial_{\mu}A_{\nu}^{(\alpha)\, a}-\partial_{\nu}A_{\mu}^{(\alpha)\, a}-g_{\alpha}f^{(\alpha)\, abc}A_{\mu}^{(\alpha)\, b}A_{\nu}^{(\alpha)\, c}. 
\end{align}
$ T_{a,\, ij}^{(\alpha)}, \alpha=1,\,2,\,3 $ are the generators of the Standard Model gauge groups 
\begin{equation}
 \begin{split}
  T_{a,\, ij}^{(1)} &=Y_i\,\delta_{ij}\,, \\
  T_{a,\, ij}^{(2)} &=\frac{1}{2}\,\sigma_{a,\, ij}\,, \\
  T_{a,\, ij}^{(3)} &=\frac{1}{2}\,\lambda_{a,\, ij}\,, 
 \end{split}
 \label{generators}
\end{equation}
where $ Y_i $ is the hypercharge. The Pauli sigma matrices $ \sigma_a $ are given in (\ref{PauliMatrix}) and $ \lambda_a $ are the eight Gell-Mann matrices which will not be needed in this work. $ f^{(\alpha)\, abc} $ are the totally antisymmetric structure constants of the corresponding gauge group. 

The Feynman rules can be extracted from the interaction Lagrangian in the usual way. Since the gravitino and the gauginos are Majorana fields for which exist Wick contractions different from those of Dirac fermions, amplitudes will contain charge conjugation matrices and there may arise ambiguities concerning the relative sign of interfering diagrams. 

Therefore we use a method that introduces a continuous fermion flow~\cite{Denner:1992vza}. The direction of this fermion flow in a process can be chosen arbitrarily, if the corresponding Feynman rules are used. Amplitudes are then written down in the direction opposite to the continuous fermion flow. This method avoids the appearance of charge conjugation matrices and gives the correct relative signs of different diagrams contributing to a single process. 

The complete set of Feynman rules that are relevant for this work is provided in Appendix~\ref{decaywidths}. 

\subsection{Gravitino Cosmology}
The addition of the gravitino to the particle spectrum can lead to several cosmological problems. We want to discuss shortly two of the main problems and see what restrictions for gravitino theories arise therefrom. 

\subsubsection*{Thermal Gravitino Production}
If gravitinos are in thermal equilibrium in the early universe, their relic density leads to overclosure of the universe. That means they contribute a density $ \Omega_{3/2}=\varrho_{3/2}/\varrho_c>1 $. This problem can be circumvented in an inflationary universe, since during inflation any initial abundance of gravitinos is diluted due to the exponential expansion of the universe. Gravitinos can then be reproduced by scattering processes in the thermal plasma after the universe has been reheated (cf. Section~\ref{cosmology}). Thermal gravitino production is enhanced due to the contribution from the less suppressed interactions of the goldstino component. The gravitino relic density in that case becomes~\cite{Bolz:2000fu} 
\begin{equation}
\Omega_{3/2}h^2\simeq 0.27\left( \frac{T_R}{10^{10}\usk\unit{GeV}}\right) \left( \frac{100\usk\unit{GeV}}{m_{3/2}}\right) \left( \frac{m_{\tilde{g}}}{1\usk\unit{TeV}}\right) ^2, 
\end{equation}
where $ T_R $ is the reheating temperature of the universe after inflation and $ m_{\tilde{g}} $ is the gluino mass. For reasonable values of the gluino mass and a gravitino mass in the $ \mathcal{O}(100) $ GeV range, thermally produced gravitinos can amount to the observed dark matter density, i.e. $ \Omega_{3/2}\simeq\Omega_{DM} $. This constrains the reheating temperature to be $ T_R\approx\mathcal{O}(10^{10})\usk\unit{GeV} $, which is compatible with the constraint $ T_R\gtrsim 10^9\usk\unit{GeV} $ from thermal leptogenesis (cf. Section~\ref{cosmology}).

\subsubsection*{Impact on Big Bang Nucleosynthesis}
Long-lived supersymmetric particles can affect the successful predictions of BBN if their lifetimes are longer than $ \sim 1 $\usk s, i.e. if they decay during or after primordial nucleosynthesis. The particles produced in these late decays may induce hadronic and electromagnetic showers and thereby lead to hadro- and photo-dissociation of light elements. 

The gravitino---in cases where it is not the LSP---decays into lighter particles with a lifetime~\cite{Weinberg:1982zq} 
\begin{equation}
\tau_{\text{3/2}}\approx 3\usk\unit{years}\left( \frac{100\usk\unit{GeV}}{m_{3/2}}\right) ^3. 
\end{equation}
The late decay of the gravitino then implies an upper bound on the reheating temperature to ensure that the BBN predictions are not significantly altered~\cite{Kawasaki:2004qu}: $ T_R\lesssim 10^5\usk\unit{GeV} $. This low value for the reheating temperature is, however, not compatible with the value favored by thermal leptogenesis and the paradigm of gravitino dark matter. 

On the other hand, if the gravitino is the LSP and stable, the lifetime of the next-to-lightest supersymmetric particle decaying into a gravitino and Standard Model particles is given by~\cite{Martin:1997ns}
\begin{equation}
\tau_{\text{NLSP}}\approx 9\usk\unit{days}\left( \frac{m_{3/2}}{10\usk\unit{GeV}}\right) ^2\left( \frac{150\usk\unit{GeV}}{m_{\text{NLSP}}}\right) ^5. 
\end{equation}
Thus, the NLSP is present during or after BBN and its late decays may spoil the predictions of standard BBN. For instance, the hadronic decays of a neutralino NLSP can dissociate the primordial generated light elements, whereas a long-lived stau NLSP can form a bound state with $ ^4 $He and catalyze the production of $ ^6 $Li. This can lead to an overproduction of $ ^6 $Li by a factor 300--600~\cite{Hamaguchi:2007mp}. Other possible NLSPs like sneutrinos or stops do not substantially affect the standard BBN predictions~\cite{Kanzaki:2006hm, DiazCruz:2007fc}. 

However, there is a more general solution to this problem. The introduction of a small $ R $-parity violation causes the NLSP to decay into Standard Model particles before the onset of BBN. Due to the double suppression of the gravitino couplings to SM particles by the Planck mass and the small $ R $-parity violation, the gravitino remains very long-lived and therefore a viable candidate for the cold dark matter~\cite{Takayama:2000uz}. This paradigm we will further discuss in the following chapter.  

\chapter{Gravitino Decays via Bilinear \texorpdfstring{\textit{R}}{R}-Parity Breaking}
\label{model}
In this chapter we will shortly introduce bilinear $ R $-parity breaking models of the type discussed thoroughly in~\cite{Buchmuller:2007ui} and then discuss in detail the decay channels of the LSP gravitino via such $ R $-parity violating couplings. 

\section{Bilinear \texorpdfstring{\textit{R}}{R}-Parity Breaking}
In models with bilinear $ R $-parity breaking, the MSSM superpotential contains the additional term 
\begin{equation}
 W_{\slashed{R}_p}=\mu_i\hat{L}_i\hat{H}_u\,, 
\end{equation}
where $ \mu_i $ are $ R $-parity violating higgsino--lepton mixing masses and summation over the generation indices $ i=1,\,2,\,3 $ is assumed~(cf. Section~\ref{susy}). 

From equation (\ref{Rparity}) we see that $ R $-parity breaking implies $ B-L $ breaking. Thus, it can be shown that $ \mu_i $ is suppressed compared to the supersymmetric higgsino mass parameter $ \mu $ by $ \mu_i/\mu\sim v^2_{B-L}/\MP^2$, with $ v_{B-L} $ being the scale of $ B-L $ breaking. The $ R $-parity violating bilinear coupling $ \mu_i $ can be rotated away by a redefinition of the Higgs and lepton superfields: 
\begin{equation}
 \hat{H}_d=\hat{H}'_d-\frac{\mu_i}{\mu}\hat{L}'_i\qquad\text{and}\qquad\hat{L}_i=\hat{L}'_i+\frac{\mu_i}{\mu}\hat{H}'_d\,. 
\end{equation}
This mixing of the Higgs and lepton superfields induces trilinear $ R $-parity breaking Yukawa couplings 
\begin{equation}
 \lambda_{ijk}=\frac{\mu_k}{\mu}\lambda_{ij}^e\qquad\text{and}\qquad\lambda'_{ijk}=\frac{\mu_k}{\mu}\lambda_{ij}^d\,, 
\end{equation}
that are suppressed by $ \mathcal{O}(\mu_i/\mu) $. The baryon number violating couplings $ \lambda^{\prime\prime}_{ijk} $ are suppressed at higher order in this model. Thus, the model is compatible with the bound from the proton lifetime. In order to be also compatible with cosmology, there are additional constraints on the Yukawa couplings: As already mentioned in Section~\ref{susy}, the viability of baryogenesis via leptogenesis requires $ \lambda, \lambda'<10^{-7} $. Otherwise the generated baryon asymmetry is washed out before the electroweak phase transition. On the other hand, the couplings must be large enough to ensure that the NLSP decays before the onset of BBN. A sufficiently short lifetime can be achieved with $ \lambda, \lambda'>10^{-14} $~\cite{Buchmuller:2007ui}. 

Apart from the supersymmetric term discussed above, the corresponding soft bilinear supersymmetry and $ R $-parity breaking terms $ B_i $ and $ m_{L_iH_d}^2 $ arise in the Lagrangian: 
\begin{equation}
 \begin{split}
  -\mathscr{L}_{soft}= &\:m_{H_d}^2|H_d|^2+m_{H_u}^2|H_u|^2+m_{\tilde{L}_i}^2|\tilde{L}_i|^2 \\
  &\qquad+\left( BH_dH_u+B_i\tilde{L}_iH_u+m_{L_iH_d}^2\tilde{L}_iH_d^*+h.c.\right) +\ldots\,. 
 \end{split}
 \label{HiggsSlepton}
\end{equation}
Minimization of the scalar potential yields besides the Higgs VEVs a nonvanishing vacuum expectation value along the sneutrino field direction: 
\begin{equation}
 \left\langle \tilde{\nu}_i\right\rangle \simeq\frac{B_iv_u+m_{L_iH_d}^2v_d^*}{m_{\tilde{\nu}_i}^2}\,. 
\end{equation}
This sneutrino VEV explicitly breaks lepton number and generates not only one neutrino mass, but also nonvanishing mixing between neutralinos and neutrinos, as well as between charginos and charged leptons \cite{Barbier:2004ez}. In addition, the sneutrino VEV can be used to parameterize the Higgs--slepton mixing. For later convenience we thus rewrite the relation for the sneutrino VEV in the following form: 
\begin{equation}
 B_i\sin{\beta}+m_{L_iH_d}^2\cos{\beta}\simeq m_{\tilde{\nu}_i}^2\frac{\left\langle \tilde{\nu}_i\right\rangle }{v}\,. 
 \label{sneutrinoVEV}
\end{equation}

The above mentioned mixings are responsible for the two-body decays of the gravitino into gauge/Higgs boson and neutrino, which are the main source of neutrino flux in our scenario. These decays are also possible at the one-loop level if only trilinear $ R $-parity breaking terms are considered~\cite{Lola:2007rw}, but we do not consider that case here. Since the neutralino--neutrino mixing takes place along the zino component, the branching ratios into the different gauge/Higgs boson channels are fixed by the neutralino mixing matrix once the gravitino mass is specified. 

In this analysis we assume that the sneutrino acquires a VEV only along the $\tilde{\nu}_{\tau} $ direction, since we expect the $ R $-parity violating couplings to be largest for the third generation. The produced neutrinos then have tau flavor. However, we will see later that this is not a crucial assumption, since neutrino oscillations change any pure neutrino flavor into a mixed state. In particular, due to maximal atmospheric mixing, the flux of tau and muon neutrinos turns out to be identical (see Chapter~\ref{neutrino}).

\section{Gravitino Decay Channels}
\label{widths}
In this section we compute the relevant decay channels of the LSP gravitino in the above introduced model, where the neutralino--neutrino mixing, as well as the chargino--charged lepton mixing, is generated through a nonvanishing tau sneutrino VEV $ \left\langle \tilde{\nu}_{\tau}\right\rangle  $. The primary decay channels of the gravitino into neutrinos and other Standard Model particles are 
\begin{equation}
 \begin{split}
  \psi_{3/2} &\rightarrow\gamma\,\nu_{\tau}\,, \\
  \psi_{3/2} &\rightarrow W^{\pm}\tau^{\mp}, \\
  \psi_{3/2} &\rightarrow Z^0\nu_{\tau}\,, \\
  \psi_{3/2} &\rightarrow h\,\nu_{\tau}\,. 
 \end{split}
\end{equation}
The first decay mode is practically always allowed, since the tau neutrino mass is very small. On the other hand, the decay modes into $ W^{\pm}\tau^{\mp} $ and $ Z^0\nu_{\tau} $ are only accessible for a gravitino heavier than the weak gauge bosons, while the channel into $ h\,\nu_{\tau} $ requires a gravitino mass above the threshold for the production of the lightest Higgs boson. 

The total decay width of the gravitino determines the gravitino lifetime according to 
\begin{equation}
 \tau_{3/2}=\frac{1}{\Gamma_{3/2}}\,. 
\end{equation}
However, the partial decay widths of the gravitino will contain the sneutrino VEV from bilinear R-parity breaking whose value is not known. Thus, we will not be able to fix the absolute lifetime of the gravitino. Nevertheless, from the relative decay widths of the different decay channels we are able to determine the branching ratios for these channels. This is a crucial input for the prediction of spectra of gravitino decay products.

\subsection{Calculation of Decay Widths}
The gravitino decay widths can be computed from the Feynman rules given in Appendix~\ref{decaywidths}. Here we will only present the detailed calculation for the decay width of the process $ \psi_{3/2}\rightarrow\gamma\,\nu_{\tau} $\,. The calculation of the other decay widths is given in Appendix~\ref{decaywidths}.

\subsection*{$ \boldsymbol{\psi_{3/2}\rightarrow\gamma\,\nu_{\tau}} $}
At tree level only one Feynman diagram contributes to the decay of the LSP gravitino into the photon and the tau neutrino: 
\begin{equation*}
 \parbox{4.7cm}{
  \begin{picture}(135,137) (15,-18)
   \SetWidth{0.5}
   \SetColor{Black}
   \Line(100,50)(115,24)
   \DashArrowLine(125,41)(115,24){4}
   \Vertex(115,24){2.5}
   \Text(136,19)[]{\normalsize{\Black{$q$}}}
   \LongArrow(126,24)(135,8)
   \ArrowLine(115,24)(130,-2)
   \Photon(100,50)(115,24){5}{2}
   \Text(102,84)[]{\normalsize{\Black{$k$}}}
   \LongArrow(103,73)(112,88)
   \Photon(100,50)(130,102){5}{4}
   \Text(30,50)[]{\normalsize{\Black{$\psi_{\mu}$}}}
   \Text(70,65)[]{\normalsize{\Black{$p$}}}
   \LongArrow(61,59)(79,59)
   \Line(40,52)(100,52)\Line(40,48)(100,48)
   \Vertex(100,50){2.5}
   \Text(130,50)[]{\normalsize{\Black{$\left\langle\tilde{\nu}_{\tau}\right\rangle$}}}
   \Text(135,-10)[]{\normalsize{\Black{$\nu_{\tau}$}}}
   \Text(135,111)[]{\normalsize{\Black{$\gamma$}}}
   \ArrowArcn(51.67,-33.07)(66.09,91.45,28.04)
   \Text(97,30)[]{\normalsize{\Black{$\tilde{\chi}^0$}}}
  \end{picture}
 }. 
\end{equation*}
There is no direct coupling of the photino to the tau neutrino, since it is electromagnetically neutral. However, the photino mixes with the zino through the neutralino mass eigenstates and the zino couples to the tau neutrino via the tau sneutrino VEV. Using the rotation of the neutral gauge eigenstates into the neutralinos (cf. Section~\ref{susy}) 
\begin{equation}
 \tilde{\gamma}=\sum_{i,\,\alpha=1}^4S^*_{\tilde{\gamma}i}P_{i\alpha}^*\tilde{\chi}_{\alpha}^0\qquad\text{and}\qquad\tilde{Z}^0=\sum_{i,\,\alpha=1}^4S^*_{\tilde{Z}i}P_{i\alpha}^*\tilde{\chi}_{\alpha}^0\,, 
\end{equation}
and the relation between the gauge eigenstates and mass eigenstates of the electroweak gauge bosons (\ref{neutraleigenstates}) together with the generators of the corresponding gauge groups (\ref{generators}), we can write the amplitude of this process as 
\begin{equation*}
 \begin{split}
  i\mathcal{M}= &-\bar{u}^r(q)\,i\sqrt{2}\left\langle \tilde{\nu}_{\tau}\right\rangle \left( g\frac{\sigma_{3,\,11}}{2}\cos{\theta_W}-g'Y_{\nu_L}\sin{\theta_W}\right) P_R\\
  &\:\cdot\left( \sum_{i,\,j,\,\alpha=1}^4S^*_{\tilde{Z}i}P_{i\alpha}^*\frac{i\left( \slashed{q}+m_{\tilde{\chi}_{\alpha}^0}\right) }{q^2-m_{\tilde{\chi}_{\alpha}^0}^2}P_{\alpha j}S_{j\tilde{\gamma}}\right) \frac{i}{4\,\MP}\,\gamma^{\mu}\left[ \slashed{k},\,\gamma^{\rho}\right] \psi_{\mu}^{+\,s}(p)\,\epsilon_{\rho}^{\lambda\,*}(k) \\
  \simeq &-\frac{ig_Z\left\langle \tilde{\nu}_{\tau}\right\rangle }{8\sqrt{2}\,\MP}\left( \sum_{\alpha=1}^4\frac{S^*_{\tilde{Z}\alpha}S_{\alpha\tilde{\gamma}}}{m_{\tilde{\chi}_{\alpha}^0}}\right) \bar{u}^r(q)\left( 1+\gamma^5\right) \gamma^{\mu}\left[ \slashed{k},\,\gamma^{\rho}\right] \psi_{\mu}^{+\,s}(p)\,\epsilon_{\rho}^{\lambda\,*}(k)\,. 
 \end{split}
\end{equation*}
To approximate the gaugino propagator we used the fact that $ q^2=m_{\nu}^2\ll m_{\tilde{\chi}_{\alpha}^0}^2 $. Although the tau neutrino is no mass eigenstate we regard it as the final state here. Its mass is depicted as $ m_{\nu} $ and is negligible in this calculation. In addition, we introduced $ g_Z\equiv g/\cos{\theta_W} $. In the next step we will introduce $ \xi_{\tau}=\left\langle \tilde{\nu}_{\tau}\right\rangle /v $ and the photino--zino mixing parameter 
\begin{equation}
 U_{\tilde{\gamma}\tilde{Z}}=m_Z\sum_{\alpha=1}^4\frac{S^*_{\tilde{Z}\alpha}S_{\alpha\tilde{\gamma}}}{m_{\tilde{\chi}_{\alpha}^0}}\,. 
\end{equation}
Since we are looking at the decay of a gravitino and therefore work in the gravitino rest frame, the gravitino has no defined helicity. Therefore, we have to average over the initial gravitino spin states. And since we do not want to measure the polarization of the decay products we sum over their helicity states. Then we can use the polarization sums for  gravitinos (\ref{PolTensor+}), fermions (\ref{fermionPolarization}) and polarization vectors (\ref{vectorPolarization}), and the squared amplitude becomes 
\begin{equation*}
 \begin{split}
  \abs{\bar{\mathcal{M}}}^2= &\:\frac{1}{4}\sum_s\sum_r\sum_{\lambda}\mathcal{M}\mathcal{M}^* \\
  \simeq &-\frac{\xi_{\tau}^2\,g_{\rho\sigma}}{256\,\MP^2}\abs{U_{\tilde{\gamma}\tilde{Z}}}^2 \\
  &\:\cdot\Tr\left[ (\slashed{q}+m_{\nu})\left( 1+\gamma^5\right) \gamma^{\mu}\left[ \slashed{k},\,\gamma^{\rho}\right] P_{\mu\nu}^+(p)\left[ \gamma^{\sigma},\,\slashed{k}\right] \gamma^{\nu}\left( 1-\gamma^5\right) \right] \\
  = &-\frac{\xi_{\tau}^2}{128\,\MP^2}\abs{U_{\tilde{\gamma}\tilde{Z}}}^2\Tr\left[ \slashed{q}\left( 1+\gamma^5\right) \gamma^{\mu}\left[ \slashed{k},\,\gamma^{\rho}\right] P_{\mu\nu}^+(p)\left[ \gamma_{\rho},\,\slashed{k}\right] \gamma^{\nu}\right]  \\
  = &\:\frac{2\,\xi_{\tau}^2}{3\,m_{3/2}^2 \MP^2}\abs{U_{\tilde{\gamma}\tilde{Z}}}^2\left\lbrace \left( p\cdot k\right) ^2\left( p\cdot q\right) +m_{3/2}^2\left( p\cdot k\right) \left( k\cdot q\right) \right\rbrace .
 \end{split}
\end{equation*}
In the second step we used the projector property of the chirality operators (\ref{chirality}) and in the last step we already replaced squared four-momenta by the corresponding squared particle masses. The trace could be further simplified and calculated by hand exploiting the identities (\ref{gammaID1}) and (\ref{gammaID2}) for the gamma matrix structure and the constraints (\ref{polarizationIDs}) for the polarization tensor. However, here we used for the calculation of the trace the Mathematica package \textit{Feyncalc}~\cite{Mertig:1990an}. 

According to equation (\ref{twobodyscalar}), the further scalar products of the four-momenta for this process are given by 
\begin{equation*}
 \begin{split}
  \left( p\cdot k\right) &=\frac{m_{3/2}^2-m_{\nu}^2}{2}=\left( k\cdot q\right) , \\
  \left( p\cdot q\right) &=\frac{m_{3/2}^2+m_{\nu}^2}{2}\,. 
 \end{split}
\end{equation*}
Thus, the squared amplitude becomes 
\begin{equation*}
 \abs{\bar{\mathcal{M}}}^2\simeq\frac{\xi_{\tau}^2\left( m_{3/2}^2-m_{\nu}^2\right) ^2\left( 3\,m_{3/2}^2+m_{\nu}^2\right) }{12\,m_{3/2}^2\MP^2}\abs{U_{\tilde{\gamma}\tilde{Z}}}^2. 
\end{equation*}
Using equations (\ref{twobodywidth}) and (\ref{twobodymomentum}) we finally obtain the decay width: 
\begin{equation}
 \Gamma\left( \psi_{3/2}\rightarrow\gamma\,\nu_{\tau}\right) \simeq\frac{\xi_{\tau}^2}{64\pi}\frac{m_{3/2}^3}{\MP^2}\abs{U_{\tilde{\gamma}\tilde{Z}}}^2\left( 1-\frac{m_{\nu}^2}{m_{3/2}^2}\right) ^3\left( 1+\frac{1}{3}\frac{m_{\nu}^2}{m_{3/2}^2}\right) 
\end{equation}
or if we neglect the contributions of the small neutrino mass
\begin{equation}
 \Gamma\left( \psi_{3/2}\rightarrow\gamma\,\nu_{\tau}\right) \simeq\frac{\xi_{\tau}^2}{64\pi}\frac{m_{3/2}^3}{\MP^2}\abs{U_{\tilde{\gamma}\tilde{Z}}}^2. 
\end{equation}
The conjugate process $ \psi_{3/2}\rightarrow \gamma\,\bar{\nu}_{\tau} $, that produces tau antineutrinos, has the same decay width: 
\begin{equation*}
 \parbox{4.7cm}{
 \begin{picture}(135,137) (15,-18)
  \SetWidth{0.5}
  \SetColor{Black}
  \Line(100,50)(115,24)
  \DashArrowLine(115,24)(125,41){4}
  \Vertex(115,24){2.5}
  \Text(136,19)[]{\normalsize{\Black{$q$}}}
  \LongArrow(126,24)(135,8)
  \ArrowLine(130,-2)(115,24)
  \Photon(100,50)(115,24){5}{2}
  \Text(102,84)[]{\normalsize{\Black{$k$}}}
  \LongArrow(103,73)(112,88)
  \Photon(100,50)(130,102){5}{4}
  \Text(30,50)[]{\normalsize{\Black{$\psi_{\mu}$}}}
  \Text(70,65)[]{\normalsize{\Black{$p$}}}
  \LongArrow(61,59)(79,59)
  \Line(40,52)(100,52)\Line(40,48)(100,48)
  \Vertex(100,50){2.5}
  \Text(130,50)[]{\normalsize{\Black{$\left\langle\tilde{\nu}_{\tau}\right\rangle$}}}
  \Text(135,-10)[]{\normalsize{\Black{$\bar{\nu}_{\tau}$}}}
  \Text(135,111)[]{\normalsize{\Black{$\gamma$}}}
  \ArrowArcn(51.67,-33.07)(66.09,91.45,28.04)
  \Text(97,30)[]{\normalsize{\Black{$\tilde{\chi}^0$}}}
 \end{picture}
 } 
\end{equation*}
\begin{equation}
 \Gamma\left( \psi_{3/2}\rightarrow\gamma\,\bar{\nu}_{\tau}\right) \simeq\frac{\xi_{\tau}^2}{64\pi}\frac{m_{3/2}^3}{\MP^2}\abs{U_{\tilde{\gamma}\tilde{Z}}}^2. 
\end{equation}
The decay widths of the other channels are given by\footnote{These results do not exactly coincide with those presented in~\cite{Ishiwata:2008cu}: For the interference terms proportional to $ j_X $ we find a negative sign and a larger coefficient.} 
\begin{equation}
 \begin{split}
  \Gamma\left( \psi_{3/2}\rightarrow Z^0\nu_{\tau}\right) \simeq &\:\frac{\xi_{\tau}^2}{64\pi}\frac{m_{3/2}^3}{\MP^2}\,\beta_Z^2\left\lbrace \,U_{\tilde{Z}\tilde{Z}}^2f_Z-\frac{8}{3}\frac{m_Z}{m_{3/2}}\,U_{\tilde{Z}\tilde{Z}}\,j_Z+\frac{1}{6}\,h_Z\right\rbrace , \\
  \Gamma\left( \psi_{3/2}\rightarrow W^+\tau^-\right) \simeq &\:\frac{\xi_{\tau}^2}{32\pi}\frac{m_{3/2}^3}{\MP^2}\,\beta_W^2\left\lbrace \,U_{\tilde{W}\tilde{W}}^2f_W-\frac{8}{3}\frac{m_W}{m_{3/2}}\,U_{\tilde{W}\tilde{W}}\,j_W+\frac{1}{6}\,h_W\right\rbrace , \\
  \Gamma\left( \psi_{3/2}\rightarrow h\,\nu_{\tau}\right) \simeq &\:\frac{\xi_{\tau}^2}{384\pi}\frac{m_{3/2}^3}{\MP^2}\,\beta_h^4\abs{\frac{m_{\tilde{\nu}_{\tau}}^2}{m_{\tilde{\nu}_{\tau}}^2-m_h^2}+\sin{\beta}\,U_{\tilde{H}_u^0\tilde{Z}}+\cos{\beta}\,U_{\tilde{H}_d^0\tilde{Z}}}^2, 
 \end{split}
 \label{gravitinowidths}
\end{equation}
where the zino--zino, wino--wino and higgsino--zino mixing parameters are given as 
\begin{equation}
 \begin{split}
  U_{\tilde{Z}\tilde{Z}} &=m_Z\sum_{\alpha=1}^4\frac{S^*_{\tilde{Z}\alpha}S_{\alpha\tilde{Z}}}{m_{\tilde{\chi}_{\alpha}^0}}\,, \\
  U_{\tilde{W}\tilde{W}} &=\frac{m_W}{2}\sum_{\alpha=1}^2\frac{V^*_{\tilde{W}^+\alpha}U_{\alpha\tilde{W}^-}+h.c.}{m_{\tilde{\chi}_{\alpha}^{\pm}}}\,, \\
  U_{\tilde{H}_{u,\,d}^0\tilde{Z}} &=m_Z\sum_{\alpha=1}^4\frac{S^*_{\tilde{Z}\alpha}S_{\alpha\tilde{H}_{u,\,d}^0}}{m_{\tilde{\chi}_{\alpha}^0}}\,, 
 \end{split}
\end{equation}
and the kinematic functions $ \beta_X $, $ f_X $, $ j_X $ and $ h_X $ are given by 
\begin{equation}
 \begin{split}
  \beta_X &=1-\frac{m_X^2}{m_{3/2}^2}\,, \\
  f_X &=1+\frac{2}{3}\frac{m_X^2}{m_{3/2}^2}+\frac{1}{3}\frac{m_X^4}{m_{3/2}^4}\,, \\
  j_X &=1+\frac{1}{2}\frac{m_X^2}{m_{3/2}^2}\,, \\
  h_X &=1+10\,\frac{m_X^2}{m_{3/2}^2}+\frac{m_X^4}{m_{3/2}^4}\,. 
 \end{split}
\end{equation}
The decay widths for the conjugate processes $ \psi_{3/2}\rightarrow Z^0\bar{\nu}_{\tau} $, $ \psi_{3/2}\rightarrow W^-\tau^+ $ and $ \psi_{3/2}\rightarrow h\,\bar{\nu}_{\tau} $ are equal. 

The Higgs decay channel and the contribution from the non-abelian 4--vertex to the partial widths of the gravitino decay into the weak gauge bosons was previously neglected in~\cite{Ibarra:2007wg}. Therefore the contribution of the photon channel has been overestimated. On the other hand, it seems that the photon channel has been underestimated in~\cite{Ishiwata:2008cu} due to the different result for the interference term in the $ Z^0 $ and $ W $ decay channels. 

\begin{figure}
 \centering
 \includegraphics[scale=1.2]{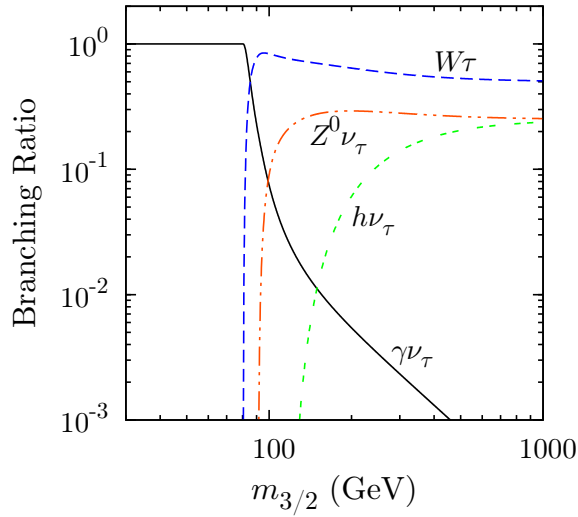} 
 \caption[Branching ratios of the gravitino decay channels.]{Branching ratios of the different gravitino decay channels as a function of the gravitino mass in the case of the MSSM decoupling limit.}
 \label{BRplot}
\end{figure}
Even if we cannot determine the absolute value for the gravitino lifetime due to the unknown value of the tau sneutrino VEV, we are still able to determine the branching ratios of the different decay channels. They are defined as the ratio of the partial decay width and the total decay width: 
\begin{equation}
 \Gamma_{\text{tot}}=\sum_X\Gamma\left( \psi_{3/2}\rightarrow X\right) \qquad\text{and}\qquad\BR\left( X\right) =\frac{\Gamma\left( \psi_{3/2}\rightarrow X\right) }{\Gamma_{\text{tot}}}\,. 
\end{equation}
Using the same set of SUSY parameters as~\cite{Ishiwata:2008cu}---i.e. we work in the MSSM decoupling limit (see Section~\ref{susy}, we use $ \mu=10\usk\unit{TeV} $ for the numerical calculation) and use $ \tan{\beta}=10 $, $ m_h=115\usk\unit{GeV} $, $ m_{\tilde{\nu}}=2\,m_{3/2} $, $ M_1=1.5\,m_{3/2} $ and the GUT relation (\ref{GUTrelation}) between $ M_1 $ and $ M_2 $ at the electroweak scale---we can calculate the branching ratios for the gravitino decay channels. For the weak mixing angle we use the ratio of the electroweak gauge boson masses 
\begin{equation}
 \cos\theta_W=\frac{m_W}{m_Z}\,. 
\end{equation}
The result for the branching ratios is given as a function of the gravitino mass in Figure~\ref{BRplot} and for several specific masses in Table~\ref{BRtable}.\footnote{Due to the different result for the interference terms in equation (\ref{gravitinowidths}), the photon channel in our result is less suppressed than in~\cite{Ishiwata:2008cu}.} As expected, the branching ratio for the decay into photon and tau neutrino drops drastically for gravitino masses above the threshold for the production of weak gauge bosons. This is explained by the $ R $-parity violating mixing of the zino and wino to the tau neutrino and the tau lepton, respectively. The photino, on the other hand, has no direct coupling and the decay width is therefore suppressed with respect to the weak gauge bosons. 
\begin{table}
 \centering
 \begin{tabular}{rcccc}
  \toprule 
  $ m_{3/2}\quad $ & $ \BR(\gamma\,\nu_{\tau}) $ & $ \BR(W^{\pm}\tau^{\mp}) $ & $ \BR(Z^0\nu_{\tau}) $ & $ \BR(h\,\nu_{\tau}) $ \\
  \midrule 
  $ 10\usk\unit{GeV} $ & $ 100\usk\% $ & --- & --- & --- \\
  $ 85\usk\unit{GeV} $ & $ 53\usk\% $ & $ 47\usk\% $ & --- & --- \\
  $ 100\usk\unit{GeV} $ & $ 7.5\usk\% $ & $ 83\usk\% $ & $ 9.0\usk\% $ & --- \\
  $ 150\usk\unit{GeV} $ & $ 1.1\usk\% $ & $ 70\usk\% $ & $ 28\usk\% $ & $ 1.2\usk\% $ \\
  $ 250\usk\unit{GeV} $ & $ 0.34\usk\% $ & $ 60\usk\% $ & $ 29\usk\% $ & $ 11\usk\% $ \\
  $ 1000\usk\unit{GeV} $ & $ 0.021\usk\% $ & $ 51\usk\% $ & $ 25\usk\% $ & $ 24\usk\% $ \\
  \bottomrule 
 \end{tabular}
 \caption[Branching ratios of the gravitino decay channels.]{Branching ratios of the different gravitino decay channels for several specific gravitino masses in the MSSM decoupling limit.}
 \label{BRtable}
\end{table}

For the above results we used a numerical calculation of the mixing parameters $ U_{\tilde{\gamma}\tilde{Z}} $, $ U_{\tilde{Z}\tilde{Z}} $, $ U_{\tilde{W}\tilde{W}} $ and $ U_{\tilde{H_{u,\,d}^0}\tilde{Z}} $. Following~\cite{Ibarra:2007wg}, we can also give an analytical approximation for some of these mixing parameters. In the MSSM decoupling limit, the photino, zino and wino are approximately mass eigenstates since the off-diagonal terms in the neutralino and chargino mixing matrices are subdominant (see Section~\ref{susy}). In this case, the zino--zino and wino--wino mixing parameters are simply given by the propagators of those particles, rescaled by the $ Z^0 $ and $ W $ boson masses, respectively: 
\begin{equation}
 \abs{U_{\tilde{Z}\tilde{Z}}}\simeq\frac{m_Z}{M'_{N,\,\tilde{Z}\tilde{Z}}}=\frac{m_Z}{M_1s_W^2+M_2c_W^2}\quad\text{and}\quad\abs{U_{\tilde{W}\tilde{W}}}\simeq\frac{m_W}{X_{\tilde{W}\tilde{W}}}=\frac{m_W}{M_2}\,. 
\end{equation}
As the photino has no direct coupling to the neutrino, the mixing parameter is approximated by the product of the photino propagator, the photino--zino mixing and the zino propagator, rescaled by the $ Z^0 $ boson mass: 
\begin{equation}
 \abs{U_{\tilde{\gamma}\tilde{Z}}}\simeq\frac{m_Z\,M'_{N,\,\tilde{\gamma}\tilde{Z}}}{M'_{N,\,\tilde{\gamma}\tilde{\gamma}}M'_{N,\,\tilde{Z}\tilde{Z}}}=\frac{m_Z\left( M_2-M_1\right) s_Wc_W}{\left( M_1c_W^2+M_2s_W^2\right) \left( M_1s_W^2+M_2c_W^2\right) }\,. 
\end{equation}
These approximations differ by less than 10\usk\% from the numerical calculation. The ratios of the mixing parameters are then given approximately as 
\begin{equation}
 \abs{U_{\tilde{\gamma}\tilde{Z}}}:\abs{U_{\tilde{Z}\tilde{Z}}}:\abs{U_{\tilde{W}\tilde{W}}}\simeq 1:3.2:2.6\,,
\end{equation}
where we only had to imply the GUT relation (\ref{GUTrelation}). For the higgsino--zino mixing parameters this method cannot be used since $ \tilde{H}_u^0 $ and $ \tilde{H}_d^0 $ are no mass eigenstates. However, in the MSSM decoupling limit the higgsino--zino mixing parameters are suppressed due to the large $ \mu $ parameter. In that case, the diagram with Higgs--slepton mixing dominates the decay channel. For tau sneutrino masses that are not degenerate with the lightest Higgs mass we have $ m_{\tilde{\nu}_{\tau}}^2/(m_{\tilde{\nu}_{\tau}}^2-m_h^2)=\mathcal{O}(1) $, and since the tau sneutrino has to be heavier than the LSP gravitino, we conclude that there is no strong dependence of the Higgs channel on $ m_{\tilde{\nu}_{\tau}} $. 

We see that the branching ratios given in Figure~\ref{BRplot} and Table~\ref{BRtable} are not only valid for the specific choice of parameters listed above, but approximately also for a large parameter space. In fact, the branching ratios for the electroweak gauge bosons dominantly depend only on the gravitino mass, while the Higgs channel additionally depends on the Higgs mass by means of the phase space factor.

\subsection[Fragmentation of the \texorpdfstring{\textit{Z}$^0$, \textit{W} and \textit{h}}{Z, W and h} Bosons]{Fragmentation of the \textit{Z}$ ^0 $, \textit{W} and \textit{h} Bosons}
Except for the photons and neutrinos, the decay products of the above discussed gravitino decay modes are not stable. Since we are interested in gravitino decays at astrophysical distances, we have to determine the spectra of stable particles at the end of the fragmentation processes of the $ Z^0 $, $ W $ and $ h $ bosons and the decay of the $ \tau $ lepton. The stable final state particles are photons, electrons, protons, neutrinos and their corresponding antiparticles. 

\begin{figure}
 \centering
 \includegraphics[scale=0.6]{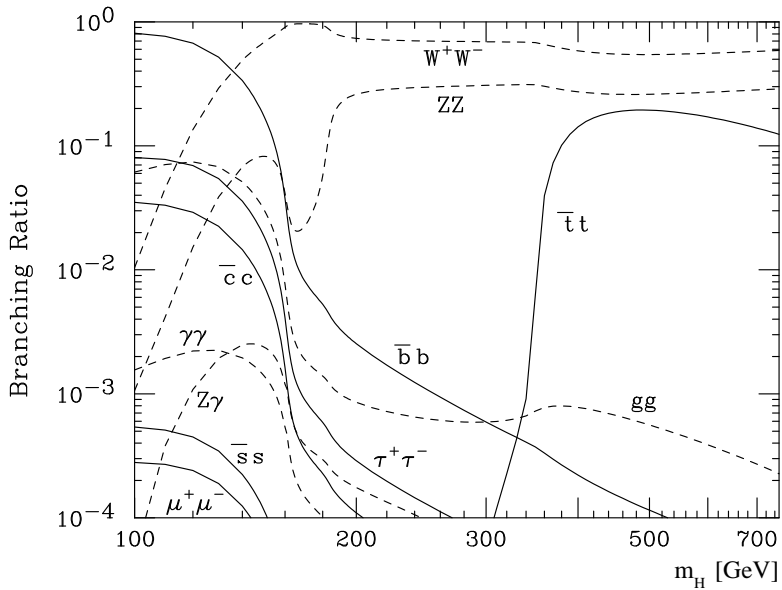} 
 \caption[Branching ratios of the SM Higgs boson.]{Branching ratios for the dominant decay channels of the Standard Model Higgs boson. Figure taken from~\cite{Carena:2002es}. }
 \label{higgsfragmentation}
\end{figure}

The dominant decay modes and the corresponding branching ratios for the first fragmentation step are presented in Table~\ref{fragmentationmodes} and in Figure~\ref{higgsfragmentation}. Since we work in the MSSM decoupling limit, the lightest Higgs boson $ h $ has the same interactions as the Standard Model Higgs boson. 
\begin{table}
 \centering
 \begin{tabular}{rclc}
  \toprule 
  \multicolumn{3}{c}{Decay mode} & Branching Ratio \\
  \midrule 
  $ Z^0 $ & $ \rightarrow $ & $ e^+\,e^- $ & $ (3.363\pm 0.003)\usk\% $ \\
  & & $ \mu^+\,\mu^- $ & $ (3.366\pm 0.007)\usk\% $ \\
  & & $ \tau^+\,\tau^- $ & $ (3.370\pm 0.008)\usk\% $ \\
  & & invisible & $ (20.00\pm 0.06)\usk\% $ \\
  & & hadrons & $ (69.91\pm 0.06)\usk\% $ \\
  \midrule 
  $ W^- $ & $ \rightarrow $ & $ e^-\,\bar{\nu}_e $ & $ (10.75\pm 0.13)\usk\% $ \\
  & & $ \mu^-\,\bar{\nu}_{\mu} $ & $ (10.57\pm 0.15)\usk\% $ \\
  & & $ \tau^-\,\bar{\nu}_{\tau} $ & $ (11.25\pm 0.20)\usk\% $ \\
  & & hadrons & $ (67.60\pm 0.27)\usk\% $ \\
  \midrule 
  $ \tau^- $ & $ \rightarrow $ & $ \mu^-\,\bar{\nu}_{\mu}\,\nu_{\tau} $ & $ (17.36\pm 0.05)\usk\% $ \\
  & & $ e^-\,\bar{\nu}_e\,\nu_{\tau} $ & $ (17.85\pm 0.05)\usk\% $ \\
  & & $ \pi^-\,\nu_{\tau} $ & $ (10.91\pm 0.07)\usk\% $ \\
  & & $ \pi^-\,\pi^0\,\nu_{\tau} $ & $ (25.52\pm 0.10)\usk\% $ \\
  & & $ \pi^-\,\pi^0\,\pi^0\,\nu_{\tau} $ & $ (9.27\pm 0.12)\usk\% $ \\
  & & $ \pi^-\,\pi^+\,\pi^-\,\nu_{\tau} $ & $ (8.99\pm 0.06)\usk\% $ \\
  \bottomrule 
 \end{tabular}
 \caption[Decay modes of the $ Z^0 $ and $ W $ bosons, and the $ \tau $ lepton.]{Dominant decay modes of the $ Z^0 $ and $ W $ bosons, and the $ \tau $ lepton. The conjugate decay modes have equal branching ratios. Figures taken from~\cite{Amsler:2008zz}. }
 \label{fragmentationmodes}
\end{table}

In fact, the complete calculation of fragmentation processes is very complicated. To obtain the decay products from the $ Z^0\nu_{\tau} $, $ W^{\pm}\tau^{\mp} $ and $ h\,\nu_{\tau} $ channels and their energies, we therefore mimic the decay of a resting gravitino with the event generator PYTHIA 6.4~\cite{Sjostrand:2006za}. A description of the used PYTHIA program is presented in Appendix~\ref{pythia}. 

In order to infer the neutrino spectra from the neutrino energy distribution obtained from PYTHIA, we need a large number of generated events for the simulation. Thus we use $ 10^6 $ events for most neutrino channels. For the strongly peaked tau neutrino channels of the gravitino decay into $ Z^0\nu_{\tau} $ and $ h\,\nu_{\tau} $ we use twice the number of events. 

The numbers of neutrinos of different flavors, generated in the fragmentation of the different decay products, are summarized in Table~\ref{neutrinonumbers}. 
\begin{table}
 \centering
 \begin{tabular}{rclccc}
  \toprule 
  \multicolumn{3}{c}{Decay mode} & $ \nu_e+\bar{\nu}_e $ & $ \nu_{\mu}+\bar{\nu}_\mu $ & $ \nu_{\tau}+\bar{\nu}_\tau $ \\
  \midrule 
  $ \psi_{3/2} $ & $ \rightarrow $ & $ Z^0\nu_{\tau} $ & $ 16.2 $ & $ 29.6 $ & $ 1.2 $ \\
  & & $ W^{\pm}\tau^{\mp} $ & $ 15.9 $ & $ 28.9 $ & $ 1.2 $ \\
  & & $ h\,\nu_{\tau} $ & $ 29.5 $ & $ 54.1 $ & $ 1.2 $ \\
  \bottomrule 
 \end{tabular}
 \caption[Numbers of neutrinos and antineutrinos from fragmentation.]{Numbers of neutrinos and antineutrinos from the fragmentation of the different decay products simulated with PYTHIA.}
 \label{neutrinonumbers}
\end{table}
We note that a large number of muon and electron neutrinos is produced in the fragmentation. At lower energies their ratio is 2:1 as in the case of atmospheric neutrinos (cf. Chapter~\ref{neutrino}). 

\subsection{Neutrino Spectra from Gravitino Decays}
The injection spectrum of neutrinos from gravitino decays is composed of a series of contributions. The gravitino decay into a photon and a tau neutrino is a two-body decay with a very long lifetime and virtually vanishing daughter particle masses. Therefore, the resulting neutrino spectrum from this channel is just a line at half the gravitino mass: 
\begin{equation}
 \frac{dN_{\nu_{\tau}}}{dE}\left( \psi_{3/2}\rightarrow\gamma\,\nu_{\tau}\right) \simeq\delta\left( E-\frac{m_{3/2}}{2}\right) . 
\end{equation}
By contrast, the decays into the heavy gauge bosons and the Higgs boson contribute a continuous spectrum. Anyway, the main features of the fragmentation spectra are also lines. The decays of the gravitino into $ Z^0\nu_{\tau} $ and $ h\,\nu_{\tau} $ result in lines centered at 
\begin{equation}
 E_{\nu_{\tau}}=\frac{m_{3/2}}{2}\left( 1-\frac{m_{Z,h}^2}{m_{3/2}^2}\right) . 
\end{equation}
However, these lines are not monoenergetic, but have a shape described by a normalized Breit--Wigner profile: 
\begin{equation}
 \frac{dN_{\nu_{\tau}}}{dE}\left( \psi_{3/2}\rightarrow Z^0/h\,\nu_{\tau}\right) =\frac{\left( \int_{0}^{\infty}dE\big/\left[ \left( E^2-E_{\nu_{\tau}}^2\right) ^2+E_{\nu_{\tau}}^2\Gamma_{\nu_{\tau}}^2\right] \right) ^{-1}}{\left( E^2-E_{\nu_{\tau}}^2\right) ^2+E_{\nu_{\tau}}^2\Gamma_{\nu_{\tau}}^2}\,, 
 \label{BreitWigner}
\end{equation}
where the widths are given in terms of the $ Z^0/h $ boson widths according to 
\begin{equation}
 \Gamma_{\nu_{\tau}}=\abs{\frac{\partial E_{\nu_{\tau}}}{\partial m_{Z,h}}}\Gamma_{Z,h}=\frac{m_{Z,h}}{m_{3/2}}\,\Gamma_{Z,h}\,. 
\end{equation}
The leptonic decays $ W\rightarrow l\,\nu_l $, $Z^0\rightarrow\nu_l\,\bar{\nu}_l $ and $ h\rightarrow l\,\nu_l $ also produce, in the rest frame of the decaying particle, monoenergetic neutrinos in all flavors. However, due to the boost of the bosons in different directions, the lines smear out almost completely in the Earth's rest frame, giving just an additional contribution to the continuous part of the spectrum. 

\begin{figure}
 \centering
 \includegraphics[scale=1.2]{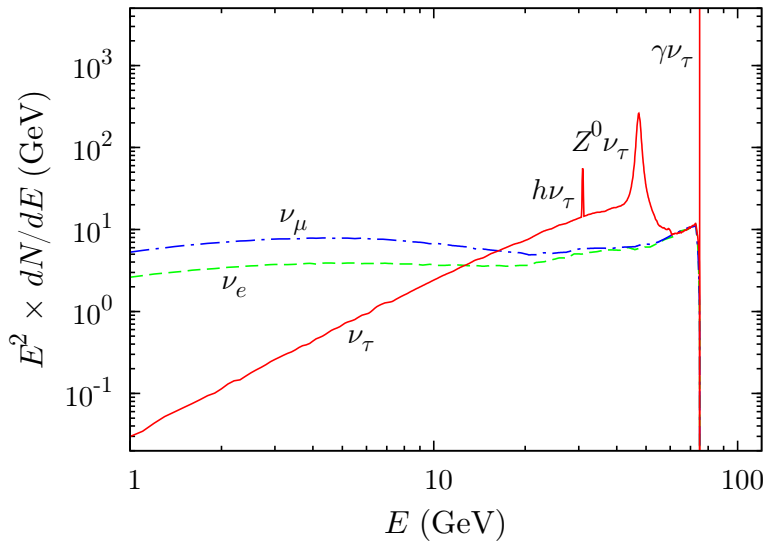} 
 \caption[Neutrino injection spectra from gravitino decay.]{Neutrino injection energy spectra from gravitino decay for the different flavors in the case of $ m_{3/2}=150\usk\unit{GeV} $. The dominant $ Z^0 $ line is clearly visible in the tau neutrino spectrum at an energy of 47\usk GeV, while the photon line is located at the end of the spectrum. Considering a Higgs mass of 115\usk GeV, the peak at $\sim 30 $\usk GeV is strongly suppressed by the phase space factor.}
 \label{energyspec}
\end{figure}
As already mentioned in the previous section, we extracted the continuous fragmentation spectra for the different neutrino flavors in the $ W^{\pm}\tau^{\mp} $, $ Z^0\nu_{\tau} $ and $ h\,\nu_{\tau} $ channels from the PYTHIA simulation. We denote these spectra by $ dN_{\nu_l}^{W\tau}/dE $, $ dN_{\nu_l}^{Z\nu}/dE $ and $ dN_{\nu_l}^{h\nu}/dE $, respectively. Weighting the constituent channels with the corresponding branching ratios, the total spectra for the different neutrino flavors are given by 
\begin{equation}
 \begin{split}
  \frac{dN_{\nu_e}}{dE}= &\BR\left( W^{\pm}\tau^{\mp}\right) \frac{dN_{\nu_e}^{W\tau}}{dE}+\BR\left( Z^0\nu_{\tau}\right) \frac{dN_{\nu_e}^{Z\nu}}{dE}+\BR\left( h\,\nu_{\tau}\right) \frac{dN_{\nu_e}^{h\nu}}{dE}\,, \\
  \frac{dN_{\nu_{\mu}}}{dE}= &\BR\left( W^{\pm}\tau^{\mp}\right) \frac{dN_{\nu_{\mu}}^{W\tau}}{dE}+\BR\left( Z^0\nu_{\tau}\right) \frac{dN_{\nu_{\mu}}^{Z\nu}}{dE}+\BR\left( h\,\nu_{\tau}\right) \frac{dN_{\nu_{\mu}}^{h\nu}}{dE}\,, \\
  \frac{dN_{\nu_{\tau}}}{dE}= &\BR\left( \gamma\,\nu_{\tau}\right) \delta\left( E-\frac{m_{3/2}}{2}\right) +\BR\left( W^{\pm}\tau^{\mp}\right) \frac{dN_{\nu_{\tau}}^{W\tau}}{dE} \\
  &\qquad+\BR\left( Z^0\nu_{\tau}\right) \frac{dN_{\nu_{\tau}}^{Z\nu}}{dE}+\BR\left( h\,\nu_{\tau}\right) \frac{dN_{\nu_{\tau}}^{h\nu}}{dE}\,. 
 \end{split}
\end{equation}
The resulting energy spectra for the different neutrino flavors are shown in Figure~\ref{energyspec}. In this case, and in most parts of this analysis, we consider a gravitino mass of $ m_{3/2}=150 $\usk GeV, motivated by the interpretation of anomalies in other cosmic ray channels as mentioned in the preface. Furthermore, at lower gravitino masses, the detection of any signal is much more difficult due to the lower neutrino yield and higher background fluxes. 
 
For this value of the gravitino mass, both the $ Z^0 $ and Higgs lines are visible in the spectrum in addition to the photon line. We see that the tau neutrino spectrum shows a very characteristic signature of two or three distinctive peaks (three in the region above the lightest Higgs threshold, as in this case) in addition to a continuum which is suppressed at low energies. From Figure \ref{BRplot} we see that the dominant line from the decay into $ Z^0\nu_{\tau} $ always has a branching ratio larger than 25\usk\%. The spectra of the other two flavors are very similar to each other, consisting only of a continuum contribution practically following a power law behavior $\propto E^{-2}$ at energies below the sharp threshold at half the gravitino mass.  

\chapter{Neutrinos as a Probe of Gravitino Dark Matter}
\label{neutrino}
In the previous chapter we have seen that the decay of the LSP gravitino in $ R $-parity breaking models produces stable Standard Model particles with characteristic energy spectra. The consequences of astrophysical gravitino dark matter decays have already been studied in several works: For instance, the gamma-ray signal from gravitino decays is studied in~\cite{Ibarra:2007wg, Bertone:2007aw}, while the antimatter signals (i.e. fluxes of positrons and antiprotons) are discussed in~\cite{Ibarra:2008qg, Ishiwata:2008cu, Tran:2008zz}. 

In this work we consider the neutrino signal at the Earth arising from gravitino dark matter decays and discuss the implications for the detection of these neutrino fluxes. In addition, we will use the neutrino signal to place constraints on the gravitino parameter space.

\section{Neutrino Signals}

\subsection{Neutrino Fluxes}
In this section we determine the neutrino flux at the Earth originating from gravitino decays. In this context the flux is defined as the number of particles per unit area and unit time. In fact, however, in most parts we will work with the differential flux per unit solid angle: 
\begin{equation}
 J\equiv\frac{d\phi}{d\Omega}\equiv\frac{dN}{dA\,dt\,d\Omega}\,. 
\end{equation}
There are two sources for a diffuse neutrino background: Gravitinos decaying in the dark matter halo of the Milky Way and those decaying in the halos of other galaxies at cosmological distances. 

\paragraph{Extragalactic Contribution}
Gravitino decays at cosmological distances give rise to a redshifted spectrum. A neutrino, generated in a gravitino decay at the comoving distance $ \chi(z) $ with energy $ E_0 $, is observed on Earth with an energy $ E=E_0/(1+z) $ due to the redshift of relativistic particles in an expanding universe. The number of generated neutrinos per unit energy and time in a comoving volume element located at the comoving coordinates $ \vec{\chi}=\left( \chi,\,\theta,\,\phi \right) $ is given by 
\begin{equation}
 \frac{dN_0(\vec{\chi})}{dE_0\,dt}=\frac{n_{3/2}(\vec{\chi})}{\tau_{3/2}}\,\frac{dN_{\nu}}{dE_0}\,d^3\chi=\frac{n_{3/2}(\vec{\chi})}{\tau_{3/2}}\,\frac{dN_{\nu}}{dE_0}\,\chi^2\sin\theta\, d\chi\, d\theta\, d\phi\,. 
\end{equation}
Here, the number density of gravitinos $ n_{3/2}(\vec{\chi})=\Omega_{3/2}\varrho_c/m_{3/2} $ is assumed to be constant on cosmological scales and $ dN_{\nu}/dE_0 $ is the neutrino energy spectrum from a single gravitino decay (cf. previous chapter). Since the gravitino lifetime is much larger than the age of the universe, the decrease of $ n_{3/2} $ through decays can be neglected. 

An observer at the origin of the coordinate system receives a differential flux per unit energy of 
\begin{equation}
 \frac{dJ_{eg}}{dE}=\int_0^{\infty}\frac{d\chi}{4\pi\chi^2}\frac{dN_0(\vec{\chi})}{dE\,dt\,d\Omega}=\frac{\Omega_{3/2}\varrho_c}{4\pi\,\tau_{3/2}\,m_{3/2}}\int_0^{\infty}\frac{dN_{\nu}}{d\left( E\left( 1+z\right) \right) }\,d\chi\,. 
\end{equation}
In order to perform the integration over the comoving distance, we have to use the dynamics of the universe presented in Section~\ref{cosmology}. Neutrinos propagate very similar to light since they are neutral and highly relativistic. Thus, we have for the line element according to the Friedmann--Robertson--Walker metric~(\ref{FRW}) 
\begin{equation}
 0\simeq ds^2=dt^2-a^2(t)\,d\chi^2\,,\quad\text{so that}\qquad d\chi=\frac{dt}{a(t)}\,. 
\end{equation}
Using the definition of the redshift (\ref{redshift}), we can turn the dependence on time into a dependence on the redshift: 
\begin{equation}
 \frac{dz}{dt}=\frac{d}{dt}\left( \frac{a_0}{a}\right) =-\frac{a_0}{a^2}\,\dot{a}=-\frac{a_0}{a}\,H\,.
\end{equation}
The Hubble parameter is described by the Friedmann equation (\ref{friedmann}). In accordance with observations we assume a spatially flat matter- and cosmological constant-dominated universe with $ \Omega_{\Lambda}+\Omega_m=1 $ and $ \kappa\simeq\Omega_{\Lambda}/\Omega_m\approx 3 $. These parameters describe very well the cosmology of our universe up to $ z=\mathcal{O}(1000) $, where the radiation content plays no dominant role. In this case, the Hubble parameter is given as 
\begin{equation}
 H(z)=H_0\left( 1+z\right) ^{3/2}\sqrt{\Omega_m\left( 1+\kappa\left( 1+z\right) ^{-3}\right) }
\end{equation}
and the differential of the comoving distance becomes 
\begin{equation}
 d\chi=\frac{(1+z)^{-3/2}\,dz}{a_0H_0\sqrt{\Omega_m\left( 1+\kappa\left( 1+z\right) ^{-3}\right) }}\,. 
\end{equation}

Consequently, we obtain for the extragalactic contribution to the differential neutrino flux per unit energy 
\begin{equation}
 \frac{dJ_{eg}}{dE}=A_{eg}\int_1^{1+z_{dec}}dy\frac{dN_{\nu}}{d(E\,y)}\frac{y^{-3/2}}{\sqrt{1+\kappa\, y^{-3}}}\,, 
\end{equation}
where $ y\equiv 1+z $ and $ z_{dec} $ is the time when the neutrinos decouple from the thermal plasma and start to propagate freely.\footnote{In fact, the expression for the redshift integral is valid only from the beginning of the matter dominated universe onwards. However, since the contributions from high $ z $ are negligible, we can actually perform the integration up to $ y=\infty $.} The coefficient is given by 
\begin{equation}
 \begin{split}
  A_{eg} &=\frac{\Omega_{DM}\varrho_c}{4\pi\,\tau_{3/2}\,m_{3/2}H_0\Omega_m^{1/2}} \\
  &=1.1\times 10^{-7}\usk(\unit{cm^2\usk s\usk sr})^{-1}\left( \frac{\tau_{3/2}}{1.3\times 10^{26}\usk\unit{s}}\right) ^{-1}\left( \frac{m_{3/2}}{150\usk\unit{GeV}}\right) ^{-1}. 
 \end{split}
\end{equation}
To obtain this result we have taken the gravitino density to equal the cold dark matter density, $ \Omega_{3/2}h^2=\Omega_{DM}h^2\simeq 0.1 $, and the other constants as $ \varrho_c\simeq 1.05\, h^2\times 10^{-5}\usk\unit{GeV\usk cm^{-3}} $, $ \Omega_m\simeq 0.25 $ and $ h\simeq 0.73 $. 

For a monochromatic spectrum, the redshift integral can be solved analytically and results in~\cite{Buchmuller:2007ui} 
\begin{equation}
 \frac{dJ_{eg}}{dE}=A_{eg}\,\frac{2}{m_{3/2}}\left[ 1+\kappa\left( \frac{2\,E}{m_{3/2}}\right) ^3\right] ^{-1/2}\left( \frac{2\,E}{m_{3/2}}\right) ^{1/2}\Theta\left( 1-\frac{2\,E}{m_{3/2}}\right) , 
\end{equation}
where $ \Theta(x) $ denotes the Heaviside step function. This expression produces a characteristic 'triangular' contribution to the spectrum that terminates abruptly at the threshold.

\paragraph{Halo Contribution}
In addition to the isotropic extragalactic signal there exists an anisotropic signal from gravitino decays in the Milky Way halo. Thanks to the small extension of the Milky Way compared to cosmological scales, we can work with distances in a flat space-time. Since we are interested in the signal observed from Earth, it is convenient to use galactic coordinates, that have their origin in the Sun. In these coordinates spatial positions are specified by the distance $ s $ to the Sun, the galactic latitude $ b $ and the galactic longitude $ l $. Analogous to the extragalactic case, the number of neutrinos from gravitino decay per unit energy and time in a volume element located at the coordinates $ \vec{l}=\left( s,\,b,\,l \right) $ is given by
\begin{equation}
 \frac{dN(\vec{l})}{dE\,dt}=\frac{n_{3/2}(\vec{l})}{\tau_{3/2}}\,\frac{dN_{\nu}}{dE}\, s^2\cos b\, ds\, db\, dl\,. 
\end{equation}
However, in this case the gravitino number density depends on the position in the galactic halo: 
\begin{equation}
 n_{3/2}(\vec{l})=\frac{\varrho_{halo}(r(\vec{l}))}{m_{3/2}}\,, 
\end{equation}
where we assume that the Milky Way dark matter halo consists only of gravitinos. The halo density distribution depends on the halo model but is spherically symmetric with respect to the galactic center in any case. A list of established halo models is given in Section~\ref{darkmatter}. 

The distance $ r $ in the density profiles is given with respect to the galactic center, so we need to express it in terms of galactic coordinates. We find 
\begin{equation}
 r(s,\,b,\,l)=\sqrt{s^2+r_{\odot}^2-2\,s\,r_{\odot}\cos{b}\cos{l}}\,, 
\end{equation}
where $ r_{\odot}\simeq 8.5\usk\unit{kpc} $ is the radius of the solar orbit around the galactic center~\cite{Bertone:2007aw}. Due to the far distance to the galactic center the observed signal is anisotropic. It is given by 
\begin{equation}
 \frac{dJ_{halo}(b,\,l)}{dE}=A_{halo}(b,\,l)\,\frac{dN_{\nu}}{dE}\,, 
\end{equation}
where the halo flux parameter is given by the line-of-sight integral 
\begin{equation}
 A_{halo}(b,\,l)=\frac{1}{4\pi\,\tau_{3/2}\,m_{3/2}}\int_0^{s_{halo}}\varrho_{halo}(r(s,\,b,\,l))\,ds\,. 
\end{equation}
In this expression $ s_{halo} $ is the extension of the dark matter halo in the corresponding direction.\footnote{For the density profiles given in Section~\ref{darkmatter} the halo radius can safely taken to be infinite since the profiles are falling off at least proportional to $ r^{-2} $ for large radii.} Since the anisotropy is not very strong if the galactic center is excluded, we will use in the following analysis only an averaged signal. The averaged full-sky halo flux parameter is computed via\footnote{Due to the symmetry of the halo profile we actually only integrate over one hemisphere in the galactic latitude.} 
\begin{equation}
 A_{halo}=\frac{1}{2\pi}\int_0^{2\pi}dl\int_0^{\pi/2}A_{halo}(b,\,l)\,\cos{b}\,db\,, 
\end{equation}
and the averaged halo flux parameter excluding the galactic disk (galactic latitude $ -10^{\circ}\leq b\leq 10^{\circ} $ excluded) is given by 
\begin{equation}
 A_{excl. disk}=\left( \int_0^{2\pi}dl\int_{\pi/18}^{\pi/2}A_{halo}(b,\,l)\,\cos{b}\,db\right) \times\left( 2\pi\int_{\pi/18}^{\pi/2}\cos{b}\,db\right) ^{-1}. 
\end{equation}

It is also interesting to see how much the signal is increased in the direction of the galactic center. In order to perform an integration over a cone around the galactic center, it is convenient to change to new angular coordinates $ \Theta $ and $ \Phi $ by the transformations 
\begin{equation}
 \begin{split}
  \cos{b}\cos{l}\quad &\rightarrow\quad\cos{\Theta}\,, \\
  \cos{b}\,db\,dl\quad &\rightarrow\quad\sin{\Theta}\,d\Theta\,d\Phi\,. 
 \end{split}
\end{equation}
The angle $ \Theta $ describes the declination of the line of sight to the direction towards the galactic center and the angle $ \Phi $ gives the position in the rotationally symmetric direction around the connecting line between the Sun and the galactic center. The averaged halo flux parameter from a cone towards the galactic center with a half-cone opening angle of $ 5^{\circ} $ can then be computed via 
\begin{equation}
 A_{GC}=\left( \int_0^{2\pi}d\Phi\int_0^{5\pi/180}A_{halo}(\Theta,\,\Phi)\,\sin{\Theta}\,d\Theta\right) \times\left( 2\pi\int_0^{5\pi/180}\sin{\Theta}\,d\Theta\right) ^{-1}. 
\end{equation}

\begin{table}
 \centering 
 \begin{tabular}{lccc}
  \toprule
  & $ A_{halo} $ & $ A_{excl. disk} $ & $ A_{GC} $ \\
  \midrule
  NFW & $ 6.8\times 10^{-8} $ & $ 6.5\times 10^{-8} $ & $ 4.2\times 10^{-7} $ \\
  NFW$ _c $ & $ 7.1\times 10^{-8} $ & $ 6.4\times 10^{-8} $ & $ 9.4\times 10^{-7} $ \\
  Moore & $ 6.9\times 10^{-8} $ & $ 6.4\times 10^{-8} $ & $ 7.6\times 10^{-7} $ \\
  Moore$ _c $ & $ 7.2\times 10^{-8} $ & $ 6.4\times 10^{-8} $ & $ 1.3\times 10^{-6} $ \\
  Kra & $ 5.8\times 10^{-8} $ & $ 5.6\times 10^{-8} $ & $ 1.9\times 10^{-7} $ \\
  iso & $ 7.0\times 10^{-8} $ & $ 6.7\times 10^{-8} $ & $ 2.5\times 10^{-7} $ \\
  \bottomrule
 \end{tabular}
 \caption[Halo flux parameters for different halo models.]{Halo flux parameters for different halo models for the averaged full-sky signal, the signal excluding the galactic disk and the signal from the galactic center, respectively. The numerical values for the parameters are given in units of $ (\unit{cm^2\usk s\usk sr})^{-1}(1.3\times 10^{26}\usk\unit{s}/\tau_{3/2})(150\usk\unit{GeV}/m_{3/2}) $.}
 \label{haloflux}
\end{table}
The numerical halo flux parameters obtained from different halo profiles are presented in Table~\ref{haloflux}. As expected from the different slopes in the inner parts of the Milky Way (see Figure~\ref{rotationcurve}), the results differ most for the cone directed towards the galactic center. By contrast, the results for the averaged halo flux parameters excluding the galactic disk agree within a few percent. 

Thus, in the following sections we will only present the results obtained using the Navarro--Frenk--White profile that is given by (cf. Section~\ref{darkmatter}) 
\begin{equation}
 \varrho_{NFW}(r)=\frac{\varrho_0}{\left( r/r_c\right) \left[ 1+\left( r/r_c\right) \right] ^2}\,, 
\end{equation}
with $ r_c=20\usk\unit{kpc} $ and a normalization $ \varrho_0=0.26\usk\unit{GeV\usk cm^{-3}} $. The halo flux parameter for the averaged signal excluding the galactic disk is given, in this case, by 
\begin{equation}
 A^{\text{NFW}}_{excl. disk}=6.5\times 10^{-8}\usk(\unit{cm^2\usk s\usk sr})^{-1}\left( \frac{1.3\times 10^{26}\usk\unit{s}}{\tau_{3/2}}\right) \left( \frac{150\usk\unit{GeV}}{m_{3/2}}\right) . 
\end{equation}
This value will be used in most parts of the further analysis.

\subsection{Neutrino Propagation and Neutrino Oscillations}
After being produced in gravitino decays, the neutrinos propagate over astrophysical distances until they can be observed on Earth. While propagating the neutrinos undergo flavor oscillations. Following the review in~\cite{Strumia:2006db} we will thus shortly summarize the relevant consequences of neutrino oscillations. 

The neutrino gauge eigenstates $ \nu_e $, $ \nu_{\mu} $ and $ \nu_{\tau} $ do not coincide with the mass eigenstates $ \nu_i,\;i=1,\,2,\,3 $. Since the neutrino masses are negligible but different from each other, there arise oscillations between the different neutrino flavors. The quantum mechanical treatment of neutrino oscillations leads to formulae for the probability that a neutrino with specific flavor is converted into a neutrino with another flavor. There are several general constraints for these probabilities: $ CPT $-invariance implies 
\begin{equation}
 P(\nu_l\rightarrow\nu_{l'})=P(\bar{\nu}_{l'}\rightarrow\bar{\nu}_l)\,. 
 \label{CPTconstraint}
\end{equation}
If we assume $ CP $-invariance (and thus also $ T $-invariance), we have 
\begin{equation}
 \begin{split}
  P(\nu_l\rightarrow\nu_{l'})=P(\bar{\nu}_l\rightarrow\bar{\nu}_{l'})\,, \\
  P(\nu_l\rightarrow\nu_{l'})=P(\nu_{l'}\rightarrow\nu_l)\,. 
  \label{CPconstraint}
 \end{split}
\end{equation}
The conversion probabilities for the different flavors are then given by~\cite{Strumia:2006db}
\begin{align}
 P(\nu_e\leftrightarrow\nu_{\mu}) &=s_{23}^2\sin^22\theta_{13}S_{23}+c_{23}^2\sin^22\theta_{12}S_{12}\,, \nonumber\\
 P(\nu_e\leftrightarrow\nu_{\tau}) &=c_{23}^2\sin^22\theta_{13}S_{23}+s_{23}^2\sin^22\theta_{12}S_{12}\,, \\
 P(\nu_{\mu}\leftrightarrow\nu_{\tau}) &=c_{13}^4\sin^22\theta_{23}S_{23}-s_{23}^2c_{23}^2\sin^22\theta_{12}S_{12}\,, \nonumber
 \intertext{while the survival probabilities are} 
 P(\nu_e\rightarrow\nu_e) &=1-\sin^22\theta_{13}S_{23}-c_{13}^4\sin^22\theta_{12}S_{12}\,, \nonumber\\  P(\nu_{\mu}\rightarrow\nu_{\mu}) &=1-4\,c_{13}^2s_{23}^2(1-c_{13}^2s_{23}^2)S_{23}-c_{23}^4\sin^22\theta_{12}S_{12}\,, \\
 P(\nu_{\tau}\rightarrow\nu_{\tau}) &=1-4\,c_{13}^2c_{23}^2(1-c_{13}^2c_{23}^2)S_{23}-s_{23}^4\sin^22\theta_{12}S_{12}\,. \nonumber
\end{align}
In the above expressions we used\index{} the abbreviations 
\begin{equation}
 s_{ij}=\sin\theta_{ij}\,,\quad c_{ij}=\cos\theta_{ij}\qquad\text{and}\qquad S_{ij}=\sin^2\left( \frac{\Delta m_{ij}^2L}{4\,E}\right) . 
\end{equation}
Using convenient units, we have for the last expression 
\begin{equation}
 S_{ij}\simeq\sin^2\left( 1.27\,\frac{\Delta m_{ij}^2\usk(\unit{eV}^2)\,L\usk(\unit{km})}{E\usk(\unit{GeV})}\right) . 
\end{equation}
From the roots of this function we can infer the oscillation length for two-flavor oscillations: 
\begin{equation}
 \lambda_{ij}=\frac{4\pi E}{\Delta m_{ij}^2}\simeq 2.48\usk\unit{km}\,\frac{E\usk(\unit{GeV})}{\Delta m_{ij}^2\usk(\unit{eV^2})}\,. 
 \label{oscillationlength}
\end{equation}

The experimental best-fit values for the neutrino mixing angles $ \theta_{12} $, $ \theta_{23} $ and $ \theta_{13} $, and the neutrino mass differences $ \Delta m_{ij}^2=m_j^2-m_i^2 $ are given in Table~\ref{oscillationparameter}. 
\begin{table}
 \centering 
 \begin{tabular}{ccccc}
  \toprule
  $ \sin^2{\theta_{12}} $ & $ \sin^2{\theta_{23}} $ & $ \sin^2{\theta_{13}} $ & $ \Delta m_{12}^2\usk(\unit{eV^2}) $ & $ \abs{\Delta m_{13}^2}\usk(\unit{eV^2}) $ \\
  \midrule
  $ 0.304 $ & $ 0.50 $ & $ 0.01 $ & $ 7.65\times 10^{-5} $ & $ 2.40\times 10^{-3} $ \\
  \bottomrule
 \end{tabular}
 \caption[Neutrino oscillation parameters.]{Neutrino oscillation parameters. Figures taken from~\cite{Schwetz:2008er}. }
 \label{oscillationparameter}
\end{table}
From the value of $ \theta_{23} $ we see that there is maximal mixing between muon and tau neutrinos. \smallskip

The cosmological dark matter density is spatially constant and the dark matter density in the galactic halo varies only on scales that are large compared to the oscillation lengths at energies in the GeV range. This leads to an averaging of the neutrino oscillations from the viewpoint of the observer. The huge path length combined with the finite energy resolution of detectors adds to this effect. Therefore, we are in the so-called long baseline limit, where $ S_{12}=S_{23}=1/2 $. 

Neglecting the small mixing angle $ \theta_{13} $, the resulting oscillation probabilities are 
\begin{equation}
 \begin{split}
  P(\nu_e\rightarrow\nu_e) &=0.56\,, \\
  P(\nu_e\leftrightarrow\nu_{\mu})=P(\nu_e\leftrightarrow\nu_{\tau}) &=0.22\,, \\
  P(\nu_{\mu}\rightarrow\nu_{\mu})=P(\nu_{\mu}\leftrightarrow\nu_{\tau})=P(\nu_{\tau}\rightarrow\nu_{\tau}) &=0.39\,. 
 \end{split}
\label{oscprob}
\end{equation}
All other oscillation probabilities are determined by the relations (\ref{CPTconstraint}) and (\ref{CPconstraint}). We note that even when the primary neutrino flux is originally mainly composed of tau neutrinos, the flavor oscillations during the propagation produce comparable fluxes in all neutrino flavors. In particular, due to the maximal mixing between muon and tau flavor, the signals for the muon and tau neutrinos are identical after propagation. Thus, the initial assumption that the $ R $-parity breaking coupling is generated through a sneutrino VEV in the tau generation is not crucial. 

\begin{figure}
 \centering
 \includegraphics[scale=1.2]{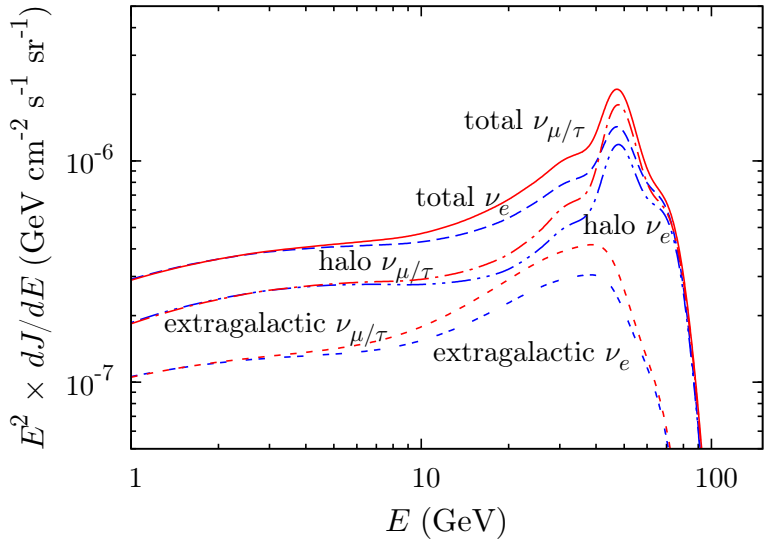} 
 \caption[Neutrino fluxes after propagation.]{Total neutrino fluxes and the extragalactic and halo contributions for the different neutrino flavors after propagation using an energy resolution of $ 10 $\usk\%. The gravitino mass and lifetime are chosen to be $ m_{3/2}=150 $\usk GeV and $ \tau_{3/2}=1.3\times 10^{26} $\usk s.}
 \label{eghalo}
\end{figure}
The fluxes for the different neutrino flavors and their extragalactic and halo contributions are shown in Figure \ref{eghalo}. As mentioned before, we use the averaged halo flux excluding the galactic disk. Additionally, we employ an energy resolution of $ \Delta E/E=10\usk\% $. Therefore, we convolute the neutrino flux with a Gaussian distribution 
\begin{equation}
 \Phi(E,E')=\frac{1}{\sqrt{2\pi}\,\sigma}\,e^{-\frac{1}{2}\left( \frac{E-E'}{\sigma}\right) ^2} 
\end{equation}
in the following way: 
\begin{equation}
 \frac{dJ_{\nu,\,\text{Gauss}}}{dE}=\int_{-\infty}^{\infty}\Phi(E,E')\,\frac{dJ_{\nu}}{dE'}\,dE'\,. 
\end{equation}
The standard deviation of the Gaussian distribution is chosen according to the energy resolution: $ \sigma=\Delta E/E $.
 
Note that even with this optimistic assumption for the energy resolution the lines from the decay into $ \gamma\,\nu_\tau $ and $ h\,\nu_\tau $ become practically indistinguishable from the continuum whereas the line from the decay into $ Z^0\nu_\tau $ can be resolved. Its position could allow a determination of the gravitino mass, even without determining the endpoint of the spectrum. 

In addition, we note that the extragalactic contribution to the neutrino fluxes is subdominant. Nevertheless, we will include this contribution in the numerical results.

\section{Neutrino Background}
The detection of a possible neutrino signal from gravitino decay is hindered by considerable neutrino backgrounds. Namely, in the energy range of 1--1000\usk GeV, there exist large background neutrino fluxes produced by interactions of cosmic rays with the Earth's atmosphere or with the solar corona, as well as neutrino fluxes from distant galactic sources. Let us briefly discuss these backgrounds separately.

\subsection{Atmospheric Neutrinos}
Collisions of energetic cosmic rays with the Earth's atmosphere initiate cascades that generate atmospheric neutrinos. When the primary cosmic rays hit the nuclei of air in the upper part of the atmosphere, mostly pions and some kaons are produced. The charged pions decay dominantly into muons and muon neutrinos via 
\begin{equation*}
 \pi^+\rightarrow\mu^+\nu_{\mu}\qquad\text{and}\qquad\pi^-\rightarrow\mu^-\bar{\nu}_{\mu}\,, 
\end{equation*}
while the decay into electrons and electron neutrinos is suppressed by $ m_e^2/m_{\mu}^2 $. Before decaying, the muons produced in the pion decays travel an average distance 
\begin{equation*}
 \bar{d}_{\mu}\approx c\,\tau_{\mu}\gamma_{\mu}\approx 1\usk\unit{km}\,\frac{E_{\mu}}{0.3\usk\unit{GeV}}\,, 
\end{equation*}
where $ \tau_{\mu} $ is the muon lifetime and $ \gamma_{\mu}=E_{\mu}/m_{\mu} $ is the relativistic dilatation factor. If the muons decay via 
\begin{equation*}
 \mu^-\rightarrow e^-\,\bar{\nu}_e\,\nu_{\mu}\qquad\text{and}\qquad\mu^+\rightarrow e^+\,\nu_e\,\bar{\nu}_{\mu} 
\end{equation*}
before they hit the Earth's surface, one obtains a flux ratio of 2:1 for muon and electron neutrinos and antineutrinos. However, for energies above a few GeV the muons hit the surface before decaying. Therefore, at higher energies the ratio $ (\nu_{\mu}+\bar{\nu}_{\mu})/(\nu_e+\bar{\nu}_e) $ is larger than 2. 

In order to determine the fluxes of atmospheric neutrinos one has to consider several effects. First of all, the spectrum of cosmic rays and their chemical composition has to be known. For energies above a few GeV the cosmic ray flux has an energy dependence $ d\phi/dE\propto E^{-3.7} $ and most of its constituent particles are protons. At low energies ($ \lesssim 10\usk\unit{GeV} $) the cosmic ray flux is modulated by the solar wind and is therefore dependent on the eleven-year-cycle of solar activity. 

Another important effect on lower energy cosmic rays comes from the geomagnetic field. Since the cosmic ray particles travel along the field lines at the poles and perpendicular to the field lines at the equator, this effect is dependent on the position on Earth. So there exists a position-dependent geomagnetic cutoff in the magnetic rigidity $ \mathcal{R}=p/(Ze) $, where $ p $ is the momentum and $ Ze $ the charge of the cosmic rays. Particles with a magnetic rigidity below the cutoff do not reach the atmosphere. 

These effects and the details of the interactions and decays in the atmosphere can be incorporated into a Monte Carlo procedure, making it possible to calculate the atmospheric neutrino fluxes with high accuracy. The largest uncertainties stem from the initial cosmic ray spectrum and the hadronic interactions, leading to a $ \sim 20 $\usk\% uncertainty in the absolute neutrino fluxes. However, the ratios of the fluxes of the different neutrino flavors are accurate to better than 5\usk\%. 

At energies above 1\usk GeV the fluxes of the atmospheric neutrinos obey an approximate power law: 
\begin{equation}
 \frac{d\phi_{\nu_e}}{dE}\propto E^{-3.5}\qquad\text{and}\qquad\frac{d\phi_{\nu_{\mu}}}{dE}\propto E^{-3}
\end{equation}
for electron and muon neutrinos, respectively. At lower energies the slope is shallower due to geomagnetic effects and solar modulation. \smallskip

In this work we use the fluxes of atmospheric electron and muon neutrinos computed by Battistoni \textit{et al.} with the Monte Carlo package FLUKA (Fluktuierende Kaskade) assuming massless neutrinos~\cite{Battistoni:1999at}. For the energy range 0.1--112\usk GeV we use the fluxes for the Kamioka site including the geomagnetic cutoff~\cite{Battistoni:2001url}. To be conservative, we use the larger fluxes from the minimum solar activity sample. For the energy range 25--10000\usk GeV only the muon neutrino flux is available~\cite{Battistoni:2003ju}. 

The main tau neutrino background stems from atmospheric muon neutrinos that oscillate into tau neutrinos on the way to the detector. From equation (\ref{oscillationlength}) we see that the oscillation length on Earth is too short for a significant conversion of electron neutrinos. Therefore, we apply two-flavor neutrino oscillations between muon and tau neutrinos to the initial fluxes. Since the mixing between muon and tau neutrinos is maximal ($ \sin2\theta_{23}\simeq1 $), the conversion probability is given by 
\begin{equation}
 P(\nu_{\mu}\rightarrow\nu_{\tau})\simeq\sin^2\left( 1.27\,\frac{\abs{\Delta m_{13}^2}\!(\unit{eV}^2)\,L\usk(\unit{km})}{E\usk(\unit{GeV})}\right) . 
\end{equation}
In this expression $ E $ is the neutrino energy and $ L $ their propagation length after the production in the atmosphere. The latter depends on the zenith angle and is given by 
\begin{equation}
 L=\sqrt{(R_{\oplus}\cos\theta)^2+2R_{\oplus}h+h^2}-R_{\oplus}\cos\theta\,, 
\end{equation}
with $ R_{\oplus}\simeq 6.4\times 10^6\usk\unit{m} $ being the mean Earth radius and $ h\simeq 15\usk\unit{km} $ being the mean altitude at which the muon neutrinos are produced~\cite{Athar:2004uk}. Moreover, the relevant neutrino parameters for atmospheric oscillations are given in Table~\ref{oscillationparameter}.

\subsection{Further Neutrino Background Sources}
In addition to the flux of tau neutrinos originating from the conversion of muon neutrinos, there exists an intrinsic contribution from the decay of charmed particles that are produced in the atmosphere in cosmic ray interactions. This flux is suppressed by a factor of about $ 10^6 $ compared to the flux of electron and muon neutrinos from pion and kaon decay. 

This intrinsic contribution has been computed by Pasquali and Reno~\cite{Pasquali:1998xf} in a PYTHIA based calculation and can be parameterized as 
\begin{equation}
 \frac{dJ_{\nu_{\tau}}}{dE}=\frac{1}{E^3}\,10^{-A+Bx-Cx^2-Dx^3}\usk\unit{GeV^2\usk cm^{-2}\usk s^{-1}\usk sr^{-1}}, 
\end{equation}
where $ x=\log_{10}E\usk(\unit{GeV}) $, $ A=6.69 $, $ B=1.05 $, $ C=0.150 $ and $ D=-0.00820 $. The next-to-leading-order perturbative QCD calculation also presented in the paper gives lower fluxes at energies below several TeV and is therefore less conservative. \smallskip

Analogous to the production of neutrinos in the Earth's atmosphere, neutrinos are produced in the solar corona by cosmic ray collisions. This neutrino flux has been studied by Ingelman and Thunman in~\cite{Ingelman:1996mj}, who found that the flux of electron and muon neutrinos integrated over the solar disk can be described by the following parameterization: 
\begin{equation}
 \frac{d\phi_{\nu_{e,\,\mu}}}{dE}=\unit{GeV^{-1}\usk cm^{-2}\usk s^{-1}}\left\lbrace 
 \begin{array}{cc}
 \frac{N_0\left( E\usk(\unit{GeV})\right) ^{-\gamma-1}}{1+AE\usk(\unit{GeV})}, & E<E_0\,, \\
 \frac{N_0'\left( E\usk(\unit{GeV})\right) ^{-\gamma'-1}}{1+AE\usk(\unit{GeV})}, & E>E_0\,. 
 \end{array}\right. 
\end{equation}
It is valid for $ 10^2\usk\unit{GeV} \leq E \leq 10^8\usk\unit{GeV} $ and the numerical values of the parameters are given in Table~\ref{coronatab} for both neutrino flavors. 
\begin{table}
 \centering
 \begin{tabular}{ccccccc}
  \toprule 
  Flavor & $ N_0 $ & $ \gamma $ & $ A $ & $ E_0\usk(\unit{GeV}) $ & $ \gamma' $ & $ N_0' $ \\
  \midrule 
  $ \nu_e+\bar{\nu}_e $ & $ 7.4\times 10^{-6} $ & $ 2.03 $ & $ 8.5\times 10^{-6} $ & $ 1.2\times 10^6 $ & $ 2.33 $ & $ 5.0\times 10^{-4} $ \\
  $ \nu_{\mu}+\bar{\nu}_{\mu} $ & $ 1.3\times 10^{-5} $ & $ 1.98 $ & $ 8.5\times 10^{-6} $ & $ 3.0\times 10^6 $ & $ 2.38 $ & $ 5.1\times 10^{-3} $ \\
  \bottomrule 
 \end{tabular}
 \caption[Values for the parameterization of the corona neutrino flux.]{Values for the parameterization of the corona electron and muon neutrino flux.}
 \label{coronatab}
\end{table}

The electron and muon neutrinos and antineutrinos produced in the solar corona oscillate during their propagation to Earth. Considering the large distance traveled (compared to the oscillation lengths), the conversion
and survival probabilities can be averaged, and the fluxes on Earth in the different flavors can be obtained from equation~(\ref{oscprob}). \smallskip

Last, the fluxes of tau neutrinos that originate from galactic sources are discussed by Athar, Lee and Lin in~\cite{Athar:2004um}. For the tau neutrino flux from the galactic plane in the presence of neutrino oscillations they find the parameterization 
\begin{equation}
 \frac{dJ_{\nu_{\tau}}}{dE}=9\times 10^{-6}\usk\unit{GeV^{-1}\usk cm^{-2}\usk s^{-1}\usk sr^{-1}}\left( E\usk(\unit{GeV})\right) ^{-2.64}, 
\end{equation}
that is valid in the energy range $ 1\usk\unit{GeV}\leq E\leq 10^3\usk\unit{GeV} $.

\section{Reduction of the Neutrino Background}
\begin{figure}
 \centering
 \includegraphics[scale=1.2]{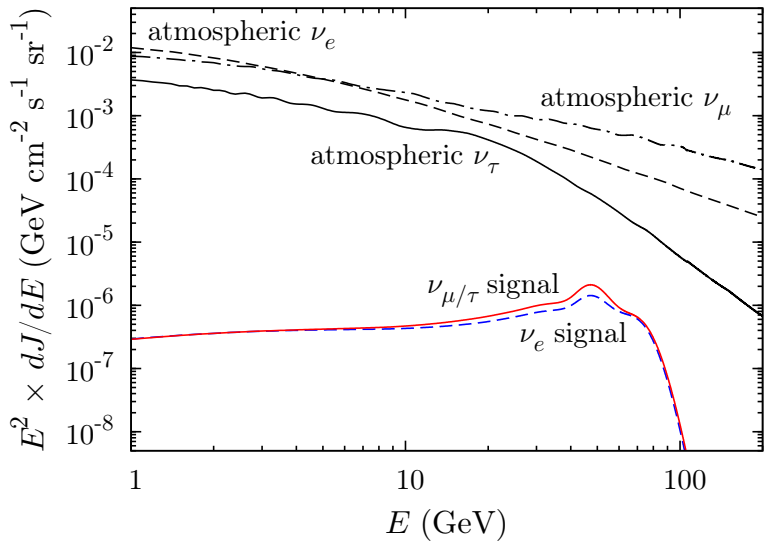} 
 \caption[Averaged full-sky neutrino fluxes expected at the Super-K site.]{Averaged full-sky neutrino fluxes expected at the Super-K site. The gravitino parameters and the energy resolution are chosen as in Figure~\ref{eghalo}.}
 \label{nutotal}
\end{figure}
The averaged full-sky signal for the neutrinos from gravitino decay is shown in Figure~\ref{nutotal} together
with the results for the atmospheric background from FLUKA. The signal lies several orders of magnitude below the expected atmospheric background for all flavors. Therefore, we find that the interpretation of the EGRET and HEAT anomalies in terms of gravitino decay is compatible with neutrino flux measurements, as it does not lead to an overproduction of neutrinos. 

Going beyond this consistency check, we will examine in the following section the possibility of detecting this extraordinary flux contribution in neutrino experiments. Due to the low signal-to-background ratio, the signal cannot be detected directly. It will therefore be necessary to find strategies for effectively reducing the background in order to have any chance of detecting the signal. As is apparent from Figure~\ref{nutotal}, the tau neutrino channel appears to be the most promising of the three flavors, since it has the lowest background. 

In general, the neutrino spectrum from gravitino decay has some very characteristic features that could allow to
distinguish it from the featureless backgrounds, but the question is whether neutrino detectors are able to reach sufficient sensitivity to resolve these features.

\subsection{Electron and Muon Neutrinos}
For the electron and muon neutrinos, that are more easily detected in neutrino observatories than tau neutrinos, the signal-to-background ratio is very small ($ \sim 10^{-3} $--$ 10^{-4} $) for gravitino lifetimes that are not already excluded by bounds from the gamma-ray~\cite{Ibarra:2007wg, Bertone:2007aw} or antimatter signals~\cite{Ibarra:2008qg, Ishiwata:2008cu}. As can be seen from Figure~\ref{nutotal}, even the peak of the spectrum is three orders of magnitude below the background. This would make distinguishing an extraordinary signal from the background appear extremely difficult. 

Unfortunately, we could not find a suitable strategy to sufficiently reduce this background, for example by exploiting directionality. In general, the atmospheric neutrino baseline ($ L\lesssim 2\,R_{\oplus} $) is too short for a complete conversion of muon neutrinos into another flavor at energies on the order of 50--100\usk GeV (cf. equation (\ref{oscillationlength})). Therefore, a detection of the signal without having prior knowledge of the position of the peak in the neutrino spectrum from gravitino decay seems hopeless. If information on the spectrum is available, for instance from the detection of a monochromatic gamma-ray line in the Fermi Gamma-ray Space Telescope (FGST, formerly known as the Gamma-ray Large Area Space Telescope (GLAST)~\cite{fgst:2008url}), one could envisage strategies to disentangle the neutrino signal from the background. Anyway, that would probably require a much better knowledge of the atmospheric neutrino flux at the relevant energies and a better energy resolution of neutrino detectors than presently available.

\subsection{Tau Neutrinos}
For the tau neutrinos the signal-to-background ratio is more promising since it lies above $ \sim 10^{-2} $ at the peak energy (cf. Figure~\ref{nutotal}). Moreover, most of the tau neutrino background from atmospheric oscillations can be effectively reduced by exploiting directionality. Since we are interested in energies of $ \mathcal{O} $(10--100)\usk GeV, we are in the region where the oscillation length for two-flavor oscillations between muon and tau neutrinos is larger than the Earth diameter. Thus, the conversion probability for down-going muon neutrinos (i.e. neutrinos from above the horizon) is approximately given by 
\begin{equation}
 P(\nu_{\mu}\rightarrow\nu_{\tau})\simeq\left( 1.27\,\frac{\abs{\Delta m_{13}^2}\!(\unit{eV}^2)\,L\usk(\unit{km})}{E\usk(\unit{GeV})}\right) ^2. 
\end{equation}
We have seen earlier that the atmospheric muon neutrino flux has an energy dependence $ \propto E^{-3} $ in the range above 1\usk GeV. Therefore, we expect for the down-going tau neutrinos a flux with an energy dependence 
\begin{equation}
 \frac{d\phi_{\nu_{\tau}}}{dE}\propto E^{-5}. 
\end{equation}

Due to the steeper slope, the previously subdominant tau neutrino background sources become important at higher energies. Anyway, these backgrounds can also be effectively reduced by exploiting directionality. In fact, since the solar corona neutrinos and the galactic neutrinos come from the direction of the Sun or from the galactic plane, these directions could simply be excluded from the search to reduce the background. This is why we excluded the galactic disk in the calculation of the halo contribution to the neutrino flux. 

\begin{figure}
 \centering
 \includegraphics[scale=1.05]{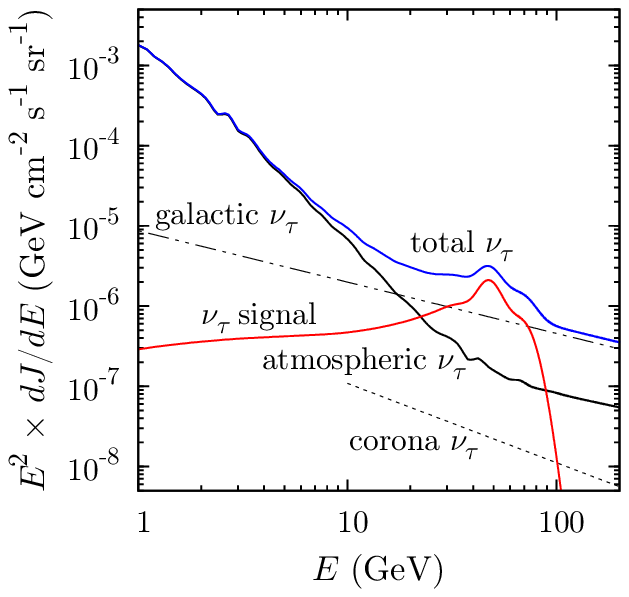} 
 \hspace{1pt}
 \includegraphics[scale=1.05]{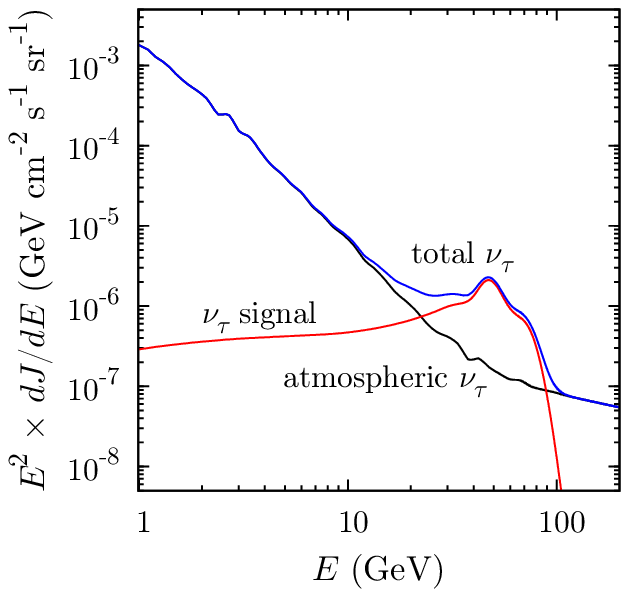} 
 \caption[Down-going tau neutrino fluxes expected at the Super-K site.]{Down-going tau neutrino fluxes expected at the Super-K site with (left) and without (right) the contribution from the galactic disk and the solar corona. The gravitino parameters and the energy resolution are chosen as in Figure~\ref{eghalo}.}
 \label{nutaudown}
\end{figure}
Thus---regarding only down-going tau neutrinos---the background can be reduced by several orders of magnitude. In Figure~\ref{nutaudown} we show the fluxes for down-going tau neutrinos at the Super-Kamiokande site. We see that in this case the signal can overcome the simulated background from FLUKA, even without cutting away the Sun or the galactic plane.

\section{Detection Prospects}
In this section we discuss the prospects for the detection of the extraordinary signal in the tau neutrino flavor in present and future neutrino experiments. For that purpose, we first briefly describe the detection technique of water \v{C}erenkov detectors following the reviews in~\cite{Strumia:2006db, GonzalezGarcia:2002dz}. 

\subsection{Water \texorpdfstring{\v{C}erenkov}{Cerenkov} Detectors}
Water \v{C}erenkov detectors like \textit{Super-Kamiokande} or \textit{IceCube} consist of pure water or ice and a large number of photomultiplier tubes (PMTs) that detect in real time the \v{C}erenkov light emitted by relativistic charged particles. 

Charged leptons can be generated by neutrinos, that cross the detector, via charged current (CC) scatterings off nuclei in the water ($ \nu_l\,N\rightarrow l\,N' $) and in the case of electron neutrinos also by elastic scatterings off electrons. Following these interactions, \v{C}erenkov light is emitted if the charged particle's velocity is larger than the speed of light in water. 

This technique works well for electron and muon neutrinos, which generate an electron and a muon in the CC interaction, respectively. Muons produce a sharp \v{C}erenkov ring, while electrons produce a diffuse ring. Thus, the neutrino flavor can be identified using the \v{C}erenkov light signature. On the other hand, it is not possible to distinguish particles from antiparticles. 

\begin{figure}
 \centering
 \includegraphics[scale=1.2]{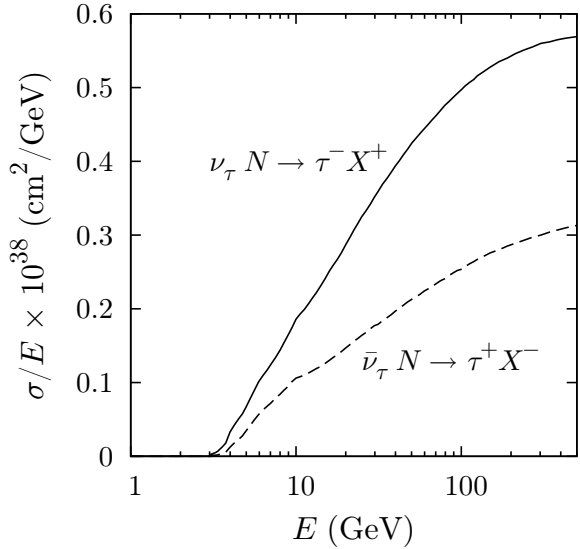} 
 \caption[Cross sections of CC tau neutrino interactions.]{Total cross sections of CC interactions for $ \nu_{\tau} $ (solid) and $ \bar{\nu}_{\tau} $ (dashed) calculated in~\cite{Kato:2007re} using the program NEUT. }
 \label{CCcrosssection}
\end{figure}
Tau leptons, generated from tau neutrinos, decay very fast via hadronic channels and therefore do not produce a clear \v{C}erenkov light signature. The cross section for CC interactions of tau neutrinos and antineutrinos with nuclei has been calculated with the neutrino interaction simulation program NEUT in~\cite{Kato:2007re}. The result is presented in Figure~\ref{CCcrosssection}. Clearly visible is the threshold for $ \tau $ lepton production in the reaction $ \nu_{\tau}\,n\rightarrow\tau\,p $ at $ E_{\nu_{\tau}}>m_{\tau}+m_{\tau}^2/2\,m_n\simeq 3.5\usk\unit{GeV} $. 

By the measurement of the \v{C}erenkov light with PMTs, it is possible to reconstruct the energy and the direction of the charged lepton. However, this is not sufficient to reconstruct the neutrino energy without knowing the direction of the incoming neutrino, which differs from the lepton direction by $ \Delta\theta\approx 30^{\circ}\sqrt{1/E_{\nu}\usk(\unit{GeV})} $~\cite{PalomaresRuiz:2007ry}. Therefore, the neutrino events are usually classified according to the signal topology. 

Electron or muon events for which the particle track starts and ends inside the detector are called \textit{fully contained} events. In this case, the lepton energy can be reconstructed. This event type is further divided into sub-GeV and multi-GeV events, depending on whether the lepton energy is below or above $ \sim 1.4 $\usk GeV. On average, sub-GeV events originate from neutrinos of several hundred MeV, and multi-GeV events arise from neutrinos of few GeV energy. 

If the lepton is produced inside the detector but escapes from it, the event is called \textit{partially contained}. In this case, only the visible energy of the lepton can be reconstructed. Since these events originate from neutrinos with energies comparable to those of fully contained events, they are usually grouped together. 

When a muon is produced in the rock below the detector and its track ends inside the detector, one speaks of \textit{up-going stopping} muon events. The typical energy of parent neutrinos is $ \sim 10 $\usk GeV. However, this event topology does not occur for electrons, as they produce electromagnetic showers in the surrounding material before reaching the detector. We note that down-going muons are usually not taken into account, since they cannot be distinguished from the background of cosmic ray muons. 

If a muon track crosses the whole detector from below, the event is called \textit{up-going through-going} muon. These events originate from neutrinos with energies of 10--1000\usk GeV. 

\begin{figure}
 \centering
 \includegraphics[scale=0.6]{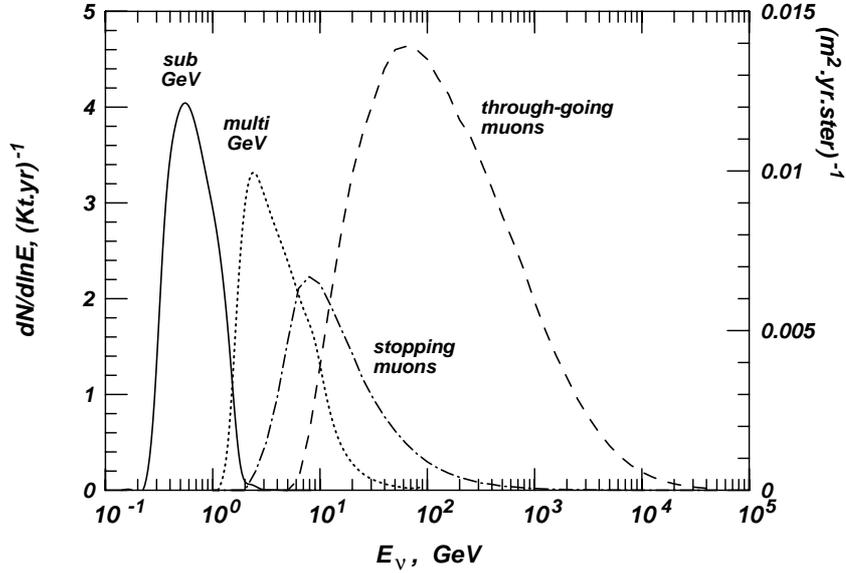} 
 \caption[Neutrino energies and event topologies.]{Contributions of neutrinos to distinct event topologies as a function of the neutrino energy. Figure taken from~\cite{GonzalezGarcia:2002dz}.}
 \label{bins}
\end{figure}
The contribution of neutrinos to the different event topologies described above can be simulated and is presented in Figure~\ref{bins} as a function of the neutrino energy. From this discussion we see that it is only possible to reconstruct the neutrino energy on a statistical basis using the event topology and the constraint that the neutrino energy must be larger than the visible energy of the charged lepton. 

Given this fact, the employed energy resolution of 10\usk\% for the neutrino flux in all figures is very optimistic and way beyond feasibility in present neutrino detectors.

\subsection{Observability in Super-Kamiokande}
Super-Kamiokande is a cylindrical 50\usk kton water \v{C}erenkov detector located in the Kamioka mine in central Japan. The fiducial mass of the detector is 22.5\usk kton (i.e. the effective detector mass that can be used for analyses). 

The Super-Kamiokande collaboration has developed a statistical method to discriminate tau neutrinos from the background of other flavors~\cite{Abe:2006fu}. Using two different strategies, namely a likelihood analysis and a neural network, they find an efficiency of 43.1\usk\% and 39.0\usk\%, respectively, to identify tau neutrinos correctly. However, they still misidentify 3.8\usk\% and 3.4\usk\%, respectively, of the electron and muon background neutrinos as tau neutrinos. 

Due to the large number of atmospheric electron and muon neutrino events, the sample of tau neutrinos is thus dominated by misidentified neutrinos. The actual tau neutrino events can therefore only be extracted on a statistical
basis, using Monte Carlo methods. In the end, the data is found to be consistent with the expected atmospheric tau neutrino flux in the presence of neutrino oscillations: The full-sky atmospheric tau neutrino signal results in fact in 78 events from CC interactions in the fiducial volume during the Super-K I period and 43 events during the Super-K II period~\cite{Kato:2007re, Abe:2006fu}. 

However, the above mentioned analysis does not exploit the information about the spectral shape of the signal, apart from setting a threshold for $ \tau $ lepton production. So this kind of data analysis could certainly be
improved in order to search for a signal with a peak above the continuum, as in our case.

\begin{figure}[ht]
 \centering
 \includegraphics[scale=1.2]{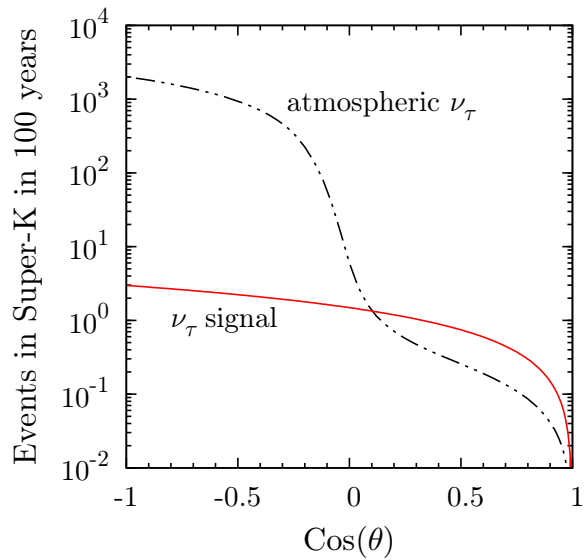} 
 \caption[Tau neutrino events expected at Super-K.]{Number of events per century of observation at Super-K due to CC interactions of $ \nu_{\tau} $ and $ \bar{\nu}_{\tau} $ from the atmosphere (dot-dashed) and the gravitino decay (solid), integrating the signal from the zenith direction to the angle $ \theta $. The gravitino mass and lifetime are chosen as in Figure~\ref{eghalo}. }
 \label{CCevents}
\end{figure}
Apart from the experimental difficulties, it is worthwhile to examine the theoretically expected signal in the tau channel. Figure~\ref{CCevents} shows the expected number of tau neutrino and antineutrino events per century of observation at Super-Kamiokande within a zenith angle integrated from $ \cos\theta $ to 1. If only down-going neutrinos are selected, the signal from the gravitino decay lies above the atmospheric background, especially for higher energies. However, the fluxes are extremely low and result only in a few events per century, making it practically impossible to discriminate them from the other flavors using statistical methods. 

One detector specifically optimized for measuring tau neutrinos above 17\usk GeV event by event is OPERA, which is already active in Gran Sasso and will measure tau neutrino appearance in a muon neutrino beam produced at CERN. Unfortunately, the detector's effective mass is more than a factor 10 smaller than that of Super-K and thus---even neglecting the issue of directionality---would be able to observe only one event in more than 1000 years. 

We therefore conclude that present detectors are not able to detect the signal, either because they do not have sufficient efficiency for identifying tau neutrinos, or because they are too small for the low intensity of our signal, or both. 

\newpage
\subsection{Future Detectors}
We will discuss briefly in this section the prospects of detection for the future.

\subsubsection*{Hyper-Kamiokande}
The prospects for Hyper-Kamiokande can be easily obtained by considering that its mass is planned to be a factor of 10 (for the 0.5\usk Mton project) to 20 (for a 1\usk Mton case) larger than Super-K. Assuming that the rest of the detector characteristics are unchanged, we expect it to have at most $ \mathcal{O} $(20--40) events from our signal per century from the upper hemisphere.\footnote{This number could be increased if the fiducial volume of Hyper-Kamiokande is larger than $ \sim 1/2 $ of the total volume, as it is in Super-Kamiokande.} 

This number of events might be still too small to allow for statistical analysis. On the other hand, we expect most of the events to appear within the peak region or near the threshold. Therefore, using an appropriate energy binning---especially if the signal has been detected already in the gamma-ray channel---could even allow to collect a significant number of events above the background in a specific energy bin on shorter timescales. 

Nevertheless, it is clear that a sufficiently good energy resolution is a key requirement for extracting the events from the line, and it remains uncertain how and if the statistical tau flavor discrimination analysis can be applied to a sample of such few events.

\subsubsection*{IceCube and km$ ^3 $ Detectors}
Detectors of km$ ^3 $ dimensions have in principle sufficient size to collect enough events to detect the signal from gravitino decay within a reasonable time span. Even considering that IceCube is actually looking downwards and not at the upper hemisphere, from the horizontal direction and the CC cross section for tau neutrinos we estimate $ \mathcal{O}(100) $ events per year for the completed experiment. Of course, the effective area depends on the neutrino energy: Taking the effective area, given in~\cite{Collaboration:2007rk} for up-going events, for our signal from the opposite direction and assuming most of the signal is above 100\usk GeV, we have instead $ \mathcal{O}(10) $ events per year. 

In general, it would be desirable to decrease the energy threshold of IceCube below 100\usk GeV in order to cover the energy range favored by the EGRET and HEAT anomalies. The combination of IceCube with the Antarctic Muon And Neutrino Detector Array (AMANDA) already allows to lower the threshold to 30\usk GeV. Additionally, plans are being considered for adding another, denser subdetector at a deeper location to improve the sensitivity to dark matter annihilations~\cite{Collaboration:2007rk}. Such a configuration could probably also be useful for investigating the present scenario and, more generally, other decaying dark matter candidates. 

As we have stated before, in the case of water \v{C}erenkov detectors, the discrimination of tau neutrinos from other neutrino flavors is generally difficult, and for IceCube strategies for tau flavor identification have been proposed only for neutrinos well above TeV energies~\cite{Cowen:2007ny}. It could therefore be more favorable to improve the energy resolution and exploit the muon neutrino final state instead.

\section{Constraints on the Gravitino Parameters}
In this final section we want to present constraints on the gravitino lifetime as a function of its mass from the non-observation of the neutrino signal from gravitino dark matter decay. To achieve this, we compare the neutrino flux from gravitino decay to the atmospheric background flux. 

In particular, we only take the halo contribution of the photon and $ Z^0 $ lines discussed in Section~\ref{widths} to approximate the signal, 
\begin{equation}
 \begin{split}
  \frac{dJ_{\gamma}}{dE} &\simeq A^{\text{NFW}}_{excl. disk}\,\BR(\gamma\,\nu_{\tau})\,\delta\left( E-\frac{m_{3/2}}{2}\right) , \\
  \frac{dJ_Z}{dE} &\simeq A^{\text{NFW}}_{excl. disk}\,\BR(Z^0\nu_{\tau})\,\frac{\left( \int_{0}^{\infty}dE\big/\left[ \left( E^2-E_{\nu_{\tau}}^2\right) ^2+E_{\nu_{\tau}}^2\Gamma_{\nu_{\tau}}^2\right] \right) ^{-1}}{\left( E^2-E_{\nu_{\tau}}^2\right) ^2+E_{\nu_{\tau}}^2\Gamma_{\nu_{\tau}}^2}\,, 
 \end{split}
\end{equation}
and consider either the signal in all neutrino flavors or only the tau flavor. According to the unitarity of the conversion probabilities for neutrino oscillations, the signal in all neutrino flavors is given by the expressions above, whereas for the tau signal we have to apply the survival probability $ P(\nu_{\tau}\rightarrow\nu_{\tau}) $. 

To allow for the poor energy resolution of neutrino detectors, we adopt a nominal energy resolution of $ \sigma=0.3 $ in $ \log_{10}E\usk(\unit{GeV}) $ around the peak position~\cite{PalomaresRuiz:2007ry}. Thus, we convolute the neutrino fluxes with a Gaussian distribution in the logarithm of the neutrino energy: 
\begin{equation}
 \Phi(E,\,E')=\frac{e^{-\frac{1}{2}(\sigma\ln10)^2}}{\sqrt{2\pi}\sigma\ln10}\,\frac{1}{E'}\,e^{-\frac{1}{2}\left( \frac{\log_{10}E/E'}{\sigma}\right) ^2}, 
\end{equation}
where the altered coefficients take care of the correct normalization and the convolution is computed according to 
\begin{equation}
 \frac{dJ_{\nu,\,\text{Gauss}}}{dE}=\int_0^{\infty}\Phi(E,E')\,\frac{dJ_{\nu}}{dE'}\,dE'\,. 
\end{equation}

\begin{figure}
 \centering
 \includegraphics[scale=1.2]{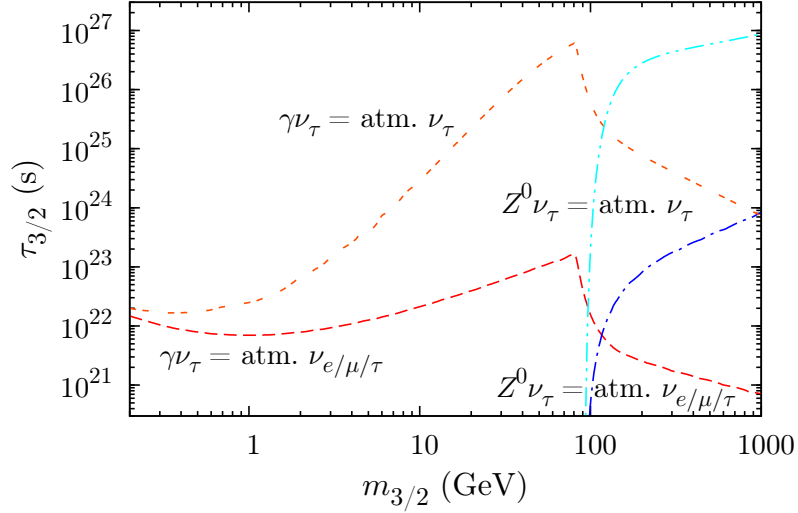} 
 \caption[Constraints on the gravitino parameter space.]{Region of the gravitino lifetime where the line signals
from the two-body decays into $\gamma\,\nu_{\tau} $ and $ Z^0\nu_{\tau} $ overcome the atmospheric background either for all neutrino flavors or for down-going tau neutrinos only. We see that considering the tau flavor and exploiting directionality can improve the constraints by three orders of magnitude for large gravitino masses. We use here the branching ratios shown in Figure~\ref{BRplot} for the two lines. Note that in the region above the $ W^{\pm} $ threshold the continuum spectrum from its fragmentation may be used to close the gap, but we do not consider this possibility here.}
 \label{LifetimeMass}
\end{figure}

The actual lower bound on the gravitino lifetime is determined by the criterion that the flux at the peak position equals the simulated background either for all flavors or for down-going tau neutrinos only. The result is presented in Figure~\ref{LifetimeMass}. For the flux in all neutrino flavors, it is similar to that presented in~\cite{PalomaresRuiz:2007ry}, except that we are weighting the channels with the gravitino branching ratios and that we have only one neutrino produced in the lines instead of two. 

Note that since the signal from gravitino decay is proportional to $ 1/\tau_{3/2} $, requiring the peak flux to be
larger than the background by a particular factor only rescales the exclusion curves by the inverse of this factor. 

We clearly see that again the tau neutrino channel in the down-going direction allows to constrain the gravitino lifetime a few orders of magnitude better than the muon or electron neutrino channels. On the other hand, similar plots for the gamma-ray channel are even more sensitive and give bounds on the order of $ 10^{27} $\usk s for gravitino masses below the $ W^\pm $ and $ Z^0 $ thresholds~\cite{Bertone:2007aw}. 

For masses above $ \mathcal{O}(100) $\usk GeV, the tau neutrino channel even starts to compete in sensitivity with the photon channel. Admittedly, we must point out that the presented bounds are only hypothetical since we neglect the difficulties connected with the measurement of such a low flux in neutrino detectors and the identification of the neutrino flavor.  

\chapter{Conclusions}
The scenario of gravitino dark matter with slightly broken $ R $-parity is theoretically well motivated as it leads to a consistent thermal history of the universe including both baryogenesis via thermal leptogenesis and standard Big Bang nucleosynthesis. 

Although the above scenario was originally devised to reconcile the thermal history of the universe with supersymmetric dark matter, it has been pointed out in the recent literature that the decay of gravitino dark matter with a lifetime of $ \sim 10^{26} $\usk s and a mass of $ \sim 150 \usk \unit{GeV} $ into massive gauge bosons may account for the anomalies observed in the diffuse extragalactic gamma-ray spectrum as measured by EGRET, as well as in the positron fraction as measured by HEAT.\footnote{The existence of an excess in the positron fraction seems to be supported by preliminary results of the PAMELA (a Payload for Antimatter Matter Exploration and Light-nuclei Astrophysics) satellite mission~\cite{Pamela:2008url}.} 

In this thesis we examined the neutrino signal from the decay of unstable gravitino dark matter in a scenario with bilinear $ R $-parity breaking. Motivated by the possible explanation of anomalies in other cosmic ray channels, we recomputed the gravitino decay widths, taking into account also the Feynman diagrams with the non-abelian 4--vertex for the $ W^{\pm} $ and $ Z^0 $ channels and the decay into the lightest Higgs boson. Using the resulting branching ratios for the gravitino decay channels, we determined the neutrino signal on Earth originating from gravitino dark matter decay. 

As a consistency check of the scenario, we employed the same gravitino parameters as those that were used to account for the anomalies in other cosmic ray channels and found that the neutrino signal is compatible with results from present neutrino experiments. In addition, we have examined the detectability of this extraordinary component of the neutrino flux to find an independent way to test this scenario. 

While the signal in the neutrino spectrum with two or more distinct peaks, resulting from two-body gravitino decays into gauge/Higgs boson and neutrino, is very characteristic, it will be challenging to detect these features in neutrino experiments. On one side, present neutrino detectors do not achieve a sufficient energy resolution to resolve peaks in the spectrum, and on the other side, the event rate is expected to be so small that the background of atmospheric neutrinos overwhelms the signal in all flavors. 

The most promising signal-to-background ratio is found for tau neutrinos, especially when considering only the flux from above the horizon, since there the atmospheric tau neutrino flux is substantially reduced. However, tau neutrinos are difficult to identify in water \v{C}erenkov detectors and probably only an event-by-event identification could allow detecting the signal with such extremely limited statistics. Therefore, we conclude that it is currently not possible to detect this contribution due to present technological limitations. 

The ideal detector for testing the present scenario would be one of megaton mass with the ability to identify and measure tau neutrinos event by event. Should such a detector ever become available, it could be worthwhile to look for this extraordinary component of the neutrino flux by employing strategies for background reduction such as the ones discussed here, especially if the anomalous signatures in the positron fraction and the diffuse extragalactic gamma-ray spectrum are confirmed by PAMELA and FGST, respectively. 

In principle, the non-observation of the neutrino signal can be used to infer bounds on the gravitino signal, but we found that these bounds presumably cannot compete with the bounds from other decay channels, like the decay into photons, positrons or antiprotons. 

On the other hand, the detection of a signal in neutrinos, that is compatible with signals in the other indirect detection channels, would in fact bring significant support to the scenario of decaying dark matter, possibly consisting of gravitinos that are unstable due to bilinear $ R $-parity breaking.  

{\appendix

\chapter{Units and Physical Constants}
\label{constants}
Here we will shortly summarize the values of all the physical and astrophysical constants that appear in the thesis. At times, we work in units where the reduced Planck constant, the speed of light and the Boltzmann constant obey $ \hbar=c=k=1 $. In this case, we have the conversion factors 
\begin{equation*}
 \begin{split}
  1\usk\unit{eV} &=1.160\,4505(20)\times 10^4\usk\unit{K}\,, \\
  1\usk\unit{eV} &=1.602\,176\,487(40)\times 10^{-19}\usk\unit{J}\,, \\
  1\usk\unit{GeV} &=1.782\,661\,758(44)\times 10^{-27}\usk\unit{kg}\,, \\
  1\usk\unit{GeV}^{-1} &=1.973\,269\,631(49)\times 10^{-16}\usk\unit{m}\,, \\
  1\usk\unit{GeV}^{-1} &=6.582\,118\,99(16)\times 10^{-25}\usk\unit{s}\,. 
 \end{split}
\end{equation*}
All figures are taken from the \textit{The Review of Particle Physics}~\cite{Amsler:2008zz}, if not marked otherwise. Numbers in parentheses represent the one standard deviation uncertainty in the last digits. \medskip

\noindent
\begin{tabular*}{\linewidth}{lllll}
 \toprule
 Quantity & $\qquad$ & Symbol & $\quad$ & Value \\
 \midrule
 electron mass & & $ m_e $ & & $ 510.998\usk910(13)\usk\unit{keV} $ \\
 muon mass & & $ m_{\mu} $ & & $ 105.658\usk3668(38)\usk\unit{MeV} $ \\
 muon mean life & & $ \tau_{\mu} $ & & $ 2.197\usk019(21)\times 10^{-6}\usk\unit{s} $ \\
 tau mass & & $ m_{\tau} $ & & $ 1.776\usk84(17)\usk\unit{GeV} $ \\
 tau mean life & & $ \tau_{\tau} $ & & $ 290.6(1.0)\times 10^{-15}\usk\unit{s} $ \\
 $ W^{\pm} $ boson mass & & $ m_W $ & & $ 80.398(25)\usk\unit{GeV} $ \\
 $ Z^0 $ boson mass & & $ m_Z $ & & $ 91.1876(21)\usk\unit{GeV} $ \\
 $ Z^0 $ boson decay width & & $ \Gamma_Z $ & & $ 2.4952(23)\usk\unit{GeV} $ \\
 neutron mass & & $ m_n $ & & $ 939.565\usk360(81)\usk\unit{MeV} $ \\
 neutron mean life & & $ \tau_n $ & & $ 885.7(8)\usk\unit{s} $ \\
 proton mass & & $ m_p $ & & $ 938.272\usk013(23)\usk\unit{MeV} $ \\
 gravitational constant & & $ G_N $ & & $ 6.708\,81(67)\times 10^{-39}\usk\unit{GeV^{-2}} $ \\
 \bottomrule
\end{tabular*}
 
\noindent
\begin{tabular*}{\linewidth}{lll}
 \toprule
 Quantity & Symbol & Value \\
 \midrule
 parsec & $ pc $ & $ 3.085\usk6776\times 10^{16}\usk\unit{m} $ \\
 Solar distance from GC & $ r_{\odot} $ & $ 8.0(5)\usk\unit{kpc} $ \\
 Earth mean equatorial radius & $ R_{\oplus} $ & $ 6.378 137\times 10^6\usk\unit{m} $ \\
 local halo density & $ \varrho_{\text{loc}} $ & 0.1--0.7\usk\unit{GeV}\usk\unit{cm^{-3}} \\
 present-day CMB temperature & $ T_0 $ & $ 2.725(1)\usk\unit{K} $ \\
 normalized Hubble constant & $ h $ & $ 0.73(3) $ \\
 critical density & $ \varrho_c $ & $ 1.053\usk68(11)\times 10^{-5}\usk h^2\usk\unit{GeV}\usk\unit{cm^{-3}} $ \\
 matter density & $ \Omega_mh^2 $ & $ 0.128(8) $ \\
 baryon density & $ \Omega_bh^2 $ & $ 0.0223(7) $ \\
 dark matter density & $ \Omega_{DM}h^2 $ & $ 0.105(8) $ \\
 dark energy density & $ \Omega_{\Lambda} $ & $ 0.73(3) $ \\
 total energy density & $ \Omega_{tot} $ & $ 1.011(12) $ \\
 baryon-to-photon ratio & $ \eta $ & $ 6.12(19)\times 10^{-10} $ \\
 \midrule
 solar neutrino mixing angle & $ \sin^2{\theta_{12}} $ & $ 0.304(_{-16}^{+22}) $ \hfill\cite{Schwetz:2008er} $ \, $ \\
 atmospheric mixing angle & $ \sin^2{\theta_{23}} $ & $ 0.50(_{-6}^{+7}) $ \hfill\cite{Schwetz:2008er} $ \, $ \\
 third neutrino mixing angle & $ \sin^2{\theta_{13}} $ & $ 0.010(_{-11}^{+16}) $ \hfill\cite{Schwetz:2008er} $ \, $ \\
 solar neutrino mass difference & $ \Delta m_{12}^2 $ & $ 7.65(_{-20}^{+23})\times 10^{-5} $ \hfill\cite{Schwetz:2008er} $ \, $ \\
 atmospheric mass difference & $ \abs{\Delta m_{13}^2} $ & $ 2.40(_{-11}^{+12})\times 10^{-3} $ \hfill\cite{Schwetz:2008er} $ \, $ \\
 \bottomrule
\end{tabular*}

\chapter{Notation, Conventions and Formulae}
\label{notation}

\paragraph{Four-Vectors and Tensors}
Lorentz indices are depicted by small Greek letters, e.g. $ \mu=0,\,1,\,2,\,3 $\,. 
The metric of Minkowski space is chosen to be 
\begin{equation}
 g_{\mu\nu}=g^{\mu\nu}=
 \begin{pmatrix}
  +1 & 0 & 0 & 0 \\
  0 & -1 & 0 & 0 \\
  0 & 0 & -1 & 0 \\
  0 & 0 & 0 & -1
 \end{pmatrix}. 
 \label{metric}
\end{equation}
Lorentz indices of four-vectors and higher rank tensors are raised and lowered using the space-time metric: 
\begin{equation}
 a_{\mu}=g_{\mu\nu}\,a^{\nu}\qquad\text{and}\qquad b^{\mu}=g^{\mu\nu}\,b_{\nu}\,. 
\end{equation}
We use in all cases a sum convention for Lorentz indices: 
\begin{equation}
 a^{\mu}b_{\mu}\equiv\sum_{\mu=0}^3a^{\mu}b_{\mu}\,. 
\end{equation}

\paragraph{Antisymmetric Tensor}
The totally antisymmetric tensor $ \varepsilon^{\mu\nu\rho\sigma} $ is chosen such that 
\begin{equation}
 \varepsilon^{0123}=+1\,. 
\end{equation}
This implies for the totally antisymmetric tensor with lowered indices, 
\begin{equation}
 \varepsilon_{\mu\nu\rho\sigma}=g_{\mu\alpha}\,g_{\nu\beta}\,g_{\rho\gamma}\,g_{\sigma\delta}\,\varepsilon^{\alpha\beta\gamma\delta}, 
\end{equation}
that $ \varepsilon_{0123}=-1 $\,. \\
The totally antisymmetric tensor in three dimensions $ \varepsilon_{ijk} $ is chosen such that 
\begin{equation}
 \varepsilon_{123}=+1\,. 
\end{equation}

\paragraph{Pauli Matrices}
The Pauli sigma matrices read 
\begin{equation}
 \sigma_1=
 \begin{pmatrix}
  0 & 1 \\
  1 & 0
 \end{pmatrix},\quad 
 \sigma_2=
 \begin{pmatrix}
  0 & -i \\
  i & 0
 \end{pmatrix},\quad 
 \sigma_3=
 \begin{pmatrix}
  1 & 0 \\
  0 & -1
 \end{pmatrix} 
 \label{PauliMatrix}
\end{equation}
and satisfy the identity 
\begin{equation}
 \sigma_i\sigma_j=\delta_{ij}+i\varepsilon_{ijk}\sigma_k\,. 
\end{equation}

\paragraph{Gamma Matrices}
The Dirac gamma matrices form a Clifford algebra
\begin{equation}
 \left\lbrace \gamma^{\mu},\,\gamma^{\nu}\right\rbrace =2\,g^{\mu\nu}. 
 \label{clifford}
\end{equation}
In most cases it is convenient to suppress the spinor indices of the gamma matrices, so we do not write them here. Using the above algebra and the metric (\ref{metric}) it can be easily seen that 
\begin{equation}
 \left( \gamma^0\right) ^2=1\qquad\text{and}\qquad\left( \gamma^i\right) ^2=-1\,. 
\end{equation}
Hermitian conjugation of the gamma matrices yields 
\begin{equation}
 \left( \gamma^{\mu}\right) ^{\dagger} =\gamma^0\gamma^{\mu}\gamma^0. 
\end{equation}
The Lorentz indices of gamma matrices are raised and lowered by the metric, just as in the case of four-vectors. That implies 
\begin{equation}
 \gamma_0=\gamma^0\qquad\text{and}\qquad\gamma_i=-\gamma^i. 
\end{equation}
Thus, we see that raising and lowering of Lorentz indices is equivalent to hermitian conjugation. 
In addition to $ \gamma^0,\,\gamma^1,\,\gamma^2,\,\gamma^3\,, $ a fifth gamma matrix can be defined 
\begin{equation}
 \gamma^5\equiv i\gamma^0\gamma^1\gamma^2\gamma^3=-\frac{i}{4!}\,\varepsilon_{\mu\nu\rho\sigma} \gamma^{\mu}\gamma^{\nu}\gamma^{\rho}\gamma^\sigma. 
\end{equation}
Using the algebra (\ref{clifford}) it can be easily shown that $ \gamma^5 $ has the following properties: 
\begin{equation}
 \begin{split}
  \left\lbrace \gamma^5,\,\gamma^{\mu}\right\rbrace &=0\,, \\
  \left( \gamma^5\right) ^2 &= 1\,, \\
  \left( \gamma^5\right) ^{\dagger} &=\gamma^5. 
 \end{split}
\end{equation}
The matrix $ \gamma^5 $ is used to define the chirality projection operators 
\begin{equation}
 P_L\equiv\frac{1}{2}\left( 1-\gamma^5\right)\qquad\!\text{and}\qquad P_R\equiv\frac{1}{2}\left( 1+\gamma^5\right) . 
\end{equation}
These operators possess the projector properties 
\begin{equation}
 P_L^2=P_L\,,\quad P_R^2=P_R\quad\text{and}\quad P_LP_R=P_RP_L=0\,. 
 \label{chirality}
\end{equation}

A very common object in theories of particles with half-integral spin is the contraction of gamma matrices with the particle momentum. Therefore, we use the Feynman slash notation 
\begin{equation}
 \slashed{p}\equiv\gamma^{\mu}p_{\mu}\,. 
\end{equation}
In order to obtain Lorentz scalars, in the formulation of theories of particles with half-integral spin there occur products of adjoint spinors and spinors. The adjoint spinor and adjoint vector-spinor are defined as 
\begin{equation}
 \bar{\psi}\equiv\psi^{\dagger}\gamma^0\qquad\text{and}\qquad\bar{\psi}_{\mu}\equiv\psi_{\mu}^{\dagger}\gamma^0, 
\end{equation}
respectively. 

\paragraph{Majorana Spinors}
Majorana particles are invariant under charge conjugation. Thus, the Majorana condition reads 
\begin{equation}
 \chi=\chi^c\equiv C\bar{\chi}^T. 
\end{equation}
Here we use the charge-conjugation matrix $ C $ that has the properties 
\begin{equation}
 \begin{split}
  C^{\dagger} &=C^{-1}, \\
  C^T &=-C\,, \\
  C\Gamma^{i\,T}C^{\dagger} &=\eta_i\Gamma^i, 
 \end{split}
\end{equation}
with
\begin{equation}
 \eta_i =\left\lbrace \begin{array}{rl}
 1\quad\text{for}\;\Gamma^i & =1, \gamma^{\mu}\gamma^5, \gamma^5 \\
 -1\quad\text{for}\;\Gamma^i & =\gamma^{\mu}, \sigma^{\mu\nu}
 \end{array}.\right. 
\end{equation}
In the above expression we use the definition $ \sigma^{\mu\nu}\equiv\frac{i}{2}\left[ \gamma^{\mu},\,\gamma^{\nu}\right] ,\quad\mu<\nu $\,. 

\paragraph{Dirac Spinors}
The free Dirac equation 
\begin{equation}
 \left( i\slashed{\partial}-m\right) \psi(x)=0 
\end{equation}
has four linearly independent plane wave solutions: 
\begin{equation}
 \psi(x)=u^s(p)\,e^{-ip\cdot x}\qquad\text{and}\qquad\psi(x)=v^s(p)\,e^{ip\cdot x},\; s=\pm\frac{1}{2}\,, 
\end{equation}
where the spinors $ u $ and $ v $ have to obey the following constraints: 
\begin{equation}
 \left( \slashed{p}-m\right) u^s(p)=0\qquad\text{and}\qquad\left( \slashed{p}+m\right) v^s(p)=0\,. 
\end{equation}
The normalization of the Dirac spinors is given by~\cite{Peskin:1995ev}
\begin{equation}
 \bar{u}^s(p)\,u^{s'}(p)=2\,m\,\delta^{ss'}\qquad\text{and}\qquad\bar{v}^s(p)\,v^{s'}(p)=-2\,m\,\delta^{ss'},  \label{spinornorm}
\end{equation}
and the polarization sums for spin-1/2 fermions are given by~\cite{Peskin:1995ev}
\begin{equation}
 \begin{split} 
  \sum_s u^s(p)\,\bar{u}^s(p) &=\slashed{p}+m\,, \\
  \sum_s v^s(p)\,\bar{v}^s(p) &=\slashed{p}-m\,, 
 \end{split}
 \label{fermionPolarization}
\end{equation}
where the sum is performed over the fermion polarizations $ s=\pm\frac{1}{2} $\,. 

\paragraph{Polarization Vectors}
Free massless spin-1 particles in the Lorentz gauge ($ \partial^{\mu}A_{\mu}=0 $) and free massive spin-1 particles obey the Proca equation~\cite{Collins:1989kn}
\begin{equation}
 \left( \partial^{\nu}\partial_{\nu}+M^2\right) A_{\mu}=0\,. 
\end{equation}
This equation has the solutions
\begin{equation}
 A_{\mu}=\epsilon_{\mu}(p)\,e^{-ip\cdot x}, 
\end{equation}
where $ \epsilon_{\mu} $ is a polarization vector. The condition $ \partial^{\mu}A_{\mu}=0 $ demands that 
\begin{equation}
 p^{\mu}\epsilon_{\mu}=0\,, 
\end{equation}
reducing the number of independent polarization vectors to three. 

For massless spin-1 particles we have the freedom to make the additional gauge transformation 
\begin{equation}
 A_{\mu}\rightarrow A'_{\mu}=A_{\mu}+\partial^{\mu}\Lambda\qquad\text{with}\qquad\partial^{\mu}\partial_{\mu}\Lambda=0\,. 
\end{equation}
The condition for $ \Lambda $ is required by the Lorentz gauge condition. This additional gauge freedom reduces the number of independent polarization vectors for massless spin-1 particles to two. 

The polarization vectors are normalized according to~\cite{Auvil:1966iu} 
\begin{equation}
 \epsilon_{\mu}^{\lambda\,*}\epsilon^{\lambda'\,\mu}=-\delta^{\lambda\lambda'},  \label{epsilonnorm}
\end{equation}
while the polarization sums are given by~\cite{Peskin:1995ev}
\begin{equation}
 \sum_{\lambda}\epsilon_{\mu}^{\lambda\,*}(p)\,\epsilon_{\nu}^{\lambda}(p)=-g_{\mu\nu}\,, 
 \label{vectorPolarization}
\end{equation}
where the sum is performed over the two polarization states of the massless spin-1 particle $ \lambda=\pm1 $, and by 
\begin{equation}
 \sum_{\lambda}\epsilon_{\mu}^{\lambda\,*}(p)\,\epsilon_{\nu}^{\lambda}(p)=-\left( g_{\mu\nu}-\frac{p_{\mu}p_{\nu}}{m_A^2}\right) , 
\end{equation}
where the sum is performed over the three polarization states of the massive spin-1 particle $ \lambda=\pm1,\,0 $. 

\paragraph{Identities for Gamma Matrices}
In the calculation of squared amplitudes we encounter traces of gamma matrices. For these the following identities can be derived from the algebra~(\ref{clifford}): 
\begin{equation}
 \begin{split}
  \Tr\left( \mathbf{1}\right) &=4\,, \\
  \Tr\left( \text{odd number of $ \gamma $s}\right) &=0\,, \\
  \Tr\left( \gamma^{\mu}\gamma^{\nu}\right) &=4\,g^{\mu\nu}, \\
  \Tr\left( \gamma^{\mu}\gamma^{\nu}\gamma^{\rho}\gamma^{\sigma}\right) &=4\left( g^{\mu\nu}g^{\rho\sigma}-g^{\mu\rho}g^{\nu\sigma}+g^{\mu\sigma}g^{\nu\rho}\right) , \\
  \Tr\left( \gamma^5\right) &=0\,, \\
  \Tr\left( \gamma^{\mu}\gamma^{\nu}\gamma^5\right) &=0\,, \\
  \Tr\left( \gamma^{\mu}\gamma^{\nu}\gamma^{\rho}\gamma^{\sigma}\gamma^5\right) &=-4\,i\,\epsilon^{\mu\nu\rho\sigma}. 
 \end{split}
 \label{GammaTrace}
\end{equation}
We also encounter contractions of gamma matrices. Using the algebra (\ref{clifford}), we obtain the following identities: 
\begin{equation}
 \begin{split}
  \gamma^{\mu}\gamma_{\mu} &=4\,, \\
  \gamma^{\mu}\gamma^{\nu}\gamma_{\mu} &=-2\,\gamma^{\nu}, \\
  \gamma^{\mu}\gamma^{\nu}\gamma^{\rho}\gamma_{\mu} &=4\,g^{\nu\rho}, \\
  \gamma^{\mu}\gamma^{\nu}\gamma^{\rho}\gamma^{\sigma}\gamma_{\mu} &=-2\,\gamma^{\sigma}\gamma^{\rho}\gamma^{\nu}. \\
 \end{split}
 \label{GammaContraction}
\end{equation}

Any complex $ 4\times 4 $ matrix $ M $ can be expanded in terms of a basis of gamma matrices, 
\begin{equation}
 \Gamma^i=\left\lbrace \mathbf{1},\,\gamma^{\mu},\,\sigma^{\mu\nu},\,\gamma^{\mu}\gamma^5,\,\gamma^5\right\rbrace , 
\end{equation}
according to the relation~\cite{Nishi:2004st} 
\begin{equation}
 M=\sum_i\frac{1}{4}\Tr\left( M\Gamma^i\right) \Gamma_i\,, 
\end{equation}
where $ \Gamma_i $ is the dual basis. The dual basis can be determined using the orthogonality relation 
\begin{equation}
 \Tr\left( \Gamma_i\Gamma^j\right) =4\,\delta_i^j 
\end{equation}
and corresponds to the hermitian conjugate of the basis. Thus we have 
\begin{equation}
\Gamma_i=\left\lbrace \mathbf{1},\,\gamma_{\mu},\,\sigma_{\mu\nu},\,\gamma^5\gamma_{\mu},\,\gamma^5\right\rbrace . 
\end{equation}

We want to apply this expansion identity to a combination of gamma matrices that appears in the coupling of the gravitino to the gauge supermultiplet: 
\begin{align*}
 \gamma^{\mu}\left[ \gamma^{\nu},\,\gamma^{\rho}\right] =& \:\frac{1}{4}\left\lbrace \Tr\left( \gamma^{\mu}\left[ \gamma^{\nu},\,\gamma^{\rho}\right] \right) \mathbf{1}+\Tr\left( \gamma^{\mu}\left[ \gamma^{\nu},\,\gamma^{\rho}\right] \gamma^{\sigma}\right) \gamma_{\sigma}\right. \\
 &+\Tr\left( \gamma^{\mu}\left[ \gamma^{\nu},\,\gamma^{\rho}\right] \sigma^{\sigma\lambda}\right) \sigma_{\sigma\lambda} \\
 &+\left. \Tr\left( \gamma^{\mu}\left[ \gamma^{\nu},\,\gamma^{\rho}\right] \gamma^{\sigma}\gamma^5\right) \gamma^5\gamma_{\sigma}+\Tr\left( \gamma^{\mu}\left[ \gamma^{\nu},\,\gamma^{\rho}\right] \gamma^5\right) \gamma^5\right\rbrace . 
\end{align*}
Using the trace identities for gamma matrices (\ref{GammaTrace}), we note that traces of an odd number of gamma matrices vanish. For the two nonvanishing traces we obtain
\begin{align*}
 \Tr\left( \gamma^{\mu}\left[ \gamma^{\nu},\,\gamma^{\rho}\right] \gamma^{\sigma}\right) =& \:4\left( g^{\mu\nu}g^{\rho\sigma}-g^{\mu\rho}g^{\nu\sigma}+g^{\mu\sigma}g^{\nu\rho}-g^{\mu\rho}g^{\nu\sigma}\right. \\
 &\left. \qquad+g^{\mu\nu}g^{\rho\sigma}-g^{\mu\sigma}g^{\nu\rho}\right) \\
 = &\:8\,g^{\mu\nu}g^{\rho\sigma}-8\,g^{\mu\rho}g^{\nu\sigma} 
 \intertext{and} 
 \Tr\left( \gamma^{\mu}\left[ \gamma^{\nu},\,\gamma^{\rho}\right] \gamma^{\sigma}\gamma^5\right) = &-4\,i\,\varepsilon^{\mu\nu\rho\sigma}+4\,i\,\varepsilon^{\mu\rho\nu\sigma}=-8\,i\,\varepsilon^{\mu\nu\rho\sigma}. 
\end{align*}
Thus, we finally obtain the relation 
\begin{align}
 \gamma^{\mu}\left[ \gamma^{\nu},\,\gamma^{\rho}\right] = &\:2\,g^{\mu\nu}\gamma^{\rho}-2\,g^{\mu\rho}\gamma^{\nu}-2\,i\,\varepsilon^{\mu\nu\rho\sigma}\gamma^5\gamma_{\sigma}\,. \label{gammaID1}
 \intertext{Using the algebra (\ref{clifford}), we can additionally derive the following relation: } 
 \left[ \gamma^{\nu},\,\gamma^{\rho}\right] \gamma^{\mu}= &\:2\,g^{\mu\rho}\gamma^{\nu}-2\,g^{\mu\nu}\gamma^{\rho}-2\,i\,\varepsilon^{\mu\nu\rho\sigma}\gamma^5\gamma_{\sigma}\,. \label{gammaID2}
 \intertext{From both of these relations we can easily derive the following identity: }
 \gamma_{\mu}\varepsilon^{\mu\nu\rho\sigma}= &-i\gamma^5\left( \gamma^{\nu}\gamma^{\rho}\gamma^{\sigma}-g^{\rho\sigma}\gamma^{\nu}+g^{\nu\sigma}\gamma^{\rho}-g^{\nu\rho}\gamma^{\sigma}\right) . 
\end{align}

\paragraph{Gravitino Polarization Tensor} 
Due to the Lorentz structure of the spin sum for the gravitino, it must be a linear combination of the following ten tensors \cite{Lurie:1968pf}: 
\begin{equation*}
 \begin{array}{ccccc}
 m_{3/2}g_{\mu\nu}\,, & m_{3/2}\gamma_{\mu}\gamma_{\nu}\,, & \gamma_{\mu}p_{\nu}\,, & \gamma_{\nu}p_{\mu}\,, & \frac{p_{\mu}p_{\nu}}{m_{3/2}}\,, \\
 \slashed{p}g_{\mu\nu}\,, & \slashed{p}\,\gamma_{\mu}\gamma_{\nu}\,, & \slashed{p}\,\gamma_{\mu}\frac{p_{\nu}}{m_{3/2}}\,, & \slashed{p}\,\gamma_{\nu}\frac{p_{\mu}}{m_{3/2}}\,, & \slashed{p}\,\frac{p_{\mu}p_{\nu}}{m_{3/2}^2}\,, 
 \end{array}
\end{equation*}
where we have added factors of $ m_{3/2} $ to get the correct dimension of the spin sum according to equation (\ref{projector}). Therefore we have 
\begin{equation*}
 \begin{split}
  P_{\mu\nu}(p)= &\:A\,m_{3/2}\,g_{\mu\nu}+B\,m_{3/2}\gamma_{\mu}\gamma_{\nu}+C\,\gamma_{\mu}p_{\nu}+D\,\gamma_{\nu}p_{\mu}+E\,\frac{p_{\mu}p_{\nu}}{m_{3/2}} \\
  &+F\,\slashed{p}\,g_{\mu\nu}+G\,\slashed{p}\,\gamma_{\mu}\gamma_{\nu}+H\,\slashed{p}\,\gamma_{\mu}\frac{p_{\nu}}{m_{3/2}}+I\,\slashed{p}\,\gamma_{\nu}\frac{p_{\mu}}{m_{3/2}}+J\,\slashed{p}\,\frac{p_{\mu}p_{\nu}}{m_{3/2}^2}\,. 
 \end{split}
\end{equation*}
Analogous to the case of Dirac spinors, we get two polarization sums for the positive and negative frequency solutions. The coefficients can be uniquely determined using the constraints from the Rarita--Schwinger equations for the mode functions (\ref{modeRarita1}), (\ref{modeRarita2}) and from the projector property of the polarization tensor (\ref{projector}). 

For the construction of the positive frequency spin sum, application of (\ref{modeRarita1}) gives 
\begin{equation*}
 A=F\,,\quad B=G\,,\quad C=H\,,\quad D=I\,, \quad E=J
\end{equation*}
and application of (\ref{modeRarita2}) gives the further relations 
\begin{equation*}
 A+4B+D=0\,,\quad 4C+2A+E=0\,,\quad D-A-2B=0\,,\quad E-2C=0\,. 
\end{equation*}
The polarization tensor then becomes 
\begin{equation*}
 P_{\mu\nu}^+(p)=A\,\left( \slashed{p}+m_{3/2}\right) \left\lbrace g_{\mu\nu}-\frac{1}{3}\left( \gamma_{\mu}\gamma_{\nu}+\gamma_{\mu}\frac{p_{\nu}}{m_{3/2}}-\gamma_{\nu}\frac{p_{\mu}}{m_{3/2}}+2\,\frac{p_{\mu}p_{\nu}}{m_{3/2}^2}\right) \right\rbrace . 
\end{equation*}
The projector property (\ref{projector}) finally gives $ A=-1 $. Therefore, the positive frequency polarization tensor of the gravitino reads 
\begin{equation}
 P_{\mu\nu}^+(p)=-\left( \slashed{p}+m_{3/2}\right) \left\lbrace g_{\mu\nu}-\frac{1}{3}\left( \gamma_{\mu}\gamma_{\nu}+\gamma_{\mu}\frac{p_{\nu}}{m_{3/2}}-\gamma_{\nu}\frac{p_{\mu}}{m_{3/2}}+2\,\frac{p_{\mu}p_{\nu}}{m_{3/2}^2}\right) \right\rbrace , 
\end{equation}
which is equivalent to the form given in equation (\ref{PolTensor+}). 

Similarly, the negative frequency polarization tensor is determined. From (\ref{modeRarita1}), (\ref{modeRarita2}) and (\ref{projector}) we obtain 
\begin{equation*}
 \begin{split}
  A+F=0\,,\quad B+G=0\,,\quad C+H=0\,,\quad D+I=0\,, \quad E+J=0\,, \\
  A+4B-D=0\,,\quad 4C-2A-E=0\,,\quad D+A+2B=0\,,\quad E+2C=0 
 \end{split}
\end{equation*}
and $ A=1 $. Thus, the negative frequency polarization tensor of the gravitino reads 
\begin{equation}
 P_{\mu\nu}^-(p)=-\left( \slashed{p}-m_{3/2}\right) \left\lbrace g_{\mu\nu}-\frac{1}{3}\left( \gamma_{\mu}\gamma_{\nu}+\gamma_{\mu}\frac{p_{\nu}}{m_{3/2}}-\gamma_{\nu}\frac{p_{\mu}}{m_{3/2}}+2\,\frac{p_{\mu}p_{\nu}}{m_{3/2}^2}\right) \right\rbrace , 
\end{equation}
which is equivalent to the form given in equation (\ref{PolTensor-}).  

\chapter{Calculation of Decay Widths}
\label{decaywidths}

\section{Feynman Rules}
In this section we provide the necessary Feynman rules for the calculation of the gravitino decay widths. This set is taken from~\cite{Pradler:2007ne} and amended by the addition of the rules for the negative frequency solution of the gravitino. 

Gravitinos are depicted as double solid lines, chiral fermions are drawn as single solid lines and scalars are drawn as dashed lines. Gauge bosons are represented by wiggled lines and gauginos by solid lines with additional wiggled lines. 

The continuous fermion flow is independent of the fermion number flow carried by fermions and sfermions and also of the momentum direction. 

\subsubsection*{External Lines}
The momentum $ p $ flows from the left side to the right side for the external lines shown below. \smallskip

\noindent
Scalar particles: 
\begin{equation*}
 \parbox{2.0cm}{
 \begin{picture}(52,3) (-2,-1)
    \SetWidth{0.5}
    \SetColor{Black}
    \Vertex(50,1){2.5}
    \DashArrowLine(0,1)(50,1){4}
 \end{picture}
 }\quad\parbox{2.0cm}{
 \begin{picture}(52,3) (-2,-1)
    \SetWidth{0.5}
    \SetColor{Black}
    \Vertex(0,1){2.5}
    \DashArrowLine(0,1)(50,1){4}
 \end{picture}
 }\quad\parbox{2.0cm}{
 \begin{picture}(52,3) (-2,-1)
    \SetWidth{0.5}
    \SetColor{Black}
    \Vertex(50,1){2.5}
    \DashArrowLine(50,1)(0,1){4}
 \end{picture}
 }\quad\parbox{2.0cm}{
 \begin{picture}(52,3) (-2,-1)
    \SetWidth{0.5}
    \SetColor{Black}
    \Vertex(0,1){2.5}
    \DashArrowLine(50,1)(0,1){4}
 \end{picture}
 }\quad =\quad 1\,. 
\end{equation*}
Gauginos and matter fermions: 
\begin{align*}
 \parbox{2.0cm}{
 \begin{picture}(52,10) (-2,-3)
    \SetWidth{0.5}
    \SetColor{Black}
    \Photon(0,1)(50,1){3}{3}
    \Line(0,1)(50,1)
    \Vertex(50,1){2.5}
    \ArrowLine(5,8)(45,8)
 \end{picture}
 }\qquad\parbox{2.0cm}{
 \begin{picture}(52,10) (-2,-3)
    \SetWidth{0.5}
    \SetColor{Black}
    \Vertex(50,1){2.5}
    \ArrowLine(0,1)(50,1)
    \ArrowLine(5,8)(45,8)
 \end{picture}
 }\qquad\parbox{2.0cm}{
 \begin{picture}(52,10) (-2,-3)
    \SetWidth{0.5}
    \SetColor{Black}
    \Vertex(50,1){2.5}
    \ArrowLine(50,1)(0,1)
    \ArrowLine(5,8)(45,8)
 \end{picture}
 }\quad & =\quad u^s(p)\,, \\
 \parbox{2.0cm}{
 \begin{picture}(52,10) (-2,-3)
    \SetWidth{0.5}
    \SetColor{Black}
    \Photon(0,1)(50,1){3}{3}
    \Line(0,1)(50,1)
    \Vertex(0,1){2.5}
    \ArrowLine(5,8)(45,8)
 \end{picture}
 }\qquad\parbox{2.0cm}{
 \begin{picture}(50,12) (-2,-3)
    \SetWidth{0.5}
    \SetColor{Black}
    \Vertex(0,1){2.5}
    \ArrowLine(0,1)(50,1)
    \ArrowLine(5,8)(45,8)
 \end{picture}
 }\qquad\parbox{2.0cm}{
 \begin{picture}(52,12) (-2,-3)
    \SetWidth{0.5}
    \SetColor{Black}
    \Vertex(0,1){2.5}
    \ArrowLine(50,1)(0,1)
    \ArrowLine(5,8)(45,8)
 \end{picture}
 }\quad & =\quad\bar{u}^s(p)\,, \\
 \parbox{2.0cm}{
 \begin{picture}(52,10) (-2,-3)
    \SetWidth{0.5}
    \SetColor{Black}
    \Photon(0,1)(50,1){3}{3}
    \Line(0,1)(50,1)
    \Vertex(0,1){2.5}
    \ArrowLine(45,8)(5,8)
 \end{picture}
 }\qquad\parbox{2.0cm}{
 \begin{picture}(52,12) (-2,-3)
    \SetWidth{0.5}
    \SetColor{Black}
    \Vertex(0,1){2.5}
    \ArrowLine(0,1)(50,1)
    \ArrowLine(45,8)(5,8)
 \end{picture}
 }\qquad\parbox{2.0cm}{
 \begin{picture}(52,12) (-2,-3)
    \SetWidth{0.5}
    \SetColor{Black}
    \Vertex(0,1){2.5}
    \ArrowLine(50,1)(0,1)
    \ArrowLine(45,8)(5,8)
 \end{picture}
 }\quad & =\quad v^s(p)\,, \\
 \parbox{2.0cm}{
 \begin{picture}(52,10) (-2,-3)
    \SetWidth{0.5}
    \SetColor{Black}
    \Photon(0,1)(50,1){3}{3}
    \Line(0,1)(50,1)
    \Vertex(50,1){2.5}
    \ArrowLine(45,8)(5,8)
 \end{picture}
 }\qquad\parbox{2.0cm}{
 \begin{picture}(52,12) (-2,-3)
    \SetWidth{0.5}
    \SetColor{Black}
    \Vertex(50,1){2.5}
    \ArrowLine(0,1)(50,1)
    \ArrowLine(45,8)(5,8)
 \end{picture}
 }\qquad\parbox{2.0cm}{
 \begin{picture}(52,12) (-2,-3)
    \SetWidth{0.5}
    \SetColor{Black}
    \Vertex(50,1){2.5}
    \ArrowLine(50,1)(0,1)
    \ArrowLine(45,8)(5,8)
 \end{picture}
 }\quad & =\quad\bar{v}^s(p)\,. 
\end{align*}
Gauge bosons: 
\begin{align*}
 \parbox{2.8cm}{
 \begin{picture}(77,16) (-2,0)
    \SetWidth{0.5}
    \SetColor{Black}
    \Photon(25,8)(75,8){3}{3}
    \Text(10,8)[]{\normalsize{\Black{$\mu,a$}}}
    \Vertex(75,8){2.5}
 \end{picture}
 }\quad =\quad\epsilon_{\mu}^a(p)\,,\quad & \quad\parbox{2.8cm}{
 \begin{picture}(77,16) (-2,0)
    \SetWidth{0.5}
    \SetColor{Black}
    \Photon(0,8)(50,8){3}{3}
    \Text(65,8)[]{\normalsize{\Black{$\mu,a$}}}
    \Vertex(0,8){2.5}
 \end{picture}
 }\quad =\quad\epsilon_{\mu}^{*a}(p)\,. 
\end{align*}
Gravitinos: 
\begin{align*}
 \parbox{2.4cm}{
 \begin{picture}(67,16) (-2,0)
    \SetWidth{0.5}
    \SetColor{Black}
    \Line(15,9)(65,9)\Line(15,5)(65,5)
    \Vertex(65,7){2.5}
    \ArrowLine(20,14)(60,14)
    \Text(5,7)[]{\normalsize{\Black{$\mu$}}}
 \end{picture}
 }\quad =\quad\psi_{\mu}^{+\,s}(p)\,,\quad & \quad\parbox{2.4cm}{
 \begin{picture}(67,16) (-2,0)
    \SetWidth{0.5}
    \SetColor{Black}
    \Line(0,9)(50,9)\Line(0,5)(50,5)
    \Vertex(0,7){2.5}
    \ArrowLine(5,14)(45,14)
    \Text(60,7)[]{\normalsize{\Black{$\mu$}}}
 \end{picture}
 }\quad =\quad\bar{\psi}_{\mu}^{+\,s}(p)\,, \\ 
 \parbox{2.4cm}{
 \begin{picture}(67,16) (-2,0)
    \SetWidth{0.5}
    \SetColor{Black}
    \Line(0,9)(50,9)\Line(0,5)(50,5)
    \Vertex(0,7){2.5}
    \ArrowLine(45,14)(5,14)
    \Text(60,7)[]{\normalsize{\Black{$\mu$}}}
 \end{picture}
 }\quad =\quad\psi_{\mu}^{-\,s}(p)\,,\quad & \quad\parbox{2.4cm}{
 \begin{picture}(67,16) (-2,0)
    \SetWidth{0.5}
    \SetColor{Black}
    \Line(15,9)(65,9)\Line(15,5)(65,5)
    \Vertex(65,7){2.5}
    \ArrowLine(60,14)(20,14)
    \Text(5,7)[]{\normalsize{\Black{$\mu$}}}
 \end{picture}
 }\quad =\quad\bar{\psi}_{\mu}^{-\,s}(p)\,. 
\end{align*}

\subsubsection*{Propagators}
The momentum $ p $ flows from the left side to the right side for the propagators shown below. \medskip
\\
Scalar particles: 
\begin{align*}
 \parbox{2.8cm}{
 \begin{picture}(82,5) (-2,-1)
    \SetWidth{0.5}
    \SetColor{Black}
    \DashArrowLine(15,1)(65,1){4}
    \Vertex(15,1){2.5}
    \Vertex(65,1){2.5}
    \Text(5,1)[]{\normalsize{\Black{$i$}}}
    \Text(75,1)[]{\normalsize{\Black{$j$}}}
 \end{picture}
 }\quad &=\quad\frac{i}{p^2-m_{\phi}^2}\,\delta^{ij}\,. 
 \intertext{Matter fermions: }
 \parbox{2.8cm}{
 \begin{picture}(82,10) (-2,-3)
    \SetWidth{0.5}
    \SetColor{Black}
    \ArrowLine(15,1)(65,1)
    \Vertex(15,1){2.5}
    \Vertex(65,1){2.5}
    \ArrowLine(20,8)(60,8)
    \Text(5,1)[]{\normalsize{\Black{$i$}}}
    \Text(75,1)[]{\normalsize{\Black{$j$}}}
 \end{picture}
 }\quad &=\quad\frac{i\left( \slashed{p}+m_{\chi}\right) }{p^2-m_{\chi}^2}\,\delta^{ij}\,, \\
 \parbox{2.8cm}{
 \begin{picture}(82,10) (-2,-3)
    \SetWidth{0.5}
    \SetColor{Black}
    \ArrowLine(15,1)(65,1)
    \Vertex(15,1){2.5}
    \Vertex(65,1){2.5}
    \ArrowLine(60,8)(20,8)
    \Text(5,1)[]{\normalsize{\Black{$i$}}}
    \Text(75,1)[]{\normalsize{\Black{$j$}}}
 \end{picture}
 }\quad &=\quad\frac{i\left( -\slashed{p}+m_{\chi}\right) }{p^2-m_{\chi}^2}\,\delta^{ij}\,. 
 \intertext{Gauginos: }
 \parbox{2.8cm}{
 \begin{picture}(82,10) (-2,-3)
    \SetWidth{0.5}
    \SetColor{Black}
    \Photon(15,1)(65,1){3}{3}
    \Line(15,1)(65,1)
    \Vertex(15,1){2.5}
    \Vertex(65,1){2.5}
    \ArrowLine(20,8)(60,8)
    \Text(5,1)[]{\normalsize{\Black{$a$}}}
    \Text(75,1)[]{\normalsize{\Black{$b$}}}
 \end{picture}
 }\quad &=\quad\frac{i\left( \slashed{p}+m_{\lambda}\right) }{p^2-m_{\lambda}^2}\,\delta^{ab}\,. 
 \intertext{Massless and massive gauge bosons: }
 \parbox{2.8cm}{
 \begin{picture}(82,10) (-2,-3)
    \SetWidth{0.5}
    \SetColor{Black}
    \Photon(15,1)(65,1){3}{3}
    \Vertex(15,1){2.5}
    \Vertex(65,1){2.5}
    \Text(0,1)[]{\normalsize{\Black{$a,\mu$}}}
    \Text(80,1)[]{\normalsize{\Black{$b,\nu$}}}
 \end{picture}
 }\quad &=\quad-\frac{ig_{\mu\nu}}{p^2}\,\delta^{ab}\,, \\
 \parbox{2.8cm}{
 \begin{picture}(82,10) (-2,-3)
    \SetWidth{0.5}
    \SetColor{Black}
    \Photon(15,1)(65,1){3}{3}
    \Vertex(15,1){2.5}
    \Vertex(65,1){2.5}
    \Text(0,1)[]{\normalsize{\Black{$a,\mu$}}}
    \Text(80,1)[]{\normalsize{\Black{$b,\nu$}}}
 \end{picture}
 }\quad &=\quad\frac{i\left( -g_{\mu\nu}+p_{\mu}p_{\nu}/m_A^2\right) }{p^2-m_A^2}\,\delta^{ab}\,. 
\end{align*}

\subsubsection*{Gauge Vertices}
The momentum $ p $ flows into the vertex for the vertices shown below. 

\begin{align*}
 \parbox{2.3cm}{
 \begin{picture}(90,85) (0,0)
    \SetWidth{0.5}
    \SetColor{Black}
    \ArrowLine(70,17)(55,43)
    \Vertex(55,43){2.5}
    \Line(25,43)(55,43)
    \Photon(25,43)(55,43){4}{2}
    \DashArrowLine(55,43)(70,69){4}
    \Text(15,43)[]{\normalsize{\Black{$a$}}}
    \Text(75,77)[]{\normalsize{\Black{$i$}}}
    \Text(75,8)[]{\normalsize{\Black{$j$}}}
    \ArrowArcn(32.46,1.96)(35.04,89.94,30.99)
    \ArrowArc(29.96,-2.04)(35.04,30.99,89.94)
 \end{picture}
 } &=-i\sqrt{2}g_{\alpha}T_{a,\,ij}^{(\alpha)}P_L & \parbox{2.3cm}{
 \begin{picture}(90,85) (0,0)
    \SetWidth{0.5}
    \SetColor{Black}
    \ArrowLine(55,43)(70,17)
    \Vertex(55,43){2.5}
    \Line(25,43)(55,43)
    \Photon(25,43)(55,43){4}{2}
    \DashArrowLine(70,69)(55,43){4}
    \Text(15,43)[]{\normalsize{\Black{$a$}}}
    \Text(75,77)[]{\normalsize{\Black{$i$}}}
    \Text(75,8)[]{\normalsize{\Black{$j$}}}
    \ArrowArcn(32.46,1.96)(35.04,89.94,30.99)
    \ArrowArc(29.96,-2.04)(35.04,30.99,89.94)
 \end{picture}
 } &=-i\sqrt{2}g_{\alpha}T_{a,\,ji}^{(\alpha)}P_R
\end{align*}

\subsubsection*{Gravitino Vertices}
The momentum $ p $ flows into the vertex for the vertices shown below. 

\begin{align*}
 \parbox{2.3cm}{
 \begin{picture}(90,85) (0,0)
    \SetWidth{0.5}
    \SetColor{Black}
    \ArrowLine(70,17)(55,43)
    \Vertex(55,43){2.5}
    \Line(25,45)(55,45)\Line(25,41)(55,41)
    \DashArrowLine(55,43)(70,69){4}
    \Text(15,43)[]{\normalsize{\Black{$\mu$}}}
    \Text(75,77)[]{\normalsize{\Black{$i,p$}}}
    \Text(75,8)[]{\normalsize{\Black{$j$}}}
    \ArrowArc(29.96,-2.04)(35.04,30.99,89.94)
 \end{picture}
 } &=-\frac{i}{\sqrt{2}\,\MP}\,\delta_{ij}\slashed{p}\gamma^{\mu}P_L & \parbox{2.3cm}{
 \begin{picture}(90,85) (0,0)
    \SetWidth{0.5}
    \SetColor{Black}
    \ArrowLine(70,17)(55,43)
    \Vertex(55,43){2.5}
    \Line(25,45)(55,45)\Line(25,41)(55,41)
    \DashArrowLine(55,43)(70,69){4}
    \Text(15,43)[]{\normalsize{\Black{$\mu$}}}
    \Text(75,77)[]{\normalsize{\Black{$i,p$}}}
    \Text(75,8)[]{\normalsize{\Black{$j$}}}
    \ArrowArcn(29.96,-2.04)(35.04,89.94,30.99)
 \end{picture}
 } &=-\frac{i}{\sqrt{2}\,\MP}\,\delta_{ij}P_L\gamma^{\mu}\slashed{p} \\
 \parbox{2.3cm}{
 \begin{picture}(90,85) (0,0)
    \SetWidth{0.5}
    \SetColor{Black}
    \ArrowLine(55,43)(70,17)
    \Vertex(55,43){2.5}
    \Line(25,45)(55,45)\Line(25,41)(55,41)
    \DashArrowLine(70,69)(55,43){4}
    \Text(15,43)[]{\normalsize{\Black{$\mu$}}}
    \Text(75,77)[]{\normalsize{\Black{$i,p$}}}
    \Text(75,8)[]{\normalsize{\Black{$j$}}}
    \ArrowArc(29.96,-2.04)(35.04,30.99,89.94)
 \end{picture}
 } &=-\frac{i}{\sqrt{2}\,\MP}\,\delta_{ij}\slashed{p}\gamma^{\mu}P_R & \parbox{2.3cm}{
 \begin{picture}(90,85) (0,0)
    \SetWidth{0.5}
    \SetColor{Black}
    \ArrowLine(55,43)(70,17)
    \Vertex(55,43){2.5}
    \Line(25,45)(55,45)\Line(25,41)(55,41)
    \DashArrowLine(70,69)(55,43){4}
    \Text(15,43)[]{\normalsize{\Black{$\mu$}}}
    \Text(75,77)[]{\normalsize{\Black{$i,p$}}}
    \Text(75,8)[]{\normalsize{\Black{$j$}}}
    \ArrowArcn(29.96,-2.04)(35.04,89.94,30.99)
 \end{picture}
 } &=-\frac{i}{\sqrt{2}\,\MP}\,\delta_{ij}P_R\gamma^{\mu}\slashed{p} \\
 \parbox{2.3cm}{
 \begin{picture}(100,86) (5,-5)
    \SetWidth{0.5}
    \SetColor{Black}
    \Vertex(55,38){2.5}
    \Photon(30,63)(55,38){3}{2.5}
    \DashArrowLine(55,38)(80,63){4}
    \ArrowLine(80,13)(55,38)
    \Line(28.59,14.41)(53.59,39.41)\Line(31.41,11.59)(56.41,36.59)
    \ArrowArc(55.5,-2.12)(23.13,40.84,139.16)
    \Text(23,70)[]{\normalsize{\Black{$a,\rho$}}}
    \Text(87,70)[]{\normalsize{\Black{$i$}}}
    \Text(87,6)[]{\normalsize{\Black{$j$}}}
    \Text(23,6)[]{\normalsize{\Black{$\mu$}}}
 \end{picture}
 } &=-\frac{ig_{\alpha}}{\sqrt{2}\,\MP}T_{a,\,ij}^{(\alpha)}P_L\gamma^{\rho}\gamma^{\mu} & \parbox{2.3cm}{
 \begin{picture}(100,86) (5,-5)
    \SetWidth{0.5}
    \SetColor{Black}
    \Vertex(55,38){2.5}
    \Photon(30,63)(55,38){3}{2.5}
    \DashArrowLine(55,38)(80,63){4}
    \ArrowLine(80,13)(55,38)
    \Line(28.59,14.41)(53.59,39.41)\Line(31.41,11.59)(56.41,36.59)
    \ArrowArcn(55.5,-2.12)(23.13,139.16,40.84)
    \Text(23,70)[]{\normalsize{\Black{$a,\rho$}}}
    \Text(87,70)[]{\normalsize{\Black{$i$}}}
    \Text(87,6)[]{\normalsize{\Black{$j$}}}
    \Text(23,6)[]{\normalsize{\Black{$\mu$}}}
 \end{picture}
 } &=-\frac{ig_{\alpha}}{\sqrt{2}\,\MP}T_{a,\,ij}^{(\alpha)}P_L\gamma^{\mu}\gamma^{\rho} \\
 \parbox{2.3cm}{
 \begin{picture}(100,86) (5,-5)
    \SetWidth{0.5}
    \SetColor{Black}
    \Vertex(55,38){2.5}
    \Photon(30,63)(55,38){3}{2.5}
    \DashArrowLine(80,63)(55,38){4}
    \ArrowLine(55,38)(80,13)
    \Line(28.59,14.41)(53.59,39.41)\Line(31.41,11.59)(56.41,36.59)
    \ArrowArc(55.5,-2.12)(23.13,40.84,139.16)
    \Text(23,70)[]{\normalsize{\Black{$a,\rho$}}}
    \Text(87,70)[]{\normalsize{\Black{$i$}}}
    \Text(87,6)[]{\normalsize{\Black{$j$}}}
    \Text(23,6)[]{\normalsize{\Black{$\mu$}}}
 \end{picture}
 } &=-\frac{ig_{\alpha}}{\sqrt{2}\,\MP}T_{a,\,ij}^{(\alpha)}P_R\gamma^{\rho}\gamma^{\mu} & \parbox{2.3cm}{
 \begin{picture}(100,86) (5,-5)
    \SetWidth{0.5}
    \SetColor{Black}
    \Vertex(55,38){2.5}
    \Photon(30,63)(55,38){3}{2.5}
    \DashArrowLine(80,63)(55,38){4}
    \ArrowLine(55,38)(80,13)
    \Line(28.59,14.41)(53.59,39.41)\Line(31.41,11.59)(56.41,36.59)
    \ArrowArcn(55.5,-2.12)(23.13,139.16,40.84)
    \Text(23,70)[]{\normalsize{\Black{$a,\rho$}}}
    \Text(87,70)[]{\normalsize{\Black{$i$}}}
    \Text(87,6)[]{\normalsize{\Black{$j$}}}
    \Text(23,6)[]{\normalsize{\Black{$\mu$}}}
 \end{picture}
 } &=-\frac{ig_{\alpha}}{\sqrt{2}\,\MP}T_{a,\,ij}^{(\alpha)}P_R\gamma^{\mu}\gamma^{\rho} \\
 \parbox{2.3cm}{
 \begin{picture}(90,85) (0,0)
    \SetWidth{0.5}
    \SetColor{Black}
    \Line(55,43)(70,17)
    \Vertex(55,43){2.5}
    \Line(25,45)(55,45)\Line(25,41)(55,41)
    \Photon(55,43)(70,69){4}{2}
    \Photon(55,43)(70,17){4}{2}
    \Text(15,43)[]{\normalsize{\Black{$\mu$}}}
    \Text(75,77)[]{\normalsize{\Black{$b,\rho,p$}}}
    \Text(75,8)[]{\normalsize{\Black{$a$}}}
    \ArrowArc(29.96,-2.04)(35.04,30.99,89.94)
 \end{picture}
 } &=-\frac{i}{4\,\MP}\,\delta_{ab}\left[ \slashed{p}, \gamma^{\rho}\right] \gamma^{\mu} & \parbox{2.3cm}{
 \begin{picture}(90,85) (0,0)
    \SetWidth{0.5}
    \SetColor{Black}
    \Line(55,43)(70,17)
    \Vertex(55,43){2.5}
    \Line(25,45)(55,45)\Line(25,41)(55,41)
    \Photon(55,43)(70,69){4}{2}
    \Photon(55,43)(70,17){4}{2}
    \Text(15,43)[]{\normalsize{\Black{$\mu$}}}
    \Text(75,77)[]{\normalsize{\Black{$b,\rho,p$}}}
    \Text(75,8)[]{\normalsize{\Black{$a$}}}
    \ArrowArcn(29.96,-2.04)(35.04,89.94,30.99)
 \end{picture}
 } &=-\frac{i}{4\,\MP}\,\delta_{ab}\gamma^{\mu}\left[ \slashed{p}, \gamma^{\rho}\right] \\
 \parbox{2.3cm}{
 \begin{picture}(100,86) (5,-5)
    \SetWidth{0.5}
    \SetColor{Black}
    \Vertex(55,38){2.5}
    \Photon(30,63)(55,38){3}{2.5}
    \Photon(80,63)(55,38){3}{2.5}
    \Line(80,13)(55,38)
    \Photon(80,13)(55,38){3}{2.5}
    \Line(28.59,14.41)(53.59,39.41)\Line(31.41,11.59)(56.41,36.59)
    \ArrowArc(55.5,-2.12)(23.13,40.84,139.16)
    \Text(23,70)[]{\normalsize{\Black{$b,\nu$}}}
    \Text(87,70)[]{\normalsize{\Black{$c,\rho$}}}
    \Text(87,6)[]{\normalsize{\Black{$a$}}}
    \Text(23,6)[]{\normalsize{\Black{$\mu$}}}
 \end{picture}
 } &=-\frac{g_{\alpha}}{4\,\MP}f^{(\alpha)\,abc}\left[ \gamma^{\nu},\,\gamma^{\rho}\right] \gamma^{\mu}\hspace*{-10mm} & \parbox{2.3cm}{
 \begin{picture}(100,86) (5,-5)
    \SetWidth{0.5}
    \SetColor{Black}
    \Vertex(55,38){2.5}
    \Photon(30,63)(55,38){3}{2.5}
    \Photon(80,63)(55,38){3}{2.5}
    \Line(80,13)(55,38)
    \Photon(80,13)(55,38){3}{2.5}
    \Line(28.59,14.41)(53.59,39.41)\Line(31.41,11.59)(56.41,36.59)
    \ArrowArcn(55.5,-2.12)(23.13,139.16,40.84)
    \Text(23,70)[]{\normalsize{\Black{$b,\nu$}}}
    \Text(87,70)[]{\normalsize{\Black{$c,\rho$}}}
    \Text(87,6)[]{\normalsize{\Black{$a$}}}
    \Text(23,6)[]{\normalsize{\Black{$\mu$}}}
 \end{picture}
 } &=-\frac{g_{\alpha}}{4\,\MP}f^{(\alpha)\,abc}\gamma^{\mu}\left[ \gamma^{\nu},\,\gamma^{\rho}\right] 
\end{align*}

\section{Kinematics}

\subsection*{Scalar Products}

\paragraph{Two-Body Decays}
Starting from 
\begin{equation}
 p=p_1+p_2\,, 
\end{equation}
we multiply by the three different four-momenta to obtain
\begin{equation}
 \begin{split}
  \left( p\cdot p\right) &=\left( p\cdot p_1\right) +\left( p\cdot p_2\right) , \\
  \left( p\cdot p_1\right) &=\left( p_1\cdot p_1\right) +\left( p_1\cdot p_2\right) , \\
  \left( p\cdot p_2\right) &=\left( p_1\cdot p_2\right) +\left( p_2\cdot p_2\right) . 
 \end{split}
\end{equation}
Using $ \left( p\cdot p\right) =M^2 $ and $ \left( p_i\cdot p_i\right) =m_i^2 $ yields the relations 
\begin{equation}
 \begin{split}
  \left( p\cdot p_1\right) &=\frac{M^2+m_1^2-m_2^2}{2}\,, \\
  \left( p\cdot p_2\right) &=\frac{M^2-m_1^2+m_2^2}{2}\,, \\
  \left( p_1\cdot p_2\right) &=\frac{M^2-m_1^2-m_2^2}{2}\,. \label{twobodyscalar}
 \end{split}
\end{equation}

\subsection*{Decay Widths}

The partial decay width for the decay of a particle of mass $ M $ into $ n $ particles is given by~\cite{Amsler:2008zz} 
\begin{equation}
d\Gamma=\frac{(2\pi)^4}{2M}\abs{\mathcal{M}}^2d\Phi_n\left( p;p_1, \ldots , p_n\right) , 
\end{equation}
where $ d\Phi_n $ is an element of the $ n $-body phase space, given by 
\begin{equation}
d\Phi_n\left( p;p_1, \ldots , p_n\right) =\delta^4\left( p-\sum_{i=1}^np_i\right) \prod_{i=1}^n\frac{d^3p_i}{(2\pi)^3\,2\,E_i}\,. 
\end{equation}

\paragraph{Two-Body Decays} 
In this case, the partial decay width becomes 
\begin{align}
d\Gamma &=\frac{1}{32\pi^2}\abs{\mathcal{M}}^2\frac{\abs{\vec{p}\,}}{M^2}\,d\Omega \\
\intertext{or if we average over the spin states}
\Gamma &=\frac{1}{8\pi}\abs{\mathcal{M}}^2\frac{\abs{\vec{p}\,}}{M^2}\,. \label{twobodywidth}
\end{align}
The momentum of the final state particles is determined by four-momentum conservation: 
\begin{equation}
 \begin{split}
  \abs{\vec{p}\,} &=\frac{1}{2M}\sqrt{\left( M^2-\left( m_1+m_2\right) ^2\right) \left( M^2-\left( m_1-m_2\right) ^2\right) } \\
  &=\frac{1}{2M}\left( M^2-\left( m_1+m_2\right) ^2\right) \sqrt{1+\frac{4\,m_1m_2}{M^2-\left( m_1+m_2\right) ^2}}\,. \label{twobodymomentum}
 \end{split}
\end{equation}

\section{Gravitino Decay Channels}
In this section we compute the dominant decay channels of the LSP gravitino in the framework of bilinear $ R $-parity breaking. The relative decay widths of the different decay channels determine the branching ratios for these channels. Therefore, this is a crucial input for the prediction of spectra of gravitino decay products.

\subsection*{$ \boldsymbol{\psi_{3/2}\rightarrow Z^0\nu_{\tau}} $}
At tree level the decay of the gravitino into the $ Z^0 $ boson and the tau neutrino gets contributions from two Feynman diagrams: 
\begin{equation*}
 \parbox{4.7cm}{
  \begin{picture}(135,137) (15,-18)
   \SetWidth{0.5}
   \SetColor{Black}
   \Line(100,50)(115,24)
   \DashArrowLine(125,41)(115,24){4}
   \Vertex(115,24){2.5}
   \Text(136,19)[]{\normalsize{\Black{$q$}}}
   \LongArrow(126,24)(135,8)
   \ArrowLine(115,24)(130,-2)
   \Photon(100,50)(115,24){5}{2}
   \Text(102,84)[]{\normalsize{\Black{$k$}}}
   \LongArrow(103,73)(112,88)
   \Photon(100,50)(130,102){5}{4}
   \Text(30,50)[]{\normalsize{\Black{$\psi_{\mu}$}}}
   \Text(70,65)[]{\normalsize{\Black{$p$}}}
   \LongArrow(61,59)(79,59)
   \Line(40,52)(100,52)\Line(40,48)(100,48)
   \Vertex(100,50){2.5}
   \Text(130,50)[]{\normalsize{\Black{$\left\langle\tilde{\nu}_{\tau}\right\rangle$}}}
   \Text(135,-10)[]{\normalsize{\Black{$\nu_{\tau}$}}}
   \Text(135,111)[]{\normalsize{\Black{$Z^0$}}}
   \ArrowArcn(51.67,-33.07)(66.09,91.45,28.04)
   \Text(97,30)[]{\normalsize{\Black{$\tilde{\chi}^0$}}}
  \end{picture}
 }\;+\;\parbox{4.9cm}{
  \begin{picture}(155,137) (15,-18)
   \SetWidth{0.5}
   \SetColor{Black}
   \Text(30,50)[]{\normalsize{\Black{$\psi_{\mu}$}}}
   \SetWidth{0.5}
   \Text(70,65)[]{\normalsize{\Black{$p$}}}
   \LongArrow(61,59)(79,59)
   \Line(40,52)(100,52)\Line(40,48)(100,48)
   \Vertex(100,50){2.5}
   \Text(135,-10)[]{\normalsize{\Black{$\nu_{\tau}$}}}
   \DashArrowLine(130,50)(100,50){4}
   \Text(142,50)[]{\normalsize{\Black{$\left\langle\tilde{\nu}_{\tau}\right\rangle$}}}
   \Text(128,32)[]{\normalsize{\Black{$q$}}}
   \LongArrow(118,36)(127,21)
   \ArrowLine(100,50)(130,-2)
   \Text(102,84)[]{\normalsize{\Black{$k$}}}
   \LongArrow(103,73)(112,88)
   \Photon(100,50)(130,102){5}{4}
   \Text(135,111)[]{\normalsize{\Black{$Z^0$}}}
   \ArrowArcn(51.67,-33.07)(66.09,91.45,28.04)
  \end{picture}
 }. 
\end{equation*}
Apart from the diagram with the abelian 3--vertex and subsequent zino--tau neutrino mixing through the tau sneutrino VEV, we have a diagram with the non-abelian 4--vertex that reduces to an effective 3--vertex using the tau sneutrino VEV from bilinear $ R $-parity breaking. Analogous to the decay of the gravitino into the photon and the tau neutrino, the amplitude for this process reads 
\begin{equation*}
 \begin{split}
  i\mathcal{M}= &-\bar{u}^r(q)\,i\sqrt{2}\left\langle \tilde{\nu}_{\tau}\right\rangle \left( g\,\frac{\sigma_{3,\,11}}{2}\cos{\theta_W}-g'Y_{\nu_{\tau}}\sin{\theta_W}\right) P_R \\
  &\;\cdot\left( \sum_{i,\,j,\,\alpha=1}^4S^*_{\tilde{Z}i}P_{i\alpha}^*\frac{i\left( \slashed{q}+m_{\tilde{\chi}_{\alpha}^0}\right) }{q^2-m_{\tilde{\chi}_{\alpha}^0}^2}P_{\alpha j}S_{j\tilde{Z}}\right) \frac{i}{4\,\MP}\,\gamma^{\mu}\left[ \slashed{k},\,\gamma^{\rho}\right] \psi_{\mu}^{+\,s}(p)\,\epsilon_{\rho}^{\lambda\,*}(k) \\
  &-\bar{u}^r(q)\,\frac{i\left\langle \tilde{\nu}_{\tau}\right\rangle }{\sqrt{2}\,\MP}\left( g\,\frac{\sigma_{3,\,11}}{2}\cos{\theta_W}-g'Y_{\nu_{\tau}}\sin{\theta_W}\right) P_R \\
  &\;\cdot\gamma^{\mu}\gamma^{\rho}\psi_{\mu}^{+\,s}(p)\,\epsilon_{\rho}^{\lambda\,*}(k) \\
  \simeq &-\frac{ig_Z\left\langle \tilde{\nu}_{\tau}\right\rangle }{8\sqrt{2}\,\MP}\left( \sum_{\alpha=1}^4\frac{S^*_{\tilde{Z}\alpha}S_{\alpha\tilde{Z}}}{m_{\tilde{\chi}_{\alpha}^0}}\right) \bar{u}^r(q)\left( 1+\gamma^5\right) \gamma^{\mu}\left[ \slashed{k},\,\gamma^{\rho}\right] \psi_{\mu}^{+\,s}(p)\,\epsilon_{\rho}^{\lambda\,*}(k) \\
  &-\frac{ig_Z\left\langle \tilde{\nu}_{\tau}\right\rangle }{4\sqrt{2}\,\MP}\,\bar{u}^r(q)\left( 1+\gamma^5\right) \gamma^{\mu}\gamma^{\rho}\psi_{\mu}^{+\,s}(p)\,\epsilon_{\rho}^{\lambda\,*}(k)\,. 
 \end{split}
\end{equation*}
Similar to the procedure in Section~\ref{widths}, we rewrite some prefactors and introduce the zino--zino mixing parameter 
\begin{equation}
 U_{\tilde{Z}\tilde{Z}}=m_Z\sum_{\alpha=1}^4\frac{S^*_{\tilde{Z}\alpha}S_{\alpha\tilde{Z}}}{m_{\tilde{\chi}_{\alpha}^0}}\,. 
\end{equation}
We note that the zino--zino mixing parameter is real-valued. Thus, the squared amplitude for this process becomes 
\begin{equation*}
 \begin{split}
  \abs{\bar{\mathcal{M}}}^2\simeq &-\frac{\xi_{\tau}^2}{256\,\MP^2}\left( g_{\rho\sigma}-\frac{k_{\rho}k_{\sigma}}{m_Z^2}\right) \\
  &\;\cdot\left\lbrace U_{\tilde{Z}\tilde{Z}}^2\Tr\left[ (\slashed{q}+m_{\nu})\left( 1+\gamma^5\right) \gamma^{\mu}\left[ \slashed{k},\,\gamma^{\rho}\right] P_{\mu\nu}^+(p)\left[ \gamma^{\sigma},\,\slashed{k}\right] \gamma^{\nu}\left( 1-\gamma^5\right) \right] \right. \\
  &+2\,m_Z\,U_{\tilde{Z}\tilde{Z}}\Tr\left[ (\slashed{q}+m_{\nu})\left( 1+\gamma^5\right) \gamma^{\mu}\left[ \slashed{k},\,\gamma^{\rho}\right] P_{\mu\nu}^+(p)\,\gamma^{\sigma}\gamma^{\nu}\left( 1-\gamma^5\right) \right]  \\
  &+2\,m_Z\,U_{\tilde{Z}\tilde{Z}}\Tr\left[ (\slashed{q}+m_{\nu})\left( 1+\gamma^5\right) \gamma^{\mu}\gamma^{\rho}P_{\mu\nu}^+(p)\left[ \gamma^{\sigma},\,\slashed{k}\right] \gamma^{\nu}\left( 1-\gamma^5\right) \right]  \\
  &+4\,m_Z^2\left.\Tr\left[ (\slashed{q}+m_{\nu})\left( 1+\gamma^5\right) \gamma^{\mu}\gamma^{\rho}P_{\mu\nu}^+(p)\,\gamma^{\sigma}\gamma^{\nu}\left( 1-\gamma^5\right) \right] \right\rbrace \\
  = &-\frac{\xi_{\tau}^2}{128\,\MP^2}\left\lbrace U_{\tilde{Z}\tilde{Z}}^2\Tr\left[ \slashed{q}\left( 1+\gamma^5\right) \gamma^{\mu}\left[ \slashed{k},\,\gamma^{\rho}\right] P_{\mu\nu}^+(p)\left[ \gamma_{\rho},\,\slashed{k}\right] \gamma^{\nu}\right] \right. \\
  &+4\,m_Z\,U_{\tilde{Z}\tilde{Z}}\Tr\left[ \slashed{q}\left( 1+\gamma^5\right) \gamma^{\mu}\left[ \slashed{k},\,\gamma^{\rho}\right] P_{\mu\nu}^+(p)\,\gamma_{\rho}\gamma^{\nu}\right]  \\
  &+4\left.\Tr\left[ \slashed{q}\left( 1+\gamma^5\right) \gamma^{\mu}\left( m_Z^2\gamma^{\rho}P_{\mu\nu}^+(p)\,\gamma_{\rho}-\slashed{k}\,P_{\mu\nu}^+(p)\,\slashed{k}\right) \gamma^{\nu}\right] \right\rbrace \\
  = &\:\frac{\xi_{\tau}^2}{3\,m_{3/2}^2 \MP^2} \\
  &\;\cdot\left\lbrace \left( 2\left( p\cdot k\right) ^2\left( p\cdot q\right) +2\,m_{3/2}^2\left( p\cdot k\right) \left( k\cdot q\right) -m_Z^2m_{3/2}^2\left( p\cdot q\right) \right) U_{\tilde{Z}\tilde{Z}}^2\right. \\
  &-2\,m_Zm_{3/2}\left( 2\left( p\cdot k\right) \left( p\cdot q\right) +m_{3/2}^2\left( k\cdot q\right) \right) U_{\tilde{Z}\tilde{Z}} \\
  &+\left. \left( p\cdot k\right) ^2\left( p\cdot q\right) +2\,m_Z^2m_{3/2}^2\left( p\cdot q\right) \right\rbrace . 
 \end{split}
\end{equation*}
In the last step we already replaced squared four-momenta by the corresponding squared particle masses. The other scalar products of the four-momenta in this process are given by 
\begin{equation*}
 \begin{split}
  \left( p\cdot k\right) &=\frac{m_{3/2}^2+m_Z^2-m_{\nu}^2}{2}\,, \\
  \left( p\cdot q\right) &=\frac{m_{3/2}^2-m_Z^2+m_{\nu}^2}{2}\,, \\
  \left( k\cdot q\right) &=\frac{m_{3/2}^2-m_Z^2-m_{\nu}^2}{2}\,. 
 \end{split}
\end{equation*}
Therefore, we obtain for the squared amplitude 
\begin{equation*}
 \begin{split}
  \abs{\bar{\mathcal{M}}}^2\simeq &\:\frac{\xi_{\tau}^2}{12\,m_{3/2}^2\MP^2}\left\lbrace \left( 3\,m_{3/2}^6-m_Z^2m_{3/2}^4-m_Z^4m_{3/2}^2-m_Z^6+m_{\nu}^6\right. \right. \\
  &\qquad+\left. m_{\nu}^4\left( m_{3/2}^2-3\,m_Z^2\right) +m_{\nu}^2\left( 3\,m_Z^4-5\,m_{3/2}^4\right) \right) U_{\tilde{Z}\tilde{Z}}^2 \\
  &-4\,m_Zm_{3/2}\left( 2\,m_{3/2}^4-m_Z^2m_{3/2}^2-m_Z^4-m_{\nu}^4\right.  \\
  &\qquad+\left. m_{\nu}^2\left( 2\,m_Z^2-m_{3/2}^2\right) \right) U_{\tilde{Z}\tilde{Z}}\\
  &+\frac{1}{2}\left( 3\,m_{3/2}^6+9\,m_Z^2m_{3/2}^4-9\,m_Z^4m_{3/2}^2-m_Z^6+m_{\nu}^6\right. \\
  &\qquad-\left. \left. m_{\nu}^4\left( m_{3/2}^2+3\,m_Z^2\right) +m_{\nu}^2\left( 3\,m_Z^4+10\,m_Z^2m_{3/2}^2-m_{3/2}^4\right) \right) \right\rbrace . 
 \end{split}
\end{equation*}
Discarding the terms proportional to the negligible neutrino mass and taking into account the correct phase space factor, the width for the decay of the gravitino into the $ Z^0 $ boson and the tau neutrino becomes 
\begin{equation}
 \Gamma\left( \psi_{3/2}\rightarrow Z^0\nu_{\tau}\right) \simeq\frac{\xi_{\tau}^2}{64\pi}\frac{m_{3/2}^3}{\MP^2}\,\beta_Z^2\left\lbrace U_{\tilde{Z}\tilde{Z}}^2f_Z-\frac{8}{3}\frac{m_Z}{m_{3/2}}\,U_{\tilde{Z}\tilde{Z}}\,j_Z+\frac{1}{6}\,h_Z\right\rbrace , 
\end{equation}
where the kinematic factors $ \beta_Z $, $ f_Z $, $ j_Z $ and $ h_Z $ are given by 
\begin{equation}
 \begin{split}
  \beta_X &=1-\frac{m_X^2}{m_{3/2}^2}\,, \\
  f_X &=1+\frac{2}{3}\frac{m_X^2}{m_{3/2}^2}+\frac{1}{3}\frac{m_X^4}{m_{3/2}^4}\,, \\
  j_X &=1+\frac{1}{2}\frac{m_X^2}{m_{3/2}^2}\,, \\
  h_X &=1+10\,\frac{m_X^2}{m_{3/2}^2}+\frac{m_X^4}{m_{3/2}^4}\,. 
 \end{split}
 \label{kinematicalfactors}
\end{equation}
The conjugate process $ \psi_{3/2}\rightarrow Z^0\bar{\nu}_{\tau} $, that produces tau antineutrinos, has the same decay width: 
\begin{equation*}
 \parbox{4.7cm}{
  \begin{picture}(135,137) (15,-18)
   \SetWidth{0.5}
   \SetColor{Black}
   \Line(100,50)(115,24)
   \DashArrowLine(115,24)(125,41){4}
   \Vertex(115,24){2.5}
   \Text(136,19)[]{\normalsize{\Black{$q$}}}
   \LongArrow(126,24)(135,8)
   \ArrowLine(130,-2)(115,24)
   \Photon(100,50)(115,24){5}{2}
   \Text(102,84)[]{\normalsize{\Black{$k$}}}
   \LongArrow(103,73)(112,88)
   \Photon(100,50)(130,102){5}{4}
   \Text(30,50)[]{\normalsize{\Black{$\psi_{\mu}$}}}
   \Text(70,65)[]{\normalsize{\Black{$p$}}}
   \LongArrow(61,59)(79,59)
   \Line(40,52)(100,52)\Line(40,48)(100,48)
   \Vertex(100,50){2.5}
   \Text(130,50)[]{\normalsize{\Black{$\left\langle\tilde{\nu}_{\tau}\right\rangle$}}}
   \Text(135,-10)[]{\normalsize{\Black{$\bar{\nu}_{\tau}$}}}
   \Text(135,111)[]{\normalsize{\Black{$Z^0$}}}
   \ArrowArcn(51.67,-33.07)(66.09,91.45,28.04)
   \Text(97,30)[]{\normalsize{\Black{$\tilde{\chi}^0$}}}
  \end{picture}
 }\;+\;\parbox{4.9cm}{
  \begin{picture}(155,137) (15,-18)
   \SetWidth{0.5}
   \SetColor{Black}
   \Text(30,50)[]{\normalsize{\Black{$\psi_{\mu}$}}}
   \SetWidth{0.5}
   \Text(70,65)[]{\normalsize{\Black{$p$}}}
   \LongArrow(61,59)(79,59)
   \Line(40,52)(100,52)\Line(40,48)(100,48)
   \Vertex(100,50){2.5}
   \Text(135,-10)[]{\normalsize{\Black{$\bar{\nu}_{\tau}$}}}
   \DashArrowLine(100,50)(130,50){4}
   \Text(142,50)[]{\normalsize{\Black{$\left\langle\tilde{\nu}_{\tau}\right\rangle$}}}
   \Text(128,32)[]{\normalsize{\Black{$q$}}}
   \LongArrow(118,36)(127,21)
   \ArrowLine(130,-2)(100,50)
   \Text(102,84)[]{\normalsize{\Black{$k$}}}
   \LongArrow(103,73)(112,88)
   \Photon(100,50)(130,102){5}{4}
   \Text(135,111)[]{\normalsize{\Black{$Z^0$}}}
   \ArrowArcn(51.67,-33.07)(66.09,91.45,28.04)
  \end{picture}
 } 
\end{equation*}
\begin{equation}
 \Gamma\left( \psi_{3/2}\rightarrow Z^0\bar{\nu}_{\tau}\right) \simeq\frac{\xi_{\tau}^2}{64\pi}\frac{m_{3/2}^3}{\MP^2}\,\beta_Z^2\left\lbrace U_{\tilde{Z}\tilde{Z}}^2f_Z-\frac{8}{3}\frac{m_Z}{m_{3/2}}\,U_{\tilde{Z}\tilde{Z}}\,j_Z+\frac{1}{6}\,h_Z\right\rbrace . 
\end{equation}

\subsection*{$ \boldsymbol{\psi_{3/2}\rightarrow W^+\tau^-} $}
Two tree-level Feynman diagrams contribute to the decay of the gravitino into the $ W^+ $ boson and the $ \tau^- $ lepton: 
\begin{equation*}
 \parbox{4.7cm}{
  \begin{picture}(135,137) (15,-18)
   \SetWidth{0.5}
   \SetColor{Black}
   \Line(100,50)(115,24)
   \DashArrowLine(125,41)(115,24){4}
   \Vertex(115,24){2.5}
   \Text(136,19)[]{\normalsize{\Black{$q$}}}
   \LongArrow(126,24)(135,8)
   \ArrowLine(115,24)(130,-2)
   \Photon(100,50)(115,24){5}{2}
   \Text(102,84)[]{\normalsize{\Black{$k$}}}
   \LongArrow(103,73)(112,88)
   \Photon(100,50)(130,102){5}{4}
   \Text(30,50)[]{\normalsize{\Black{$\psi_{\mu}$}}}
   \Text(70,65)[]{\normalsize{\Black{$p$}}}
   \LongArrow(61,59)(79,59)
   \Line(40,52)(100,52)\Line(40,48)(100,48)
   \Vertex(100,50){2.5}
   \Text(130,50)[]{\normalsize{\Black{$\left\langle\tilde{\nu}_{\tau}\right\rangle$}}}
   \Text(135,-10)[]{\normalsize{\Black{$\tau^-$}}}
   \Text(135,111)[]{\normalsize{\Black{$W^+$}}}
   \ArrowArcn(51.67,-33.07)(66.09,91.45,28.04)
   \Text(97,30)[]{\normalsize{\Black{$\tilde{\chi}^-$}}}
  \end{picture}
 }\;+\;\parbox{4.9cm}{
  \begin{picture}(155,137) (15,-18)
   \SetWidth{0.5}
   \SetColor{Black}
   \Text(30,50)[]{\normalsize{\Black{$\psi_{\mu}$}}}
   \SetWidth{0.5}
   \Text(70,65)[]{\normalsize{\Black{$p$}}}
   \LongArrow(61,59)(79,59)
   \Line(40,52)(100,52)\Line(40,48)(100,48)
   \Vertex(100,50){2.5}
   \Text(135,-10)[]{\normalsize{\Black{$\tau^-$}}}
   \DashArrowLine(130,50)(100,50){4}
   \Text(142,50)[]{\normalsize{\Black{$\left\langle\tilde{\nu}_{\tau}\right\rangle$}}}
   \Text(128,32)[]{\normalsize{\Black{$q$}}}
   \LongArrow(118,36)(127,21)
   \ArrowLine(100,50)(130,-2)
   \Text(102,84)[]{\normalsize{\Black{$k$}}}
   \LongArrow(103,73)(112,88)
   \Photon(100,50)(130,102){5}{4}
   \Text(135,111)[]{\normalsize{\Black{$W^+$}}}
   \ArrowArcn(51.67,-33.07)(66.09,91.45,28.04)
  \end{picture}
 }. 
\end{equation*}
Using the rotation of the charged gauge eigenstates into the charginos (cf. Section~\ref{susy})
\begin{equation}
 \tilde{W}^+=\sum_{\alpha=1}^2V^*_{\tilde{W}^+\alpha}\tilde{\chi}_{\alpha}^+\qquad\text{and}\qquad\tilde{W}^-=\sum_{\alpha=1}^2U^*_{\tilde{W}^-\alpha}\tilde{\chi}_{\alpha}^-\,, 
\end{equation}
and the relation between the gauge eigenstates and mass eigenstates of the electroweak gauge bosons (\ref{chargedeigenstates}) together with the generators of the corresponding gauge groups (\ref{generators}), we can write for the amplitude of this process 
\begin{equation*}
 \begin{split}
  i\mathcal{M}= &-\bar{u}^r(q)\,i\sqrt{2}\,g\left\langle \tilde{\nu}_{\tau}\right\rangle \frac{1}{\sqrt{2}}\left( \frac{\sigma_{1,\,21}}{2}-i\,\frac{\sigma_{2,\,21}}{2}\right) P_R\left( \frac{1}{2}\sum_{\alpha=1}^2V^*_{\tilde{W}^+\alpha}\right.  \\
  &\;\cdot\left. \frac{i\left( \slashed{q}+m_{\tilde{\chi}_{\alpha}^{\pm}}\right) }{q^2-m_{\tilde{\chi}_{\alpha}^{\pm}}^2}\,U_{\alpha\tilde{W}^-}+h.c.\right) \frac{i}{4\,\MP}\,\gamma^{\mu}\left[ \slashed{k},\,\gamma^{\rho}\right] \psi_{\mu}^{+\,s}(p)\,\epsilon_{\rho}^{\lambda\,*}(k) \\
  &-\bar{u}^r(q)\,\frac{ig\left\langle \tilde{\nu}_{\tau}\right\rangle }{\sqrt{2}\,\MP}\frac{1}{\sqrt{2}}\left( \frac{\sigma_{1,\,21}}{2}-i\,\frac{\sigma_{2,\,21}}{2}\right) P_R\gamma^{\mu}\gamma^{\rho}\psi_{\mu}^{+\,s}(p)\,\epsilon_{\rho}^{\lambda\,*}(k) \\
  \simeq &-\frac{ig\left\langle \tilde{\nu}_{\tau}\right\rangle }{8\,\MP}\left( \frac{1}{2}\sum_{\alpha=1}^2\frac{V^*_{\tilde{W}^+\alpha}U_{\alpha\tilde{W}^-}+h.c.}{m_{\tilde{\chi}_{\alpha}^{\pm}}}\right) \\
  &\;\cdot\bar{u}^r(q)\left( 1+\gamma^5\right) \gamma^{\mu}\left[ \slashed{k},\,\gamma^{\rho}\right] \psi_{\mu}^{+\,s}(p)\,\epsilon_{\rho}^{\lambda\,*}(k) \\
  &-\frac{ig\left\langle \tilde{\nu}_{\tau}\right\rangle }{4\,\MP}\,\bar{u}^r(q)\left( 1+\gamma^5\right) \gamma^{\mu}\gamma^{\rho}\psi_{\mu}^{+\,s}(p)\,\epsilon_{\rho}^{\lambda\,*}(k)\,. 
 \end{split}
\end{equation*}
Rewriting some coefficients and introducing the wino--wino mixing parameter 
\begin{equation}
 U_{\tilde{W}\tilde{W}}=\frac{m_W}{2}\sum_{\alpha=1}^2\frac{V^*_{\tilde{W}^+\alpha}U_{\alpha\tilde{W}^-}+h.c.}{m_{\tilde{\chi}_{\alpha}^{\pm}}}\,, 
 \label{wino-wino}
\end{equation}
that is real-valued, the squared amplitude for this process becomes 
\begin{equation*}
 \begin{split}
  \abs{\bar{\mathcal{M}}}^2= &-\frac{\xi_{\tau}^2}{128\,\MP^2}\left( g_{\rho\sigma}-\frac{k_{\rho}k_{\sigma}}{m_W^2}\right) \\
  &\;\cdot\left\lbrace U_{\tilde{W}\tilde{W}}^2\Tr\left[ (\slashed{q}+m_{\tau})\left( 1+\gamma^5\right) \gamma^{\mu}\left[ \slashed{k},\,\gamma^{\rho}\right] P_{\mu\nu}^+(p)\left[ \gamma^{\sigma},\,\slashed{k}\right] \gamma^{\nu}\left( 1-\gamma^5\right) \right] \right. \\
  &+2\,m_W\,U_{\tilde{W}\tilde{W}}\Tr\left[ (\slashed{q}+m_{\tau})\left( 1+\gamma^5\right) \gamma^{\mu}\left[ \slashed{k},\,\gamma^{\rho}\right] P_{\mu\nu}^+(p)\,\gamma^{\sigma}\gamma^{\nu}\left( 1-\gamma^5\right) \right]  \\
  &+2\,m_W\,U_{\tilde{W}\tilde{W}}\Tr\left[ (\slashed{q}+m_{\tau})\left( 1+\gamma^5\right) \gamma^{\mu}\gamma^{\rho}P_{\mu\nu}^+(p)\left[ \gamma^{\sigma},\,\slashed{k}\right] \gamma^{\nu}\left( 1-\gamma^5\right) \right]  \\
  &\left. +4\,m_W^2\Tr\left[ (\slashed{q}+m_{\tau})\left( 1+\gamma^5\right) \gamma^{\mu}\gamma^{\rho}P_{\mu\nu}^+(p)\,\gamma^{\sigma}\gamma^{\nu}\left( 1-\gamma^5\right) \right] \right\rbrace . 
 \end{split}
\end{equation*}
The traces are completely the same as in the previously discussed decay. Only the prefactors and the kinematics differ. Adopting the correct prefactors, neglecting $ m_{\tau} $ and exchanging $ m_Z $ with $ m_W $ in the kinematic factors, the width for the decay of the gravitino into the $ W^+ $ boson and the $ \tau^- $ lepton reads 
\begin{equation}
 \Gamma\left( \psi_{3/2}\rightarrow W^+\tau^-\right) \simeq\frac{\xi_{\tau}^2}{32\pi}\frac{m_{3/2}^3}{\MP^2}\,\beta_W^2\left\lbrace U_{\tilde{W}\tilde{W}}^2f_W-\frac{8}{3}\frac{m_W}{m_{3/2}}\,U_{\tilde{W}\tilde{W}}\,j_W+\frac{1}{6}\,h_W\right\rbrace , 
\end{equation}
where the kinematic factors $ \beta_W $, $ f_W $, $ j_W $ and $ h_W $ are the same as those defined in equation (\ref{kinematicalfactors}). 

The conjugate process $ \psi_{3/2}\rightarrow W^-\tau^+ $ has the same decay width 
\begin{equation*}
 \parbox{4.7cm}{
  \begin{picture}(135,137) (15,-18)
   \SetWidth{0.5}
   \SetColor{Black}
   \Line(100,50)(115,24)
   \DashArrowLine(115,24)(125,41){4}
   \Vertex(115,24){2.5}
   \Text(136,19)[]{\normalsize{\Black{$q$}}}
   \LongArrow(126,24)(135,8)
   \ArrowLine(130,-2)(115,24)
   \Photon(100,50)(115,24){5}{2}
   \Text(102,84)[]{\normalsize{\Black{$k$}}}
   \LongArrow(103,73)(112,88)
   \Photon(100,50)(130,102){5}{4}
   \Text(30,50)[]{\normalsize{\Black{$\psi_{\mu}$}}}
   \Text(70,65)[]{\normalsize{\Black{$p$}}}
   \LongArrow(61,59)(79,59)
   \Line(40,52)(100,52)\Line(40,48)(100,48)
   \Vertex(100,50){2.5}
   \Text(130,50)[]{\normalsize{\Black{$\left\langle\tilde{\nu}_{\tau}\right\rangle$}}}
   \Text(135,-10)[]{\normalsize{\Black{$\tau^+$}}}
   \Text(135,111)[]{\normalsize{\Black{$W^-$}}}
   \ArrowArcn(51.67,-33.07)(66.09,91.45,28.04)
   \Text(97,30)[]{\normalsize{\Black{$\tilde{\chi}^+$}}}
  \end{picture}
 }\;+\;\parbox{4.9cm}{
  \begin{picture}(155,137) (15,-18)
   \SetWidth{0.5}
   \SetColor{Black}
   \Text(30,50)[]{\normalsize{\Black{$\psi_{\mu}$}}}
   \SetWidth{0.5}
   \Text(70,65)[]{\normalsize{\Black{$p$}}}
   \LongArrow(61,59)(79,59)
   \Line(40,52)(100,52)\Line(40,48)(100,48)
   \Vertex(100,50){2.5}
   \Text(135,-10)[]{\normalsize{\Black{$\tau^+$}}}
   \DashArrowLine(100,50)(130,50){4}
   \Text(142,50)[]{\normalsize{\Black{$\left\langle\tilde{\nu}_{\tau}\right\rangle$}}}
   \Text(128,32)[]{\normalsize{\Black{$q$}}}
   \LongArrow(118,36)(127,21)
   \ArrowLine(130,-2)(100,50)
   \Text(102,84)[]{\normalsize{\Black{$k$}}}
   \LongArrow(103,73)(112,88)
   \Photon(100,50)(130,102){5}{4}
   \Text(135,111)[]{\normalsize{\Black{$W^-$}}}
   \ArrowArcn(51.67,-33.07)(66.09,91.45,28.04)
  \end{picture}
 }. 
\end{equation*}
\begin{equation}
 \Gamma\left( \psi_{3/2}\rightarrow W^-\tau^+\right) \simeq\frac{\xi_{\tau}^2}{32\pi}\frac{m_{3/2}^3}{\MP^2}\,\beta_W^2\left\lbrace U_{\tilde{W}\tilde{W}}^2f_W-\frac{8}{3}\frac{m_W}{m_{3/2}}\,U_{\tilde{W}\tilde{W}}\,j_W+\frac{1}{6}\,h_W\right\rbrace . 
\end{equation}

\subsection*{$ \boldsymbol{\psi_{3/2}\rightarrow h\,\nu_{\tau}} $}
The decay of the gravitino into the lightest Higgs boson and the tau neutrino also gets contributions from two tree-level Feynman diagrams 
\begin{equation*}
 \parbox{4.5cm}{
  \begin{picture}(135,137) (15,-18)
   \SetWidth{0.5}
   \SetColor{Black}
   \Text(30,50)[]{\normalsize{\Black{$\psi_{\mu}$}}}
   \SetWidth{0.5}
   \Text(70,65)[]{\normalsize{\Black{$p$}}}
   \LongArrow(61,59)(79,59)
   \Line(40,52)(100,52)\Line(40,48)(100,48)
   \Vertex(100,50){2.5}
   \Text(135,-10)[]{\normalsize{\Black{$\nu_{\tau}$}}}
   \Text(128,32)[]{\normalsize{\Black{$q$}}}
   \LongArrow(118,36)(127,21)
   \ArrowLine(100,50)(130,-2)
   \ArrowArcn(51.67,-33.07)(66.09,91.45,28.04)
   \DashArrowLine(115,76)(100,50){4}
   \Text(109,96)[]{\normalsize{\Black{$k$}}}
   \LongArrow(110,85)(119,101)
   \DashLine(115,76)(130,102){4}
   \Line(108,74)(122,78)\Line(113,83)(117,69)
   \Text(135,111)[]{\normalsize{\Black{$h$}}}
   \Text(95,67)[]{\normalsize{\Black{$\tilde{\nu}_{\tau}^*$}}}
  \end{picture}
 }\;+\;\parbox{4.5cm}{
  \begin{picture}(135,137) (15,-18)
   \SetWidth{0.5}
   \SetColor{Black}
   \Line(100,50)(115,24)
   \DashArrowLine(125,41)(115,24){4}
   \Vertex(115,24){2.5}
   \Text(136,19)[]{\normalsize{\Black{$q$}}}
   \LongArrow(126,24)(135,8)
   \ArrowLine(115,24)(130,-2)
   \Photon(100,50)(115,24){5}{2}
   \Text(102,84)[]{\normalsize{\Black{$k$}}}
   \LongArrow(103,73)(112,88)
   \DashLine(100,50)(130,102){4}
   \Text(30,50)[]{\normalsize{\Black{$\psi_{\mu}$}}}
   \Text(70,65)[]{\normalsize{\Black{$p$}}}
   \LongArrow(61,59)(79,59)
   \Line(40,52)(100,52)\Line(40,48)(100,48)
   \Vertex(100,50){2.5}
   \Text(130,50)[]{\normalsize{\Black{$\left\langle\tilde{\nu}_{\tau}\right\rangle$}}}
   \Text(135,-10)[]{\normalsize{\Black{$\nu_{\tau}$}}}
   \Text(135,111)[]{\normalsize{\Black{$h$}}}
   \ArrowArcn(51.67,-33.07)(66.09,91.45,28.04)
   \Text(97,30)[]{\normalsize{\Black{$\tilde{\chi}^0$}}}
  \end{picture}
 }. 
\end{equation*}
In the MSSM decoupling limit the mixing angle $ \alpha $ becomes equal to $ \beta-\pi/2 $ (cf. Section~\ref{susy}). Therefore we have 
\begin{equation}
 \begin{split}
  \cos\alpha\quad &\rightarrow\quad\sin\beta\,, \\
  \sin\alpha\quad &\rightarrow\quad-\cos\beta 
 \end{split}
\end{equation}
and the decomposition of the gauge eigenstates into the mass eigenstates for the neutral Higgs bosons (\ref{HiggsStates}) becomes 
\begin{equation}
 \begin{pmatrix}
  \text{Re}\!\left[ H_u^0\right]  \\
  \text{Re}\!\left[ H_d^0\right] 
 \end{pmatrix}=
 \frac{1}{\sqrt{2}}
 \begin{pmatrix}
  \sin\beta & -\cos\beta \\
  \cos\beta & \sin\beta 
 \end{pmatrix}
\begin{pmatrix}
  h \\
  H 
 \end{pmatrix}. 
 \label{decoupleHiggs}
\end{equation}
The inversion of this equation gives for the lightest Higgs boson 
\begin{equation}
 h=\sqrt{2}\left( \text{Re}\!\left[ H_u^0\right] \sin{\beta}+\text{Re}\!\left[ H_d^0\right] \cos{\beta}\right) . 
 \label{lightesthiggs}
\end{equation}
Application of this relation to the Higgs--slepton mixing terms in equation (\ref{HiggsSlepton}) together with the relation for the sneutrino VEV (\ref{sneutrinoVEV}) results in the coupling 
\begin{equation}
 i\mathscr{L}=-\frac{i}{\sqrt{2}}\,m_{\tilde{\nu}_{\tau}}^2\frac{\left\langle \tilde{\nu}_{\tau}\right\rangle }{v}\,\tilde{\nu}_{\tau}^*\,h+h.c. 
\end{equation}
between the tau sneutrino and the lightest Higgs boson, that appears in the first Feynman diagram. For the second diagram we use the vertices of the gravitino with the Higgs gauge eigenstates $ \text{Re}\!\left[ H_{u,\,d}^0\right]  $ and the corresponding higgsino states $ \tilde{H}_{u,\,d}^0 $. Then we use the rotation of the Higgs gauge eigenstates $ \text{Re}\!\left[ H_{u,\,d}^0\right]  $ into the mass eigenstates $ h $ and $ H $, which is given above in equation (\ref{decoupleHiggs}). Since we assume that the heavier Higgs bosons are kinematically not accessible, we only consider the final states with the lightest Higgs boson. Then we use, analogous to the other gravitino decay modes, the rotation of the neutral gauge eigenstates into the neutralinos (cf. Section~\ref{susy}) 
\begin{equation}
 \tilde{H}_{u,\,d}^0=\sum_{i,\,\alpha=1}^4S^*_{\tilde{H}_{u,\,d}^0i}P_{i\alpha}^*\tilde{\chi}_{\alpha}^0\,, 
\end{equation}
and the relation between the gauge eigenstates and mass eigenstates of the electroweak gauge bosons (\ref{neutraleigenstates}) together with the generators of the corresponding gauge groups (\ref{generators}) to write for the amplitude of this process 
\begin{equation*}
 \begin{split}
  i\mathcal{M}\simeq &\:\bar{u}^r(q)\,\frac{i}{\sqrt{2}\,\MP}P_R\gamma^{\mu}\slashed{k}\,\psi_{\mu}^{+\,s}(p)\,\frac{i}{k^2-m_{\tilde{\nu}_{\tau}}^2}\frac{i}{\sqrt{2}}\,m_{\tilde{\nu}_{\tau}}^2\frac{\left\langle \tilde{\nu}_{\tau}\right\rangle }{v} \\
  &+\bar{u}^r(q)\,i\sqrt{2}\left( g\,\frac{\sigma_{3,\,11}}{2}\cos{\theta_W}-g'Y_{\nu_{\tau}}\sin{\theta_W}\right) P_R \\
  &\;\cdot\left\lbrace \frac{\sin{\beta}}{\sqrt{2}}\left( \sum_{i,\,j,\,\alpha=1}^4S^*_{\tilde{Z}i}P_{i\alpha}^*\frac{i\left( \slashed{q}+m_{\tilde{\chi}_{\alpha}^0}\right) }{q^2-m_{\tilde{\chi}_{\alpha}^0}^2}P_{\alpha j}S_{j\tilde{H}_u^0}\right) \right. \\
  &\left. \qquad+\frac{\cos{\beta}}{\sqrt{2}}\left( \sum_{i,\,j,\,\alpha=1}^4S^*_{\tilde{Z}i}P_{i\alpha}^*\frac{i\left( \slashed{q}+m_{\tilde{\chi}_{\alpha}^0}\right) }{q^2-m_{\tilde{\chi}_{\alpha}^0}^2}P_{\alpha j}S_{j\tilde{H}_d^0}\right) \right\rbrace \\
  &\;\cdot\frac{i}{\sqrt{2}\,\MP}P_R\gamma^{\mu}\slashed{k}\,\psi_{\mu}^{+\,s}(p)\left\langle \tilde{\nu}_{\tau}\right\rangle 
 \end{split}
\end{equation*}
\begin{equation*}
 \begin{split}
  \hspace*{2cm}\simeq &\:\frac{i\left\langle \tilde{\nu}_{\tau}\right\rangle }{4\,v\MP}\frac{m_{\tilde{\nu}_{\tau}}^2}{m_{\tilde{\nu}_{\tau}}^2-k^2}\,\bar{u}^r(q)\left( 1+\gamma^5\right) \gamma^{\mu}\slashed{k}\,\psi_{\mu}^{+\,s}(p) \\
  &+\frac{ig_Z\left\langle \tilde{\nu}_{\tau}\right\rangle }{4\sqrt{2}\,\MP}\left\lbrace \sin{\beta}\left( \sum_{\alpha=1}^4\frac{S^*_{\tilde{Z}\alpha}S_{\alpha\tilde{H}_u^0}}{m_{\tilde{\chi}_{\alpha}^0}}\right) +\cos{\beta}\sum_{\alpha=1}^4\frac{S^*_{\tilde{Z}\alpha}S_{\alpha\tilde{H}_d^0}}{m_{\tilde{\chi}_{\alpha}^0}}\right\rbrace \\
  &\;\cdot\bar{u}^r(q)\left( 1+\gamma^5\right) \gamma^{\mu}\slashed{k}\,\psi_{\mu}^{+\,s}(p)\,. 
 \end{split}
\end{equation*}
Introducing the higgsino--zino mixing parameters 
\begin{equation}
 U_{\tilde{H}_{u,\,d}^0\tilde{Z}}=m_Z\sum_{\alpha=1}^4\frac{S^*_{\tilde{Z}\alpha}S_{\alpha\tilde{H}_{u,\,d}^0}}{m_{\tilde{\chi}_{\alpha}^0}} 
\end{equation}
and rewriting some prefactors, we obtain for the squared amplitude 
\begin{align*}
 \abs{\bar{\mathcal{M}}}^2\simeq &\:\frac{\xi_{\tau}^2}{64\,\MP^2}\abs{\frac{m_{\tilde{\nu}_{\tau}}^2}{m_{\tilde{\nu}_{\tau}}^2-k^2}+\sin{\beta}\,U_{\tilde{H}_u^0\tilde{Z}}+\cos{\beta}\,U_{\tilde{H}_d^0\tilde{Z}}}^2 \\
 &\;\cdot\Tr\left[ (\slashed{q}+m_{\nu})\left( 1+\gamma^5\right) \gamma^{\mu}\slashed{k}\,P_{\mu\nu}^+(p)\,\slashed{k}\,\gamma^{\nu}\left( 1-\gamma^5\right) \right] \\
 = &\:\frac{\xi_{\tau}^2}{32\,\MP^2}\abs{\frac{m_{\tilde{\nu}_{\tau}}^2}{m_{\tilde{\nu}_{\tau}}^2-k^2}+\sin{\beta}\,U_{\tilde{H}_u^0\tilde{Z}}+\cos{\beta}\,U_{\tilde{H}_d^0\tilde{Z}}}^2 \\
 &\;\cdot\Tr\left[ \slashed{q}\left( 1+\gamma^5\right) \gamma^{\mu}\slashed{k}\,P_{\mu\nu}^+(p)\,\slashed{k}\,\gamma^{\nu}\right] \\
 = &\:\frac{\xi_{\tau}^2}{3\,m_{3/2}^2\MP^2}\abs{\frac{m_{\tilde{\nu}_{\tau}}^2}{m_{\tilde{\nu}_{\tau}}^2-m_h^2}+\sin{\beta}\,U_{\tilde{H}_u^0\tilde{Z}}+\cos{\beta}\,U_{\tilde{H}_d^0\tilde{Z}}}^2 \\
 &\;\cdot\left\lbrace \left( p\cdot k\right) ^2\left( p\cdot q\right) -m_h^2m_{3/2}^2\left( p\cdot q\right) \right\rbrace . 
\end{align*}
Here, we already substituted the squared four-momenta by the corresponding squared particle masses. The other scalar products of the four-momenta in this process are given by 
\begin{equation*}
 \begin{split}
  \left( p\cdot k\right) &=\frac{m_{3/2}^2+m_h^2-m_{\nu}^2}{2}\,, \\
  \left( p\cdot q\right) &=\frac{m_{3/2}^2-m_h^2+m_{\nu}^2}{2}\,, \\
  \left( k\cdot q\right) &=\frac{m_{3/2}^2-m_h^2-m_{\nu}^2}{2}\,. 
 \end{split}
\end{equation*}
Therefore, we get for the squared amplitude 
\begin{equation*}
 \begin{split}
  \abs{\bar{\mathcal{M}}}^2\simeq &\:\frac{\xi_{\tau}^2}{24\,m_{3/2}^2\MP^2}\abs{\frac{m_{\tilde{\nu}_{\tau}}^2}{m_{\tilde{\nu}_{\tau}}^2-m_h^2}+\sin{\beta}\,U_{\tilde{H}_u^0\tilde{Z}}+\cos{\beta}\,U_{\tilde{H}_d^0\tilde{Z}}}^2 \\
  &\;\cdot\left\lbrace \left( m_{3/2}^2-m_h^2\right) ^3-m_{\nu}^2\left( m_{3/2}^4+2\,m_h^2m_{3/2}^2-3\,m_h^4\right) \right. \\
  &\qquad-m_{\nu}^4\left( m_{3/2}^2+3\,m_h^2\right) +m_{\nu}^6\Big\rbrace. 
 \end{split}
\end{equation*}
Discarding the terms proportional to the negligible neutrino mass and taking into account the correct phase space factor, the width for the decay of the gravitino into the lightest Higgs boson and the tau neutrino becomes 
\begin{equation}
 \Gamma\left( \psi_{3/2}\rightarrow h\,\nu_{\tau}\right) \simeq\frac{\xi_{\tau}^2}{384\pi}\frac{m_{3/2}^3}{\MP^2}\,\beta_h^4\abs{\frac{m_{\tilde{\nu}_{\tau}}^2}{m_{\tilde{\nu}_{\tau}}^2-m_h^2}+\sin{\beta}\,U_{\tilde{H}_u^0\tilde{Z}}+\cos{\beta}\,U_{\tilde{H}_d^0\tilde{Z}}}^2, 
\end{equation}
where the kinematic factor $ \beta_h $ is given in equation (\ref{kinematicalfactors}). 

The conjugate process $ \psi_{3/2}\rightarrow h\,\bar{\nu}_{\tau} $, that produces tau antineutrinos, has the same decay width: 
\begin{equation*}
 \parbox{4.5cm}{
  \begin{picture}(135,137) (15,-18)
   \SetWidth{0.5}
   \SetColor{Black}
   \Text(30,50)[]{\normalsize{\Black{$\psi_{\mu}$}}}
   \SetWidth{0.5}
   \Text(70,65)[]{\normalsize{\Black{$p$}}}
   \LongArrow(61,59)(79,59)
   \Line(40,52)(100,52)\Line(40,48)(100,48)
   \Vertex(100,50){2.5}
   \Text(135,-10)[]{\normalsize{\Black{$\bar{\nu}_{\tau}$}}}
   \Text(128,32)[]{\normalsize{\Black{$q$}}}
   \LongArrow(118,36)(127,21)
   \ArrowLine(130,-2)(100,50)
   \ArrowArcn(51.67,-33.07)(66.09,91.45,28.04)
   \DashArrowLine(100,50)(115,76){4}
   \Text(109,96)[]{\normalsize{\Black{$k$}}}
   \LongArrow(110,85)(119,101)
   \DashLine(130,102)(115,76){4}
   \Line(108,74)(122,78)\Line(113,83)(117,69)
   \Text(135,111)[]{\normalsize{\Black{$h$}}}
   \Text(95,67)[]{\normalsize{\Black{$\tilde{\nu}_{\tau}$}}}
  \end{picture}
 }\;+\;\parbox{4.5cm}{
  \begin{picture}(135,137) (15,-18)
   \SetWidth{0.5}
   \SetColor{Black}
   \Line(100,50)(115,24)
   \DashArrowLine(115,24)(125,41){4}
   \Vertex(115,24){2.5}
   \Text(136,19)[]{\normalsize{\Black{$q$}}}
   \LongArrow(126,24)(135,8)
   \ArrowLine(130,-2)(115,24)
   \Photon(100,50)(115,24){5}{2}
   \Text(102,84)[]{\normalsize{\Black{$k$}}}
   \LongArrow(103,73)(112,88)
   \DashLine(130,102)(100,50){4}
   \Text(30,50)[]{\normalsize{\Black{$\psi_{\mu}$}}}
   \Text(70,65)[]{\normalsize{\Black{$p$}}}
   \LongArrow(61,59)(79,59)
   \Line(40,52)(100,52)\Line(40,48)(100,48)
   \Vertex(100,50){2.5}
   \Text(130,50)[]{\normalsize{\Black{$\left\langle\tilde{\nu}_{\tau}\right\rangle$}}}
   \Text(135,-10)[]{\normalsize{\Black{$\bar{\nu}_{\tau}$}}}
   \Text(135,111)[]{\normalsize{\Black{$h$}}}
   \ArrowArcn(51.67,-33.07)(66.09,91.45,28.04)
   \Text(97,30)[]{\normalsize{\Black{$\tilde{\chi}^0$}}}
  \end{picture}
 }. 
\end{equation*}
\begin{equation}
  \Gamma\left( \psi_{3/2}\rightarrow h\,\bar{\nu}_{\tau}\right) \simeq\frac{\xi_{\tau}^2}{384\pi}\frac{m_{3/2}^3}{\MP^2}\,\beta_h^4\abs{\frac{m_{\tilde{\nu}_{\tau}}^2}{m_{\tilde{\nu}_{\tau}}^2-m_h^2}+\sin{\beta}\,U_{\tilde{H}_u^0\tilde{Z}}+\cos{\beta}\,U_{\tilde{H}_d^0\tilde{Z}}}^2. 
\end{equation} 

\chapter{PYTHIA Simulation}
\label{pythia}
In this appendix, we present an example of the PYTHIA simulation used to obtain the spectra of the fragmentation products. 

PYTHIA is a Monte Carlo event generator for collider simulations. Therefore we have to mimic a gravitino decay in a collider setup. To implement the gravitino in the simulation, we use the optional fourth fermion generation and redefine the properties of the particles. 

The process starts with the exclusive production of a $ Z' $ boson with mass $ m_{Z'}=2\,m_{3/2} $ in an $ e^+\,e^- $ collision with a center of mass energy slightly above twice the gravitino mass. We redefine the decay channels of the $ Z' $ boson, so that only the decay into $ \nu'_{\tau}\,\bar{\nu}'_{\tau} $ is allowed. The fourth generation neutrinos $ \nu'_{\tau} $ and $ \bar{\nu}'_{\tau} $ are chosen to represent gravitinos with mass $ m_{3/2} $. In order to have a nonvanishing phase space for this decay, we chose above a slightly increased center of mass energy for the initial collision. 

Then we define a single decay channel for the $ \nu'_{\tau} $: Either into $ Z\nu_{\tau} $, $ W^{\pm}\tau^{\mp} $ or $ h\,\nu_{\tau} $. The introduction of only one decay channel at a time allows us to determine the spectra independent of the branching ratios for the different decay channels. Anyway, these can easily be added afterwards. 

We are interested in the stable final state particles, so we have to allow all decays of long-lived particles like muons, pions, kaons and neutrons. By default, PYTHIA regards these particles as stable, since they do not decay on collider timescales. 

In the end we print out a list with the energies of the final state neutrinos (only one flavor at a time). 

The used PYTHIA program is written in FORTRAN 77 and is compiled via the command 
\begin{quote}
 \texttt{g77 'program'.f pythia-6.4.15.f}. 
\end{quote}
This produces an executable file \texttt{a.out}. The event generation is started via 
\begin{quote}
 \texttt{a.out > 'output'.dat} 
\end{quote}
and the output data is stored into the file \texttt{'output'.dat}. The following code shows the program for the simulation of the decay and fragmentation of $ \psi_{3/2}\rightarrow Z^0\nu_{\tau} $ and the corresponding conjugate process. 
\begin{small}
 \begin{verbatim}
C...Double precision and integer declarations.
      IMPLICIT DOUBLE PRECISION(A-H,O-Z)
      IMPLICIT INTEGER(I-N)
      INTEGER PYK,PYCHGE,PYCOMP

C...EXTERNAL statement links PYDATA on most platforms.
      EXTERNAL PYDATA

C...Common blocks.
C...The event record.
      COMMON/PYJETS/N,NPAD,K(4000,5),P(4000,5),V(4000,5)
C...Parameters.
      COMMON/PYDAT1/MSTU(200),PARU(200),MSTJ(200),PARJ(200)
C...Particle properties + some flavor parameters.
      COMMON/PYDAT2/KCHG(500,4),PMAS(500,4),PARF(2000),VCKM(4,4)
C...Decay information.
      COMMON/PYDAT3/MDCY(500,3),MDME(8000,2),BRAT(8000),KFDP(8000,5)
C...Selection of hard scattering subprocesses.
      COMMON/PYSUBS/MSEL,MSELPD,MSUB(500),KFIN(2,-40:40),CKIN(200)
C...Parameters.
      COMMON/PYPARS/MSTP(200),PARP(200),MSTI(200),PARI(200)
C...Supersymmetry parameters.
      COMMON/PYMSSM/IMSS(0:99),RMSS(0:99)
      
C...Set gravitino mass.
      GMASS=150.0D0
      
C...Set number of events and CMS energy.
      NEVENTS=200000
      ECM=2*GMASS+0.003
C...Note: ECM is slightly increased so that PYTHIA doesn't complain.

C...Allow fourth fermion generation.
      MSTP(1)=4
      
C...Set Z'0 (32) mass to 2*gravitino mass.
      PMAS(32,1)=2*GMASS
      
C...Set nu'_tau (18) mass to gravitino mass.
      PMAS(18,1)=GMASS
      
C...Allow only Z'0 production e+e- --> Z'0.
      MSEL=21
      
C...Allow decays of nu'_taus and Z'0s.
      MDCY(18,1)=1
      MDCY(32,1)=1
      
C...Turn off all decay channels of the Z'0.
      END=MDCY(32,2)+MDCY(32,3)-1
C...Note: MDCY(32,2)=289, MDCY(32,2)+MDCY(32,3)-1=310
      DO 300 IDC=MDCY(32,2),END
      MDME(IDC,1)=-1
      BRAT(IDC)=0
  300 CONTINUE

C...Turn on Z'0 --> nu'_tau nu'_taubar decay channel.
      MDME(MDCY(32,2)+15,1)=1
      BRAT(MDCY(32,2)+15)=1
C...Note: MDCY(32,2)+15=304

C...Define decay channel nu'_tau --> Z0 (23) and nu_tau (16).
      KFDP(MDCY(18,2),1)=23
      KFDP(MDCY(18,2),2)=16
      
C...Set branching ratio nu'_tau --> Z0 nu_tau = 1.
      MDME(MDCY(18,2),2)=0
      BRAT(MDCY(18,2))=1
      
C...Turn off other decay channels of the nu'_tau.
      END2=MDCY(18,2)+MDCY(18,3)-1
C...Note: MDCY(18,2)=150, MDCY(18,2)+MDCY(18,3)-1=152.
      DO 200 IDC=MDCY(18,2)+1,END2
      MDME(IDC,1)=-1
      BRAT(IDC)=0
      MDME(IDC,2)=102
  200 CONTINUE

C...Add new decay channel for the neutron.
      MDCY(PYCOMP(2112),1)=1
      MDCY(PYCOMP(2112),2)=7000
      MDCY(PYCOMP(2112),3)=1
      
C...Define neutron decay channel n --> p e- nu_e.
      KFDP(MDCY(PYCOMP(2112),2),1)=2212
      KFDP(MDCY(PYCOMP(2112),2),2)=11
      KFDP(MDCY(PYCOMP(2112),2),3)=12
      MDME(7000,1)=1
      BRAT(7000)=1
      
C...Increase neutron mass slightly so that PYTHIA doesn't complain.
      PMAS(PYCOMP(2112),1)=0.942
      
C...Turn off initial state radiation.
      MSUB(1)=1
      MSTP(11)=0
      
C...Allow (21) and force (22) all unstable particles to decay.
      MSTJ(21)=1
      MSTJ(22)=1
      
C...Allow decays of muons(13), pions(211), K_L0s(130) and K+-s(321).
      MDCY(13,1)=1
      MDCY(PYCOMP(130),1)=1
      MDCY(PYCOMP(211),1)=1
      MDCY(PYCOMP(321),1)=1
      
C...Allow neutron decay.
      MDCY(PYCOMP(2112),1)=1
      
C...Uncomment to show a list of decay properties.
C      CALL PYSTAT(2)

C...Uncomment to show a list of decay channels.
C      CALL PYLIST(12)

C...Initialization of the event generation procedure.
      CALL PYINIT('CMS','e+','e-',ECM)
      
C...Generation of events.
      DO 100 NEV=1,NEVENTS
      CALL PYEVNT
      
C...Uncomment to show a summary of the event listing.
C      CALL PYLIST(1)

C...Print the energies of final state neutrinos and antineutrinos: 
nu_e=12,-12, nu_mu=14,-14, nu_tau=16,-16.
      DO 400 I=1,N
         IF(K(I,1).EQ.1.AND.((K(I,2).EQ.16).OR.(K(I,2).EQ.-16))) 
         THEN PRINT *,P(I,4)
      END IF
  400 CONTINUE
  100 CONTINUE

      END
 \end{verbatim} 
\end{small} 

}

\phantomsection
\addcontentsline{toc}{chapter}{Acknowledgements}
\chapter*{Acknowledgements}
First, let me thank my supervisor Laura Covi who suggested this work on the exciting topic of gravitino dark matter. I am grateful for good advice, continuous and valuable support and her excellent mentoring throughout the entire period I worked on this thesis. 

Furthermore, I would like to thank Jan Louis for acting as second examiner for this thesis, who thereby gave me the opportunity to carry out this thesis at the DESY Theory Group. 

Many thanks also to Alejandro Ibarra and David Tran for the very pleasant collaboration on the topic of gravitino dark matter and many fertile discussions. 

In addition, I would like to thank my office colleagues Vladimir Mitev and Jasper Hasenkamp for many profound discussions and an enjoyable time at DESY. 

Moreover, let me thank my brother Christian for valuable advice and for proofreeding this work. 

Finally, I am deeply grateful to my parents and my girlfriend Melanie who are always there for me and unremittingly supported me during my years of study. They made this work possible in the first place.  

\newpage

\phantomsection
\addcontentsline{toc}{chapter}{Revision History}
\chapter*{Revision History}
\begin{tabbing}
 \hspace{4cm} \= \kill
 September 2008 \> Version handed in at the Universität Hamburg. \\[3mm]
 December 2008 \> Correction of typographic errors and minor mistakes. 
\end{tabbing}

\newpage

\phantomsection
\addcontentsline{toc}{chapter}{Bibliography}
\bibliography{references}

\providecommand{\href}[2]{#2}\begingroup\raggedright\begin{thebibliography}{10}

\bibitem{Bertone:2004pz}
G.~Bertone, D.~Hooper, and J.~Silk, ``{Particle Dark Matter: Evidence,
  Candidates and Constraints},''
  \href{http://dx.doi.org/10.1016/j.physrep.2004.08.031}{{\em Phys. Rept.} {\bf
  405} (2005)  279--390},
\href{http://arxiv.org/abs/hep-ph/0404175}{{\tt arXiv:hep-ph/0404175}}.

\bibitem{Pagels:1981ke}
H.~Pagels and J.~R. Primack, ``{Supersymmetry, Cosmology and New TeV
  Physics},''
\href{http://dx.doi.org/10.1103/PhysRevLett.48.223}{{\em Phys. Rev. Lett.} {\bf
  48} (1982)  223}.

\bibitem{Sarkar:1995dd}
S.~Sarkar, ``{Big Bang Nucleosynthesis and Physics beyond the Standard
  Model},'' \href{http://dx.doi.org/10.1088/0034-4885/59/12/001}{{\em Rept.
  Prog. Phys.} {\bf 59} (1996)  1493--1610},
\href{http://arxiv.org/abs/hep-ph/9602260}{{\tt arXiv:hep-ph/9602260}}.

\bibitem{Buchmuller:2007ui}
W.~Buchmüller, L.~Covi, K.~Hamaguchi, A.~Ibarra, and T.~Yanagida, ``{Gravitino
  Dark Matter in R-Parity Breaking Vacua},''
  \href{http://dx.doi.org/10.1088/1126-6708/2007/03/037}{{\em JHEP} {\bf 03}
  (2007)  037},
\href{http://arxiv.org/abs/hep-ph/0702184}{{\tt arXiv:hep-ph/0702184}}.

\bibitem{Ibarra:2007wg}
A.~Ibarra and D.~Tran, ``{Gamma Ray Spectrum from Gravitino Dark Matter
  Decay},'' \href{http://dx.doi.org/10.1103/PhysRevLett.100.061301}{{\em Phys.
  Rev. Lett.} {\bf 100} (2008)  061301},
\href{http://arxiv.org/abs/0709.4593}{{\tt arXiv:0709.4593 [astro-ph]}}.

\bibitem{Ibarra:2008qg}
A.~Ibarra and D.~Tran, ``{Antimatter Signatures of Gravitino Dark Matter
  Decay},'' \href{http://dx.doi.org/10.1088/1475-7516/2008/07/002}{{\em JCAP}
  {\bf 0807} (2008)  002},
\href{http://arxiv.org/abs/0804.4596}{{\tt arXiv:0804.4596 [astro-ph]}}.

\bibitem{Ishiwata:2008cu}
K.~Ishiwata, S.~Matsumoto, and T.~Moroi, ``{High Energy Cosmic Rays from the
  Decay of Gravitino Dark Matter},''
  \href{http://dx.doi.org/10.1103/PhysRevD.78.063505}{{\em Phys.Rev.} {\bf D78}
  (2008)  063505},
\href{http://arxiv.org/abs/0805.1133}{{\tt arXiv:0805.1133 [hep-ph]}}.

\bibitem{Strong:2004ry}
A.~W. Strong, I.~V. Moskalenko, and O.~Reimer, ``{A New Determination of the
  Extragalactic Diffuse Gamma-Ray Background from EGRET Data},''
  \href{http://dx.doi.org/10.1086/423196}{{\em Astrophys. J.} {\bf 613} (2004)
  956--961},
\href{http://arxiv.org/abs/astro-ph/0405441}{{\tt arXiv:astro-ph/0405441}}.

\bibitem{Barwick:1997ig}
{\bf HEAT} Collaboration, S.~W. Barwick {\em et al.}, ``{Measurements of the
  Cosmic-Ray Positron Fraction from 1- GeV to 50-GeV},''
  \href{http://dx.doi.org/10.1086/310706}{{\em Astrophys. J.} {\bf 482} (1997)
  L191--L194},
\href{http://arxiv.org/abs/astro-ph/9703192}{{\tt arXiv:astro-ph/9703192}}.

\bibitem{Covi:2008jy}
L.~Covi, M.~Grefe, A.~Ibarra, and D.~Tran, ``{Unstable Gravitino Dark Matter
  and Neutrino Flux},''
  \href{http://dx.doi.org/10.1088/1475-7516/2009/01/029}{{\em JCAP} {\bf 0901}
  (2009)  029},
\href{http://arxiv.org/abs/0809.5030}{{\tt arXiv:0809.5030 [hep-ph]}}.

\bibitem{Trodden:2004st}
M.~Trodden and S.~M. Carroll, ``{TASI Lectures: Introduction to Cosmology},''
\href{http://arxiv.org/abs/astro-ph/0401547}{{\tt arXiv:astro-ph/0401547}}.

\bibitem{Sakharov:1967dj}
A.~D. Sakharov, ``{Violation of CP Invariance, C Asymmetry, and Baryon
  Asymmetry of the Universe},''
{\em Pisma Zh. Eksp. Teor. Fiz.} {\bf 5} (1967)  32--35.

\bibitem{Fukugita:1986hr}
M.~Fukugita and T.~Yanagida, ``{Baryogenesis without Grand Unification},''
\href{http://dx.doi.org/10.1016/0370-2693(86)91126-3}{{\em Phys. Lett.} {\bf
  B174} (1986)  45}.

\bibitem{Barbier:2004ez}
R.~Barbier {\em et al.}, ``{R-Parity Violating Supersymmetry},''
  \href{http://dx.doi.org/10.1016/j.physrep.2005.08.006}{{\em Phys. Rept.} {\bf
  420} (2005)  1--202},
\href{http://arxiv.org/abs/hep-ph/0406039}{{\tt arXiv:hep-ph/0406039}}.

\bibitem{Davidson:2002qv}
S.~Davidson and A.~Ibarra, ``{A Lower Bound on the Right-Handed Neutrino Mass
  from Leptogenesis},''
  \href{http://dx.doi.org/10.1016/S0370-2693(02)01735-5}{{\em Phys. Lett.} {\bf
  B535} (2002)  25--32},
\href{http://arxiv.org/abs/hep-ph/0202239}{{\tt arXiv:hep-ph/0202239}}.

\bibitem{Buchmuller:2004nz}
W.~Buchmüller, P.~Di~Bari, and M.~Plumacher, ``{Leptogenesis for
  Pedestrians},'' \href{http://dx.doi.org/10.1016/j.aop.2004.02.003}{{\em Ann.
  Phys.} {\bf 315} (2005)  305--351},
\href{http://arxiv.org/abs/hep-ph/0401240}{{\tt arXiv:hep-ph/0401240}}.

\bibitem{Amsler:2008zz}
{\bf Particle Data Group} Collaboration, C.~Amsler {\em et al.}, ``{Review of
  Particle Physics},''
  \href{http://dx.doi.org/10.1016/j.physletb.2008.07.018}{{\em Phys. Lett.}
  {\bf B667} (2008)  1}.
Available at \url{http://pdg.lbl.gov}.

\bibitem{Begeman:1991iy}
K.~G. Begeman, A.~H. Broeils, and R.~H. Sanders, ``{Extended Rotation Curves of
  Spiral Galaxies: Dark Haloes and Modified Dynamics},''
{\em Mon. Not. Roy. Astron. Soc.} {\bf 249} (1991)  523.

\bibitem{Navarro:1995iw}
J.~F. Navarro, C.~S. Frenk, and S.~D.~M. White, ``{The Structure of Cold Dark
  Matter Halos},'' \href{http://dx.doi.org/10.1086/177173}{{\em Astrophys. J.}
  {\bf 462} (1996)  563--575},
\href{http://arxiv.org/abs/astro-ph/9508025}{{\tt arXiv:astro-ph/9508025}}.

\bibitem{Moore:1999gc}
B.~Moore, T.~R. Quinn, F.~Governato, J.~Stadel, and G.~Lake, ``{Cold Collapse
  and the Core Catastrophe},''
  \href{http://dx.doi.org/10.1046/j.1365-8711.1999.03039.x}{{\em Mon. Not. Roy.
  Astron. Soc.} {\bf 310} (1999)  1147--1152},
\href{http://arxiv.org/abs/astro-ph/9903164}{{\tt arXiv:astro-ph/9903164}}.

\bibitem{Kravtsov:1997dp}
A.~V. Kravtsov, A.~A. Klypin, J.~S. Bullock, and J.~R. Primack, ``{The Cores of
  Dark Matter Dominated Galaxies: Theory vs. Observations},''
  \href{http://dx.doi.org/10.1086/305884}{{\em Astrophys. J.} {\bf 502} (1998)
  48},
\href{http://arxiv.org/abs/astro-ph/9708176}{{\tt arXiv:astro-ph/9708176}}.

\bibitem{Mambrini:2005vk}
Y.~Mambrini, C.~Munoz, E.~Nezri, and F.~Prada, ``{Adiabatic Compression and
  Indirect Detection of Supersymmetric Dark Matter},''
  \href{http://dx.doi.org/10.1088/1475-7516/2006/01/010}{{\em JCAP} {\bf 0601}
  (2006)  010},
\href{http://arxiv.org/abs/hep-ph/0506204}{{\tt arXiv:hep-ph/0506204}}.

\bibitem{Jungman:1995df}
G.~Jungman, M.~Kamionkowski, and K.~Griest, ``{Supersymmetric Dark Matter},''
  \href{http://dx.doi.org/10.1016/0370-1573(95)00058-5}{{\em Phys. Rept.} {\bf
  267} (1996)  195--373},
\href{http://arxiv.org/abs/hep-ph/9506380}{{\tt arXiv:hep-ph/9506380}}.

\bibitem{Bertone:2007aw}
G.~Bertone, W.~Buchmüller, L.~Covi, and A.~Ibarra, ``{Gamma-Rays from Decaying
  Dark Matter},'' \href{http://dx.doi.org/10.1088/1475-7516/2007/11/003}{{\em
  JCAP} {\bf 0711} (2007)  003},
\href{http://arxiv.org/abs/0709.2299}{{\tt arXiv:0709.2299 [astro-ph]}}.

\bibitem{Martin:1997ns}
S.~P. Martin, ``{A Supersymmetry Primer},''
\href{http://arxiv.org/abs/hep-ph/9709356}{{\tt arXiv:hep-ph/9709356}}.

\bibitem{Chung:2003fi}
D.~J.~H. Chung {\em et al.}, ``{The Soft Supersymmetry-Breaking Lagrangian:
  Theory and Applications},''
  \href{http://dx.doi.org/10.1016/j.physrep.2004.08.032}{{\em Phys. Rept.} {\bf
  407} (2005)  1--203},
\href{http://arxiv.org/abs/hep-ph/0312378}{{\tt arXiv:hep-ph/0312378}}.

\bibitem{Collins:1989kn}
P.~D.~B. Collins, A.~D. Martin, and E.~J. Squires, ``{Particle Physics and
  Cosmology},''. New York, USA: Wiley (1989) 496p.

\bibitem{Nilles:1983ge}
H.~P. Nilles, ``{Supersymmetry, Supergravity and Particle Physics},''
\href{http://dx.doi.org/10.1016/0370-1573(84)90008-5}{{\em Phys. Rept.} {\bf
  110} (1984)  1--162}.

\bibitem{Pradler:2007ne}
J.~Pradler, ``{Electroweak Contributions to Thermal Gravitino Production},''
  \href{http://arxiv.org/abs/0708.2786}{{\tt arXiv:0708.2786 [hep-ph]}}.
(2007).

\bibitem{Bolz:2000xi}
M.~Bolz, ``{Thermal Production of Gravitinos},''. DESY-THESIS-2000-013, (2000).

\bibitem{Moroi:1995fs}
T.~Moroi, ``{Effects of the Gravitino on the Inflationary Universe},''
  \href{http://arxiv.org/abs/hep-ph/9503210}{{\tt arXiv:hep-ph/9503210}}.
(1995).

\bibitem{Haag:1974qh}
R.~Haag, J.~T. Lopuszanski, and M.~Sohnius, ``{All Possible Generators of
  Supersymmetries of the S Matrix},''
\href{http://dx.doi.org/10.1016/0550-3213(75)90279-5}{{\em Nucl. Phys.} {\bf
  B88} (1975)  257}.

\bibitem{Campbell:1990fa}
B.~A. Campbell, S.~Davidson, J.~R. Ellis, and K.~A. Olive, ``{Cosmological
  Baryon Asymmetry Constraints on Extensions of the Standard Model},''
\href{http://dx.doi.org/10.1016/0370-2693(91)91795-W}{{\em Phys. Lett.} {\bf
  B256} (1991)  484--490}.

\bibitem{Fischler:1990gn}
W.~Fischler, G.~F. Giudice, R.~G. Leigh, and S.~Paban, ``{Constraints on the
  Baryogenesis Scale from Neutrino Masses},''
\href{http://dx.doi.org/10.1016/0370-2693(91)91207-C}{{\em Phys. Lett.} {\bf
  B258} (1991)  45--48}.

\bibitem{Dreiner:1992vm}
H.~K. Dreiner and G.~G. Ross, ``{Sphaleron Erasure of Primordial
  Baryogenesis},'' \href{http://dx.doi.org/10.1016/0550-3213(93)90579-E}{{\em
  Nucl. Phys.} {\bf B410} (1993)  188--216},
\href{http://arxiv.org/abs/hep-ph/9207221}{{\tt arXiv:hep-ph/9207221}}.

\bibitem{Gunion:1984yn}
J.~F. Gunion and H.~E. Haber, ``{Higgs Bosons in Supersymmetric Models. 1},''
\href{http://dx.doi.org/10.1016/0550-3213(86)90340-8}{{\em Nucl. Phys.} {\bf
  B272} (1986)  1}.

\bibitem{Steffen:2007sp}
F.~D. Steffen, ``{Supersymmetric Dark Matter Candidates - The Lightest
  Neutralino, the Gravitino, and the Axino},''
\href{http://arxiv.org/abs/0711.1240}{{\tt arXiv:0711.1240 [hep-ph]}}.

\bibitem{Bolz:2000fu}
M.~Bolz, A.~Brandenburg, and W.~Buchmüller, ``{Thermal Production of
  Gravitinos},'' \href{http://dx.doi.org/10.1016/S0550-3213(01)00132-8}{{\em
  Nucl. Phys.} {\bf B606} (2001)  518--544},
\href{http://arxiv.org/abs/hep-ph/0012052}{{\tt arXiv:hep-ph/0012052}}.

\bibitem{Rarita:1941mf}
W.~Rarita and J.~S. Schwinger, ``{On a Theory of Particles with Half Integral
  Spin},''
\href{http://dx.doi.org/10.1103/PhysRev.60.61}{{\em Phys. Rev.} {\bf 60} (1941)
   61}.

\bibitem{Auvil:1966iu}
P.~R. Auvil and J.~J. Brehm, ``{Wave Functions for Particles of Higher Spin},''
  \href{http://dx.doi.org/10.1103/PhysRev.145.1152}{{\em Phys. Rev.} {\bf 145}
  (1966)  1152--1153}.

\bibitem{Denner:1992vza}
A.~Denner, H.~Eck, O.~Hahn, and J.~Küblbeck, ``{Feynman Rules for Fermion
  Number Violating Interactions},''
\href{http://dx.doi.org/10.1016/0550-3213(92)90169-C}{{\em Nucl. Phys.} {\bf
  B387} (1992)  467--484}.

\bibitem{Weinberg:1982zq}
S.~Weinberg, ``{Cosmological Constraints on the Scale of Supersymmetry
  Breaking},''
\href{http://dx.doi.org/10.1103/PhysRevLett.48.1303}{{\em Phys. Rev. Lett.}
  {\bf 48} (1982)  1303}.

\bibitem{Kawasaki:2004qu}
M.~Kawasaki, K.~Kohri, and T.~Moroi, ``{Big-Bang Nucleosynthesis and Hadronic
  Decay of Long-Lived Massive Particles},''
  \href{http://dx.doi.org/10.1103/PhysRevD.71.083502}{{\em Phys. Rev.} {\bf
  D71} (2005)  083502},
\href{http://arxiv.org/abs/astro-ph/0408426}{{\tt arXiv:astro-ph/0408426}}.

\bibitem{Hamaguchi:2007mp}
K.~Hamaguchi, T.~Hatsuda, M.~Kamimura, Y.~Kino, and T.~T. Yanagida,
  ``{Stau-Catalyzed Li-6 Production in Big-Bang Nucleosynthesis},''
  \href{http://dx.doi.org/10.1016/j.physletb.2007.05.030}{{\em Phys. Lett.}
  {\bf B650} (2007)  268--274},
\href{http://arxiv.org/abs/hep-ph/0702274}{{\tt arXiv:hep-ph/0702274}}.

\bibitem{Kanzaki:2006hm}
T.~Kanzaki, M.~Kawasaki, K.~Kohri, and T.~Moroi, ``{Cosmological Constraints on
  Gravitino LSP Scenario with Sneutrino NLSP},''
  \href{http://dx.doi.org/10.1103/PhysRevD.75.025011}{{\em Phys. Rev.} {\bf
  D75} (2007)  025011},
\href{http://arxiv.org/abs/hep-ph/0609246}{{\tt arXiv:hep-ph/0609246}}.

\bibitem{DiazCruz:2007fc}
J.~L. Diaz-Cruz, J.~R. Ellis, K.~A. Olive, and Y.~Santoso, ``{On the
  Feasibility of a Stop NLSP in Gravitino Dark Matter Scenarios},''
  \href{http://dx.doi.org/10.1088/1126-6708/2007/05/003}{{\em JHEP} {\bf 05}
  (2007)  003},
\href{http://arxiv.org/abs/hep-ph/0701229}{{\tt arXiv:hep-ph/0701229}}.

\bibitem{Takayama:2000uz}
F.~Takayama and M.~Yamaguchi, ``{Gravitino Dark Matter without R-Parity},''
  \href{http://dx.doi.org/10.1016/S0370-2693(00)00726-7}{{\em Phys. Lett.} {\bf
  B485} (2000)  388--392},
\href{http://arxiv.org/abs/hep-ph/0005214}{{\tt arXiv:hep-ph/0005214}}.

\bibitem{Lola:2007rw}
S.~Lola, P.~Osland, and A.~R. Raklev, ``{Radiative Gravitino Decays from
  R-Parity Violation},''
  \href{http://dx.doi.org/10.1016/j.physletb.2007.09.048}{{\em Phys. Lett.}
  {\bf B656} (2007)  83--90},
\href{http://arxiv.org/abs/0707.2510}{{\tt arXiv:0707.2510 [hep-ph]}}.

\bibitem{Mertig:1990an}
R.~Mertig, M.~Bohm, and A.~Denner, ``{FEYN CALC: Computer Algebraic Calculation
  of Feynman Amplitudes},''
\href{http://dx.doi.org/10.1016/0010-4655(91)90130-D}{{\em Comput. Phys.
  Commun.} {\bf 64} (1991)  345--359}.

\bibitem{Carena:2002es}
M.~S. Carena and H.~E. Haber, ``{Higgs Boson Theory and Phenomenology},''
  \href{http://dx.doi.org/10.1016/S0146-6410(02)00177-1}{{\em Prog. Part. Nucl.
  Phys.} {\bf 50} (2003)  63--152},
\href{http://arxiv.org/abs/hep-ph/0208209}{{\tt arXiv:hep-ph/0208209}}.

\bibitem{Sjostrand:2006za}
T.~Sjöstrand, S.~Mrenna, and P.~Skands, ``{PYTHIA 6.4 Physics and Manual},''
  \href{http://dx.doi.org/10.1088/1126-6708/2006/05/026}{{\em JHEP} {\bf 05}
  (2006)  026},
\href{http://arxiv.org/abs/hep-ph/0603175}{{\tt arXiv:hep-ph/0603175}}.

\bibitem{Tran:2008zz}
D.~Tran, ``{Indirect Signatures of Gravitino Dark Matter in Models with
  R-Parity Violation},''. (2008).

\bibitem{Strumia:2006db}
A.~Strumia and F.~Vissani, ``{Neutrino Masses and Mixings and...},''
\href{http://arxiv.org/abs/hep-ph/0606054}{{\tt arXiv:hep-ph/0606054}}.

\bibitem{Schwetz:2008er}
T.~Schwetz, M.~Tortola, and J.~W. Valle, ``{Three-Flavour Neutrino Oscillation
  Update},'' \href{http://dx.doi.org/10.1088/1367-2630/10/11/113011}{{\em New
  J.Phys.} {\bf 10} (2008)  113011},
\href{http://arxiv.org/abs/0808.2016}{{\tt arXiv:0808.2016 [hep-ph]}}.

\bibitem{Battistoni:1999at}
G.~Battistoni {\em et al.}, ``{A 3-Dimensional Calculation of Atmospheric
  Neutrino Flux},'' \href{http://dx.doi.org/10.1016/S0927-6505(99)00110-3}{{\em
  Astropart. Phys.} {\bf 12} (2000)  315--333},
\href{http://arxiv.org/abs/hep-ph/9907408}{{\tt arXiv:hep-ph/9907408}}.

\bibitem{Battistoni:2001url}
G.~Battistoni, A.~Ferrari, T.~Montaruli, and P.~R. Sala, ``{A New 3-Dimensional
  Calculation of Atmospheric Neutrino Fluxes Based on the FLUKA MC Code}.''
  {Available at} \url{http://www.mi.infn.it/~battist/neutrino.html}, Apr.,
  2001.

\bibitem{Battistoni:2003ju}
G.~Battistoni, A.~Ferrari, T.~Montaruli, and P.~R. Sala, ``{High Energy
  Extension of the FLUKA Atmospheric Neutrino Flux},''
\href{http://arxiv.org/abs/hep-ph/0305208}{{\tt arXiv:hep-ph/0305208}}.

\bibitem{Athar:2004uk}
H.~Athar and C.~S. Kim, ``{GeV to TeV Astrophysical Tau Neutrinos},''
  \href{http://dx.doi.org/10.1016/j.physletb.2004.07.022}{{\em Phys. Lett.}
  {\bf B598} (2004)  1--7},
\href{http://arxiv.org/abs/hep-ph/0407182}{{\tt arXiv:hep-ph/0407182}}.

\bibitem{Pasquali:1998xf}
L.~Pasquali and M.~H. Reno, ``{Tau Neutrino Fluxes from Atmospheric Charm},''
  \href{http://dx.doi.org/10.1103/PhysRevD.59.093003}{{\em Phys. Rev.} {\bf
  D59} (1999)  093003},
\href{http://arxiv.org/abs/hep-ph/9811268}{{\tt arXiv:hep-ph/9811268}}.

\bibitem{Ingelman:1996mj}
G.~Ingelman and M.~Thunman, ``{High Energy Neutrino Production by Cosmic Ray
  Interactions in the Sun},''
  \href{http://dx.doi.org/10.1103/PhysRevD.54.4385}{{\em Phys. Rev.} {\bf D54}
  (1996)  4385--4392},
\href{http://arxiv.org/abs/hep-ph/9604288}{{\tt arXiv:hep-ph/9604288}}.

\bibitem{Athar:2004um}
H.~Athar, F.-F. Lee, and G.-L. Lin, ``{Tau Neutrino Astronomy in GeV
  Energies},'' \href{http://dx.doi.org/10.1103/PhysRevD.71.103008}{{\em Phys.
  Rev.} {\bf D71} (2005)  103008},
\href{http://arxiv.org/abs/hep-ph/0407183}{{\tt arXiv:hep-ph/0407183}}.

\bibitem{fgst:2008url}
{Information on FGST available at} \url{http://fermi.gsfc.nasa.gov}.

\bibitem{GonzalezGarcia:2002dz}
M.~C. Gonzalez-Garcia and Y.~Nir, ``{Neutrino Masses and Mixing: Evidence and
  Implications},'' \href{http://dx.doi.org/10.1103/RevModPhys.75.345}{{\em Rev.
  Mod. Phys.} {\bf 75} (2003)  345--402},
\href{http://arxiv.org/abs/hep-ph/0202058}{{\tt arXiv:hep-ph/0202058}}.

\bibitem{Kato:2007re}
T.~Kato, ``{Tau Neutrino Appearance via Neutrino Oscillations in Atmospheric
  Neutrinos},''. (2007).

\bibitem{PalomaresRuiz:2007ry}
S.~Palomares-Ruiz, ``{Model-Independent Bound on the Dark Matter Lifetime},''
  \href{http://dx.doi.org/10.1016/j.physletb.2008.05.040}{{\em Phys. Lett.}
  {\bf B665} (2008)  50--53},
\href{http://arxiv.org/abs/0712.1937}{{\tt arXiv:0712.1937 [astro-ph]}}.

\bibitem{Abe:2006fu}
{\bf Super-Kamiokande} Collaboration, K.~Abe {\em et al.}, ``{A Measurement of
  Atmospheric Neutrino Flux Consistent with Tau Neutrino Appearance},''
  \href{http://dx.doi.org/10.1103/PhysRevLett.97.171801}{{\em Phys. Rev. Lett.}
  {\bf 97} (2006)  171801},
\href{http://arxiv.org/abs/hep-ex/0607059}{{\tt arXiv:hep-ex/0607059}}.

\bibitem{Collaboration:2007rk}
T.~I. Collaboration, ``{Contributions to The 10th International Conference on
  Topics in Astroparticle and Underground Physics (TAUP) 2007, Sendai, Japan,
  Sep. 11-15, 2007},''
\href{http://arxiv.org/abs/0712.3524}{{\tt arXiv:0712.3524 [astro-ph]}}.

\bibitem{Cowen:2007ny}
{\bf IceCube} Collaboration, D.~F. Cowen, ``{Tau Neutrinos in IceCube},''
\href{http://dx.doi.org/10.1088/1742-6596/60/1/048}{{\em J. Phys. Conf. Ser.}
  {\bf 60} (2007)  227--230}.

\bibitem{Pamela:2008url}
{Information on PAMELA available at} \url{http://pamela.roma2.infn.it}.

\bibitem{Peskin:1995ev}
M.~E. Peskin and D.~V. Schroeder, ``{An Introduction to Quantum Field
  Theory},''. Reading, USA: Addison-Wesley (1995) 842 p.

\bibitem{Nishi:2004st}
C.~C. Nishi, ``{Simple Derivation of General Fierz-Like Identities},''
  \href{http://dx.doi.org/10.1119/1.2074087}{{\em Am. J. Phys.} {\bf 73} (2005)
   1160--1163},
\href{http://arxiv.org/abs/hep-ph/0412245}{{\tt arXiv:hep-ph/0412245}}.

\bibitem{Lurie:1968pf}
D.~Luri\'{e}, ``{Particles and Fields},''. New York, USA: Interscience (1968)
  518p.

\end{thebibliography}\endgroup
\bibliographystyle{utphys}

\end{document}